\pdfoutput=1
\documentclass[pra,aps,twocolumn,notitlepage,superscriptaddress,10pt]{revtex4-1}
\usepackage{graphicx,graphicx,amsthm,amsmath,amssymb}
\usepackage{wasysym}
\usepackage{color}
\usepackage{bbold}
\usepackage[caption=false]{subfig}
\usepackage{hyperref}

 \definecolor{labelkey}{rgb}{0,0,1}
\definecolor{refkey}{rgb}{1,0,1}

\usepackage{geometry}
\def\ket#1{| #1 \rangle}

\def\CJJ{{\rm CJJ}}
\def\C{{\rm C}}
\def\J{{\rm J}}
\def\L{{\rm L}}
\def\LCJJ{{\rm L_{CJJ}}}
\def\CCJJ{{\rm C_{CJJ}}}
\def\GHz{\; {\rm GHz}}
\def\AFM{{\rm AFM}}
\def\M{{\rm M}}

\def\SB{{\rm SB}}
\def\MRT{{\rm MRT}}

\def\ss{\textsc{s}}

\def\({\left(}
\def\){\right)}

\begin{document}

\title{Computational Role of Collective Tunneling in a Quantum Annealer}

\author{Sergio Boixo\textsuperscript{1}, Vadim N. Smelyanskiy\textsuperscript{2}, Alireza Shabani\textsuperscript{1}, Sergei V. Isakov\textsuperscript{1}, Mark Dykman\textsuperscript{3}, Vasil S. Denchev\textsuperscript{1}, Mohammad Amin\textsuperscript{4}, Anatoly Smirnov\textsuperscript{4}, Masoud Mohseni\textsuperscript{1}, Hartmut Neven\textsuperscript{1}\\
  {\it \textsuperscript{1}Google, Venice, CA 90291, USA}\\{\it \textsuperscript{2}NASA Ames Research Center, Moffett Field, CA 94035, USA}\\{\it \textsuperscript{3}Department of Physics and Astronomy, Michigan State University, East Lansing, MI 48824, USA
}\\{\it \textsuperscript{4}D-Wave Systems Inc., Burnaby, BC V5C 6G9, Canada}}


\begin{abstract}
  Quantum tunneling is a phenomenon in which a quantum state traverses
  energy barriers above the energy of the state
  itself~\cite{Gamow_1928,fowler1928electron}. Tunneling has been hypothesized as an
  advantageous physical resource for optimization~\cite{Finnila_1994,
    Nishimori_1998,Brooks_99,farhi_quantum_2002,Santoro_2002}. Here we
  present the first experimental evidence of a computational role of
  multiqubit quantum tunneling in the evolution of a programmable
  quantum annealer. We develop a theoretical model based on a NIBA
  Quantum Master Equation to describe the multiqubit dissipative
  tunneling effects under the complex noise characteristics of such
  quantum devices. We start by considering a computational primitive,
  an optimization problem consisting of just one global and one false
  minimum. The quantum evolutions enable tunneling to the global
  minimum while the corresponding classical paths are trapped in a
  false minimum. In our study the non-convex potentials are realized
  by frustrated networks of qubit clusters with strong intra-cluster
  coupling. We show that the collective effect of the quantum
  environment is suppressed in the ``critical'' phase during the
  evolution where quantum tunneling ``decides'' the right path to
  solution. In a later stage dissipation facilitates the multiqubit
  tunneling leading to the solution state. The predictions of the
  model accurately describe the experimental data from the D-Wave Two
  quantum annealer at NASA Ames. In our computational primitive the
  temperature dependence of the probability of success in the quantum
  model is opposite to that of the classical paths with thermal
  hopping. Specifically, we provide an analysis of an optimization
  problem with sixteen qubits, demonstrating eight qubit tunneling
  that increases success probabilities. Furthermore, we report results
  for larger problems with up to 200 qubits that contain the primitive
  as subproblems.
\end{abstract}


\maketitle

\section{Introduction}

Quantum tunneling was discovered in the late 1920s to explain
radioactive decay~\cite{Gamow_1928} and field electron emission in vacuum
tubes~\cite{fowler1928electron}. Today this phenomenon is at the core of many
essential technological innovations such as the tunnel field-effect
transistor~\cite{seabaugh2010low}, field emission displays and the
scanning tunneling microscope~\cite{binnig2000scanning}. Tunneling
also plays a key role in energy and charge transport in biological and
chemical processes~\cite{mohseni2014quantum}. Recently, tunneling
effects involving multiple quantum mechanical particles have been used
to develop single electron transistors~\cite{klein1997single} and
hypersensitive measurement instruments.

Collective tunneling phenomenon plays a central role in switching
between stable states of molecular nanomagnets
\cite{Chudnovsky:1988,Leggett:1995,Sessoli:2003,burzuri_magnetic_2011}.  These bistable
units locally connected to each other are studied as building blocks
for magnetic cellular automata \cite{Imre:2006} and digital integrated
circuits \cite{Becherer:2014}. The tunneling behavior there is displayed
by ferromagnetic clusters with large spins (of size ten
\cite{Sangregorio:1997} and more \cite{Christou:2004}) formed by
individual ion spins moving as a whole under the collective energy
barrier due to a strong exchange forces between them.

Quantum tunneling, in particular for thin but high energy barriers,
has been hypothesized as an advantageous mechanism for quantum
optimization \cite{ray_sherrington-kirkpatrick_1989,Finnila_1994,
  Nishimori_1998,Brooks_99,farhi_quantum_2002,Santoro_2002}.  In
classical simulated annealing or cooling optimization algorithms, the
corresponding temperature parameter must be raised to overcome energy
barriers. But if there are many potential local minima with smaller
energy differences than the height of the barrier, the temperature
must also be lowered to distinguish between them so the algorithm can
converge to the global minimum. Quantum tunneling is present even at
zero-temperature. Therefore, for some energy landscapes, one might
expect that quantum dynamical evolutions can converge to the global
minimum faster than classical optimization algorithms. Quantum
annealing \cite{Finnila_1994,Nishimori_1998} is a technique inspired
by classical annealing but designed to take advantage of quantum
tunneling. Single qubit quantum tunneling for a programmable annealer
has been demonstrated experimentally in
Ref.~\cite{johnson2011quantum}, and two qubit tunneling has been
detected indirectly using microscopic resonant tunneling in
Ref.~\cite{lanting_cotunneling_2010}.

In the idealized limit of quantum annealing the dynamics of the system is
unitary and evolves adiabatically 
under a slowly varying time-dependent Hamiltonian.  The system will arrive at the
final ground state of the problem Hamiltonian if the total evolution
time is large compared to the inverse minimum energy gap along the Hamiltonian
path \cite{Farhi_2001}. In this paper, we shall analyze the
performance of a quantum annealing device with superconducting flux
qubits~\cite{harris_experimental_2010,harris_experimental_2010b,johnson2011quantum}. The
qubits are coupled inductively in a connectivity graph that is formed
by a grid of cells with high internal connectivity. The qubits are
subject to interaction with the environment with the dominant noise
source being spin diffusion at the superconductor insulator interface
\cite{Martinis:2005,Ioffe:2008,Paladino:2014}. This is known to
produce control errors, energy level broadening as well as thermal
excitation and relaxation \cite{Ao:1989,Kayanuma:1998}.  The noise
characteristics of individual qubits have been studied in macroscopic
resonant tunneling experiments \cite{PhysRevB.83.180502}. We show nevertheless that even under such conditions the device performance can benefit from multiqubit  tunneling of strongly interacting qubit clusters. This is of relevance
for current programmable quantum annealers, such as the D-Wave Two chip at NASA Ames. 

The performance of D-Wave's quantum annealers has been studied in a
number of recent
works~\cite{harris_experimental_2010b,johnson2011quantum,boixo2013experimental,dickson2013thermally,mcgeoch2013experimental,dash2013note,boixo_evidence_2014,lanting2014entanglement,santra2014max,ronnow_defining_2014,vinci2014hearing,shin2014quantum,vinci2014distinguishing,mcgeoch2014adiabatic,venturelli2014quantum,albash2014reexamining,king2014algorithm}. Results
from a D-Wave quantum annealer chip are very different from models
that do not quantize the superconducting flux qubits~\cite{johnson2011quantum,boixo2013experimental,boixo_evidence_2014}. It
has also been shown that under current noise parameters it is possible
to prepare entangled states of eight qubits, using static Hamiltonians
with a gap much bigger than the
temperature~\cite{lanting2014entanglement}. A good correlation with a
classical-path model~\cite{boulatov_quantum_2003} (see
Sec.~\ref{sec:svmc}) has been observed for a benchmark of random Ising
instances~\cite{shin2014quantum}, as well as differences in
distributions of excited states or degenerate ground
states~\cite{vinci2014distinguishing,albash2014reexamining}.

Reference~\cite{dickson2013thermally} makes the interesting
observation that for a problem instance with energy gap much smaller than the temperature (and
without false minima) the probability of success increases with temperature. In these conditions, noise effects are very
strong and they destroy coherent quantum superpositions. The system
actually resides in (classical) product states making random hoppings
between them. In contrast, we will introduce instances where fast
collective tunneling processes of many qubits give rise to correlated quantum
superposition states~\cite{lanting2014entanglement}.  Multiple qubit
 tunneling will play a significant role both in the formation of the
dynamical states themselves and in giving rise to a large transition
rate between these states. For these
instances we obtain an opposite temperature dependence behavior: decreasing
probability of success with increasing temperature. This is in contrast with the limit
of incoherent tunneling or to the classical-path model.

In this work we design an Ising model implementation with 16 qubits of
a computational primitive, the simplest non-convex optimization
problem consisting of just one global and one local minimum. The final
global minimum can only be reached by traversing an energy barrier. 
We develop a NIBA Quantum Master Equation which takes high and low
frequency noise into account. Our comprehensive open quantum system
modeling shows close agreement with experiments conducted using the 
D-Wave device and demonstrates how collective tunneling can exist and play a computational role in
the presence of both Ohmic and strong $1/f$ noise affecting flux qubit coherence. Quite generally, our model predicts that the
probability to find the system in the lowest energy state should
decrease with temperature for a quantum system and increase with temperature for a classical system. Consistent with the quantum model, we show that temperature and success probability are inversely related in a series of 16 qubit D-Wave experiments. We compare with alternative physically plausible models of the
hardware that only employ product states and do not allow for
multiqubit tunneling transitions.  Experimentally, we show that the D-Wave Two processor has a higher success probability than any of these models for a series of problems.
We also explore larger problems embedded on 200 qubits
that contain multiple weak-strong cluster pairs. We observe that the success
probabilities of quantum annealing outperform the models that, for
physically motivated parameter regimes, rely on
classical paths to the solution.

\section{A primitive ``probe'' problem}\label{sec:probe}

\subsection{The quantum Hamiltonian}

\begin{figure}[h]
  \centering
  \includegraphics[width=\columnwidth]{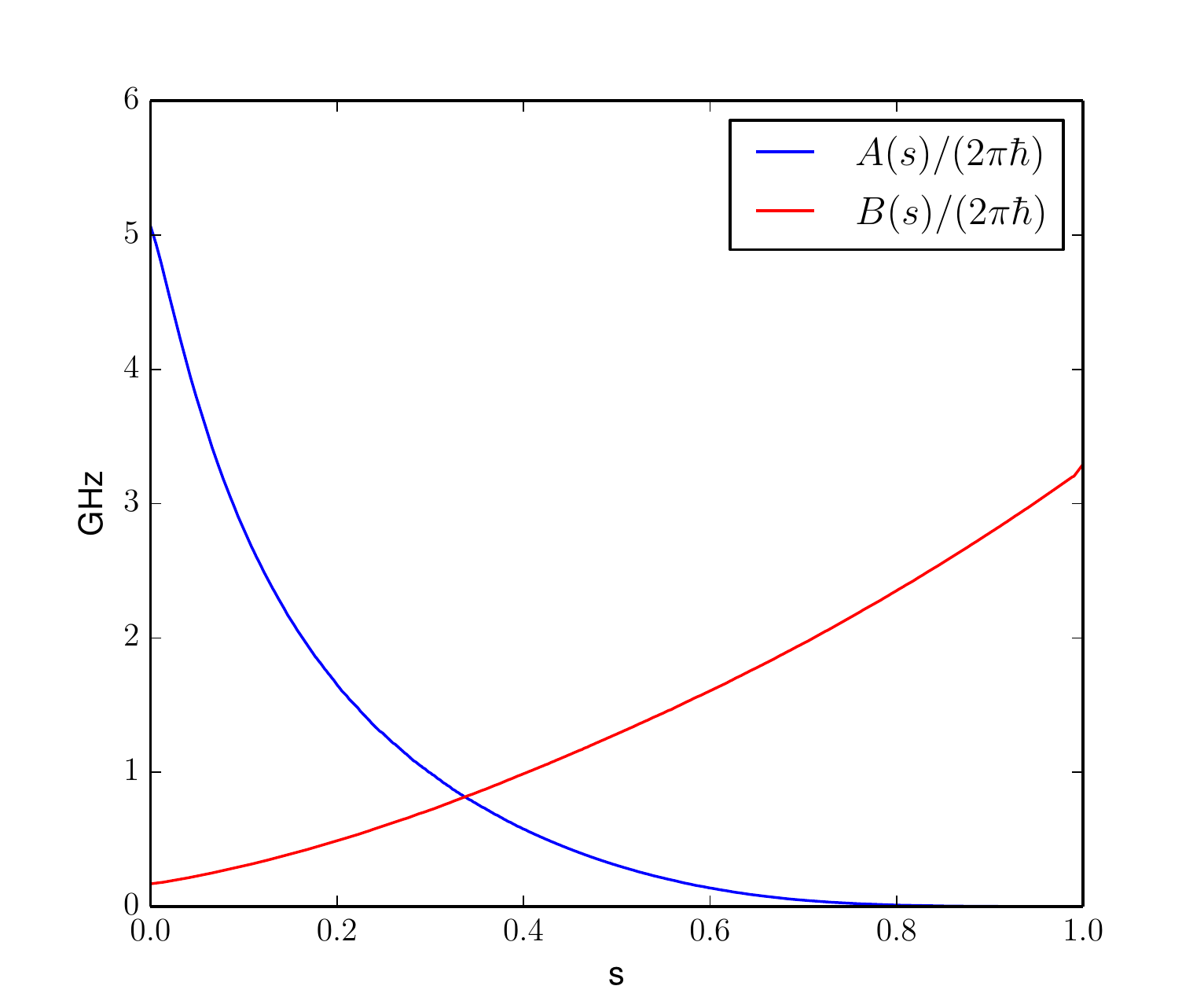}
  \caption{Quantum annealing functions $A(s)$ and $B(s)$. The function
    $A(t)$ is defined as half the median energy difference between the
    two lowest eigenstates of the experimentally superconducting flux
    qubit with zero bias. The function $B(s)$ is defined as 1.41 pico
    henries times the square of the persistent current $I_p^2(s)$, as
    explained in App.~\ref{sec:effective_potential}
    Eq.~\eqref{eq:h_annealing_explained}. }
  \label{fig:annealing_functions}
\end{figure}
The state evolution in transverse field quantum annealing is governed
by a time dependent Hamiltonian of the form~\cite{Farhi_2001}
\begin{align}\label{eq:h_annealing}
  H_0(s) &= A(s)H^{\rm D}  + B(s) H^{\rm P}\\
  H^{\rm D} &= - \sum_\mu \sigma_\mu^x \\
  H^{\rm P} &= - \sum_\mu h_\mu \sigma_\mu^z - \sum_{\mu\nu} J_{\mu\nu} \sigma_\mu^z \sigma_\nu^z\;.
\end{align}
Here $H^{\rm D}$ is the driver Hamiltonian, $H^{\rm P}$ is the problem
Hamiltonian whose ground state is the solution of an optimization
problem of interest, $\{\sigma_\mu^x, \sigma_\mu^z\}$ are Pauli
matrices acting on spin $\mu$, $s=t/t_{qa}$ is the annealing
parameter, and $t_{qa}$ is the duration of the quantum annealing
process. The functions $A(s)$ and $B(s)$ used in the rest of the paper
are shown in Fig.~\ref{fig:annealing_functions}. The Hamiltonian path
$H_0(s)$ describes an evolution of effective 2-level spin systems
(qubits) from an initial phase with a unique ground state to a final
Hamiltonian with eigenstates aligned with the $z$ quantization
axis. In the initial unique ground state all the qubits are aligned
with the effective transverse magnetic field in the $x$ direction.
See Appendix~\ref{app:single_qubit_hamiltonian} for a more complete
derivation of the single qubit Hamiltonian and the parameters of the
experimental system considered in this paper.

\subsection{The weak-strong cluster ``probe'' Hamiltonian}\label{sec:probe_hamiltonian}

The canonical primitive to study quantum tunneling is a double-well
potential: two local minima separated by an energy barrier. Our aim is
to distinguish quantum tunneling from thermal activation in a model
using classical paths. Classical paths are limited to local spin
vector dynamics over product states to traverse the barrier.
In contrast, the signature of a quantum system is that entangled
states are available as well. We utilize qubit networks of the D-Wave
Two quantum annealer at NASA Ames to design time-dependent
asymmetric double-well potentials where a classical path continuously
connects the initial global minimum to the final false minimum. In
this way, one can study how the system escapes the local minimum and
traverses the energy barrier to reach the global optimum.  We will see
that quantum tunneling results in a different final probability of
success than the corresponding dynamics over classical
paths. We compare the experimental data from the device with numerical
simulations of these classical paths and with the predictions of a
comprehensive analytical model for dissipative multiqubit tunneling.
Based on the results of this comparison, we establish a functional
role of tunneling in the evolution of a programmable quantum annealer.

\begin{figure}[h]
  \centering
  \includegraphics[width=\columnwidth]{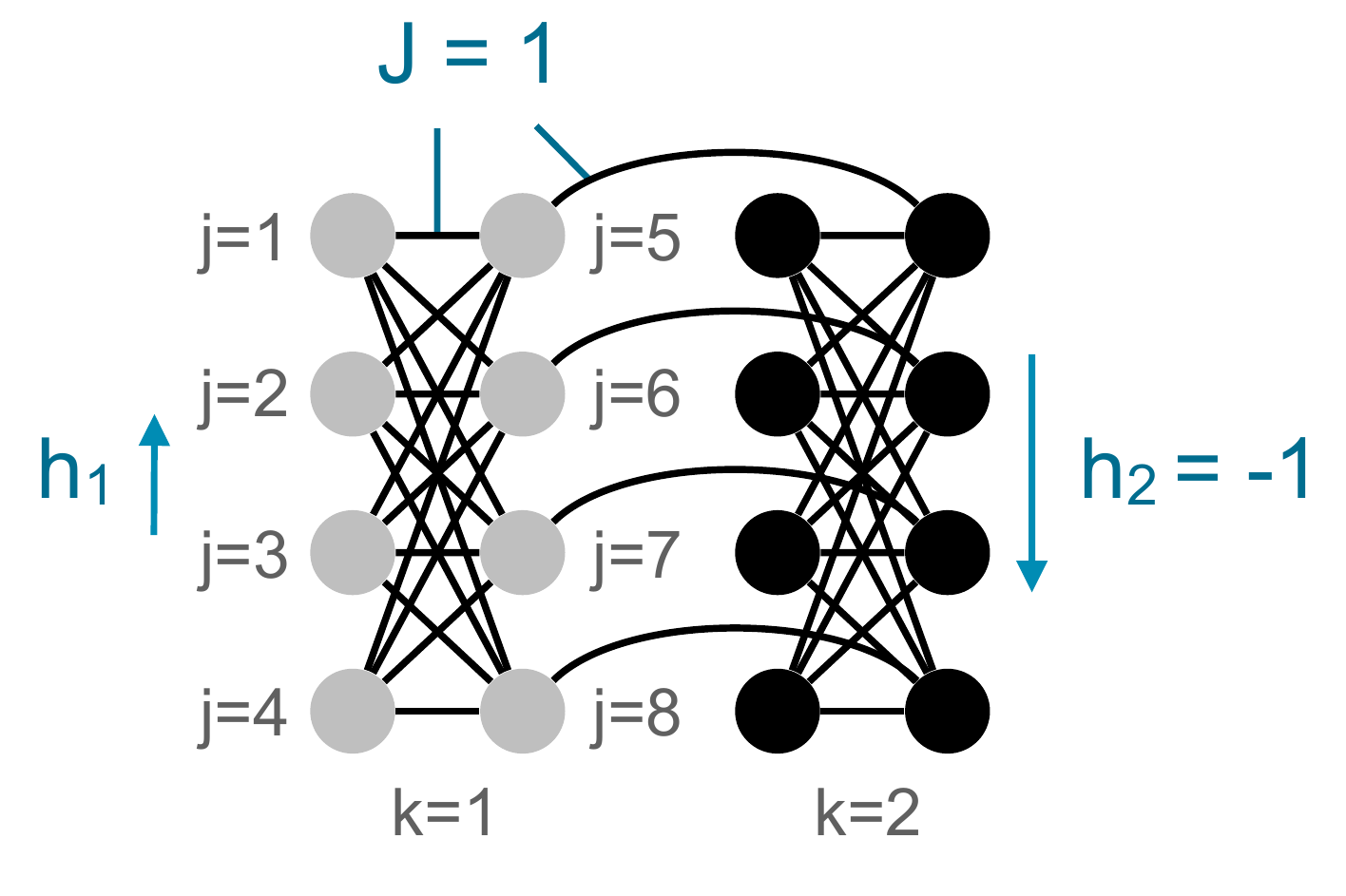}
  \caption{The probe problem under study, consisting of 16 qubits in
    two unit cells of the so-called Chimera graph. All qubits are
    ferromagnetically coupled and evolve as two distinct qubit
    clusters. At the end of the annealing evolution the right cluster
    is strongly pinned downward due to strong local fields acting on
    all qubits in that cell. However, the local magnetic field $h_1$ in the
    left cluster is weaker and serves as a bifurcation
    parameter. For $h_1 < J/2$ the left cluster will reverse its
    orientation during the annealing sweep and eventually align itself
    with the right cluster. Note the permutation symmetry in each
    column which allows us to adopt the large spin
    description. }
  \label{fig:Strong_Weak_Clusters}
\end{figure}
We now detail how the double-well potential and time evolution can be
constructed in the case of network graphs with finite connectivity. We
will focus on the particular case of the so-called Chimera graph that
connects the qubits in the current D-Wave Two architecture, although
similar constructions can be applied to more general network
architectures.  We choose our double-well primitive probe problem to
be the one depicted in Fig.~\ref{fig:Strong_Weak_Clusters}. We use two
Chimera cells, each with $n=8$ qubits. We find it useful during our
analysis to keep $n$ explicit. We will choose equal local fields for
the spins within each cell. We also choose all the couplings to be
equal and ferromagnetic. The problem Hamiltonian is of Ising
form
\begin{align}
H^{\rm P}&=H^{\rm P}_{1} + H^{\rm P}_{2} + H^{\rm P}_{1,2}\label{eq:Hp_total}\\
H^{\rm P}_{k}&= -J \sum_{\langle j, j' \rangle \in \rm intra}\sigma_{k,j}^{z}\sigma_{k,j\prime}^{z}-\sum_{j=1}^{n}h_{k}\,\sigma_{k,j}^{z} 
\label{eq:Hp1}\\
H^{\rm P}_{1,2}&= -J \sum_{j \in \rm inter} \sigma_{1,j}^{z} \sigma_{2,j}^{z} \;.   \label{eq:Hp2a}
\end{align} 
The index $k \in \{1,2\}$ denotes the Chimera cell, the first sum in (\ref{eq:Hp1})  goes over the \emph{intra}-cell couplings
depicted in Fig.~\ref{fig:Strong_Weak_Clusters}, and the second sum  goes over the \emph{inter}-cell couplings corresponding to $j\in(n/2+1,n)$ in Fig.~\ref{fig:Strong_Weak_Clusters}; $h_k$ denotes the local fields within each cell.

\begin{figure}[h]
  \centering
  \includegraphics[width=\columnwidth]{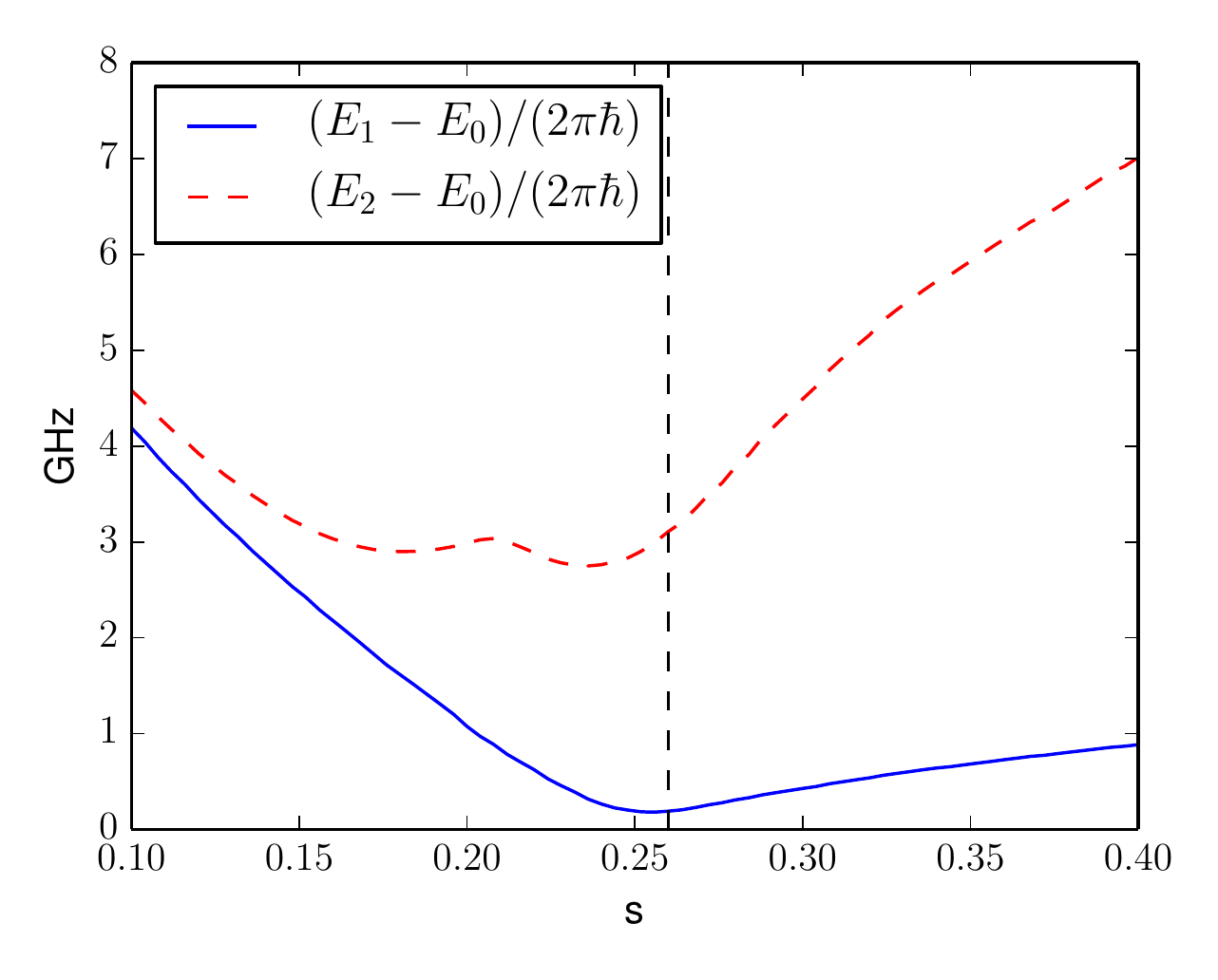}
  \caption{Gap of the quantum Hamiltonian for $h_1 = 0.44$ as a
    function of the annealing parameter. The continuous line is the
    energy difference between the ground state and the first excited
    state. The avoided crossing at $s=0.26$ (dashed vertical line) corresponds to a minimum
    gap of 180 MHz. The dashed line is the energy difference between the
      ground state and the second excited state.}
  \label{fig:gaps}
\end{figure}

Toward the end of quantum annealing the problem Hamiltonian $H^{\rm P}$ is dominating the evolution and $\langle \sigma^z_{k,j}\rangle
\simeq \pm 1$ where here $ \langle \ldots\rangle$ denotes a quantum mechanical average. There are $n^2/4$ \emph{intra}-cell couplings
and $n/2$ \emph{inter}-cell couplings. The spins within each cell tend
to move together as an homogenous cluster because flipping only one spin rises the energy by an
amount $\propto n J$ which is much greater than the energy of the
inter-cell bond $ \propto 2J$. Therefore the low lying states of $H_P$ correspond to the spins in each cluster pointing in the same direction (see Fig.~\ref{fig:gaps}).
In the first two states the second  cluster is pointing along its own local field   $h_2<0$ (largest in magnitude field).  The  difference in energy of the states  with both clusters  pointing  in the opposite  and the same   directions equals   
\begin{equation}
E_{\uparrow\downarrow}-E_{\downarrow\downarrow}=2n(J/2-h_1)\label{eq:hcond}
\end{equation}
\noindent
 If $h_1< J/2$ then the global minimum corresponds to both clusters having the same orientation as the largest magnitude field, $h_2$.
 The next energy level  (a ``false'' minimum) corresponds to the clusters oriented in the opposite directions (each along its own local field).

We now explain the onset of frustration in this system. We 
observe that at the beginning of quantum annealing
\begin{equation}
\langle \sigma^z_{k,j}\rangle \simeq  h_k\, B(s)/A(s),\qquad B(s)/A(s)
\ll 1 \label{eq:sgnz} \;.
\end{equation}
The Ising coupling terms in the problem Hamiltonian (\ref{eq:Hp1}) are
quadratic while the local field terms are linear in z-polarizations.
Therefore at the beginning of quantum annealing the effect of the local
z-fields dominates that of the \emph{inter}-cell Ising couplings.  According to
(\ref{eq:sgnz}), because $h_1$ and $h_2$ have the opposite signs so
will the z-projections of the spins $\langle \sigma^z_{k,j}\rangle$ in
the two clusters early in the evolution.
 
A key observation is that in the absence of quantum tunneling and
thermal hopping, the spin projection of the two clusters stay opposite
during the evolution.  The system would arrive to the false minimum with residual
energy relative to the global minimum equal to $n (J-2h_1)$; i.e. it will get trapped in the false minimum. To escape the false trap, all spins
in the left cluster must flip sign, which requires traversing the
barrier. At its peak, the barrier trapping the left cluster has
zero total z-polarization and therefore the barrier grows with the
ferromagnetic energy of the cluster $(n/2)^2 J$. For sufficiently
large $n$, the barrier height ${\cal O}(n^2)$ is much greater than the
residual energy ${\cal O}(n)$. It will be shown below that (for certain values of the annealing parameter $s$), all qubits in the left cluster will tunnel in a concerted motion under the energy barrier separating
the two potential wells that correspond to the opposite
z-polarizations of the cluster.

Next, we discuss an approximation which reduces the size of the
Hamiltonian matrix for the 2 unit cell problem from $2^{2n}$ to
$(n/4+1)^4$. We introduce total spin operators for each column of a
unit cell (cf. Fig.~\ref{fig:Strong_Weak_Clusters})
\begin{align}
  S_{k,1}^{\alpha}&= \frac{1}{2}\sum_{j=1}^{n/2}\sigma_{k,j}^{\alpha}\;,\quad
  S_{k,2}^{\alpha}= \frac{1}{2}\sum_{j=n/2+1}^{n}\sigma_{k,j}^{\alpha}\;,
 \end{align} 
where   $\alpha \in \{x,y,z\}$, and  $k\in \{1,2\}$ denotes the left and
right Chimera cells.  Because the intra-cell Hamiltonians
(\ref{eq:Hp1}) and the driver Hamiltonian are symmetric with respect to qubit   permutations they can be written in terms of the 
total spin operators 
 \begin{align}
H^{\rm P}_k &= - 4 J S^{z}_{k,1}S^{z}_{k,2} -2  h_{k} S_{k}^{z}\label{eq:SPD}\\
H^{\rm D} &= -2 \sum_{k,m=1,2}S_{k,m}^{x}\nonumber\;.
\end{align}
We note that the inter-cell Hamiltonian $H^{\rm P}_{1,2}$ in
Eq.~\eqref{eq:Hp2a}  does not possess the qubit permutation symmetry.
However, as explained above, the qubits in each cell tend to evolve as
homogenous clusters. Therefore one can approximate the inter-cell Hamiltonian in terms of the total spin operators for the columns 
\begin{equation}
H^{P}_{1,2}\simeq -\frac{8}{n} J S_{1,2}^{z}S_{2,2}^{z}\label{eq:S12}\;.
\end{equation}

We observe that the system Hamiltonian commutes with the total spin
operators $S_{k,m}^{2}=\sum_{\alpha\in\{x,y,z\}}(S_{k,m}^{\alpha})^2$.
Given that all qubits in the initial state are polarized along the
x-axis, this restricts the evolution to the subspace of maximum total
spin values $n/4$ for each column.  This subspace is spanned by the
basis vectors $|n/4,m_{k,m}\rangle$ corresponding to the definite
projections of column spins on the z axis $m_{k,m}=-n/4,\ldots,
n/4$. As a measure of the error incurred by this approximation, it can
be seen that the two lowest energy levels are within 0.1\% of the
exact values for the case of interest, $n=8$.

\begin{figure*}[t]
  \centering
  \includegraphics[width=0.28\textwidth]{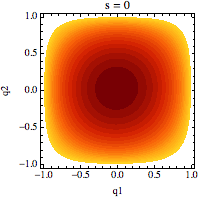}
  \includegraphics[width=0.28\textwidth]{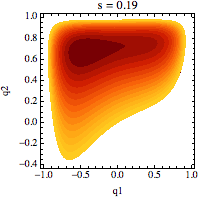}
  \includegraphics[width=0.28\textwidth]{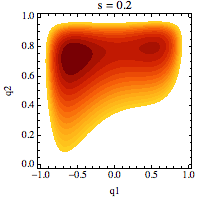}
  \includegraphics[width=0.28\textwidth]{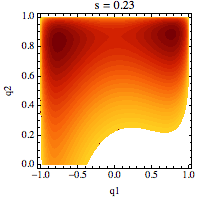}
  \includegraphics[width=0.28\textwidth]{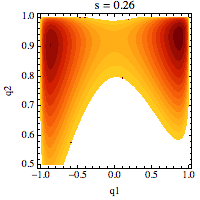}
  \includegraphics[width=0.28\textwidth]{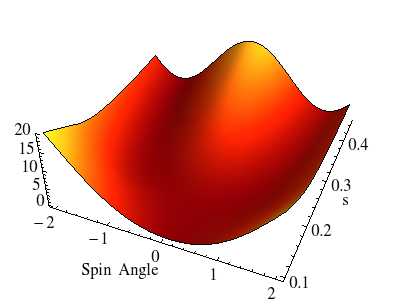}
  \caption{The five snapshots in this figure show how the energy
    landscape $U(q_1, q_2)$ evolves and a double-well potential is formed during the
    annealing schedule. The 3D plot also depicts such evolution as a
    function of an effective orientation angle for the large spins. The
    minimum that forms first would trap a classical particle moving in
    this potential. Later in the annealing evolution a second minimum
    forms and eventually becomes the global minimum. To reach this
    global minimum the system state has to traverse the energy barrier
    between them. The origin of this bifurcation is explained in the
    text.}
  \label{fig:V(q_{1,1},q_{2,1})}
\end{figure*}

\subsection{Effective energy potential for the classical paths of
  product states}\label{sec:effective_potential}

We now derive the effective potential over product states used to
study the difference between thermal hopping among classical paths, and
quantum tunneling in the quantum models.  This will also serve to
clarify the tunneling picture described above. We first transform
to a representation that contains an explicit momentum operator.
We think of each cluster $k$ as a spin-n/2 ferromagnetic 
``particle'' with the coordinate proportional to its total spin z
projection $\sum_{j=1,2} S_{k,j}^{z}$. Because the x-component of the
total spin of the cluster does not commute with the z-component it is
naturally associated with the momentum that causes the particle to
move.  This allows us to think of the z-component of a large spin as a
particle moving in a slowly time-varying potential, formalizing the
cartoon pictures sometimes drawn to illustrate quantum annealing that
show a particle escaping a local minimum in a continuous potential via
tunneling. This picture is also very similar to that of the dynamics of large spin magnetic moment  molecular materials \cite{Sessoli:2003}.

The canonically conjugate coordinate and momenta operators
 can be naturally introduced within the  WKB framework  (see App.~\ref{app:villain}) 
\begin{align}
  S_{k,1}^{z}+ S_{k,2}^{z}&=\frac{n}{2} q_k\\
 S_{k,1}^{x}+ S_{k,2}^{x}&\approx\frac{n}{2} \sqrt{1-q_k^2}\cos p_k\;
\end{align}
where $[q,p]= i (2/n)$ and $2/n\ll 1$  plays the role of Planck's constant in traditional WKB. 
To the leading order in $1/n$ the Hamiltonian becomes
\begin{align}
&H^{\rm WKB}(q_{1},q_{2},p_{1},p_{2} ,s)= - n A(s) \sum_{k=1,2}\sqrt{1-q_k^{2}} \cos p_k\nonumber\\
&-n B(s) J \sum_{k=1,2} \Big( h_{k} q_k  + n q_k^2/4\Big) - \frac n 2 B(s)  J \,q_1\, q_2 \;. \label{eq:h_villain}
\end{align}
The above Hamiltonian describes a pair of coupled ferromagnetic particles each with spin $n/2$. 
WKB theory  based on this Hamiltonian describes eigenstates and eigenvalues with logarithmic accuracy in the asymptotic limit $n\gg 1$. It also gives a reasonable  estimates already for $n=8$  (see App.~\ref{app:villain}). 

We will now consider the potential corresponding to a 
low energy description with very low momenta
\begin{equation}
U(q_1,q_2,s) =H^{\rm WKB}(q_{1},q_{2},0,0 ,s) \label{eq:U}
\end{equation}
The same potential is obtained in Ref.~\cite{farhi_quantum_2002}
projecting the Hamiltonian of large spin operators
Eqs.~\eqref{eq:SPD},~\eqref{eq:S12} over spin coherent
states, which are product states. 
The different panels in Figure~\ref{fig:V(q_{1,1},q_{2,1})} depict the
 potential $U(q_1,q_2)$  for
different values of the annealing parameter $s$ with local field
$h_1=0.44$. Initially ($s=0$) there is only a global minimum at $q_1 =q_2= 0$ corresponding to
all spins aligned with the x-direction.  As $s$ grows the minima begins to move to the left corner $(-1,1)$ corresponding to the opposite orientations of the clusters. This effect was previously mentioned in the general context.  The terms in the effective potential  corresponding to the
local fields $h_k$  are  linear in $q_k$, and dominate the Ising coupling energy (quadratic in $q_k$)
between the large spins for $|q_k|\ll 1$.
For larger values of $s$ the Ising terms begin to compete with the local fields and a plateau is formed in the vicinity of $q_1=0$ following the 
 local  bifurcation of $U$ and giving rise to a new minimum corresponding to the ferromagnetic alignment of the two clusters.
  At some value $s_c$ the two minima coexist with equal energy. For $s>s_c$  the minimum corresponding to the ferromagnetic cluster alignment  has a lower energy, smoothly connecting  to the solution state at the end of quantum annealing, corresponding to the global minimum of the potential $U(q_1,q_2,s=1)$ at $q_1=q_2=1$.

\begin{figure}[h]
  \centering
  \includegraphics[width=\columnwidth]{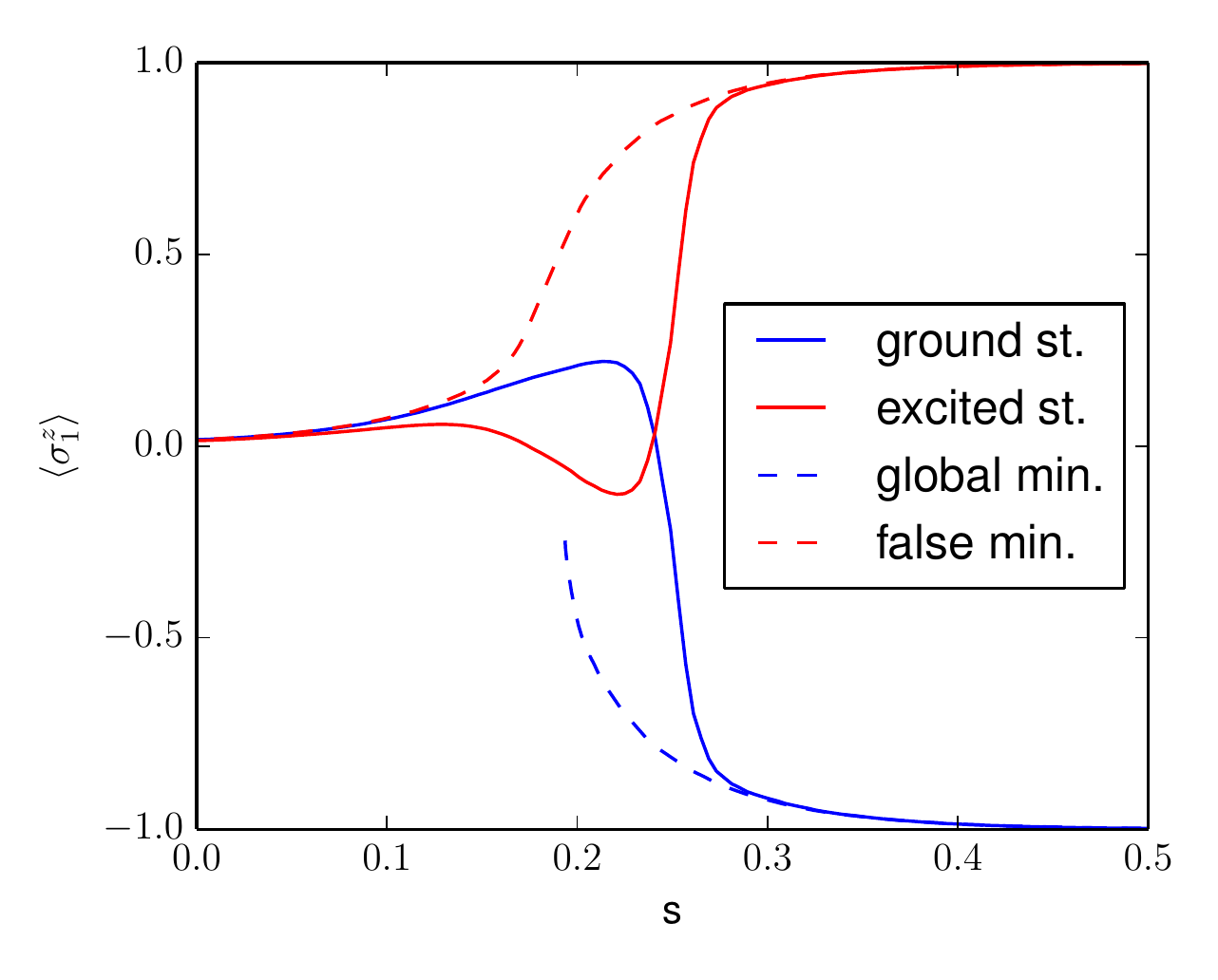}
  \caption{ Solid lines show the $z$-magnetization $\langle \sigma_1^z
    \rangle$ for the quantum ground state and first excited state as a
    function of the annealing parameter $s$ for $h_1 = 0.44$.  Dashed
    lines show the value of $q_1$ along the paths
    corresponding to the false and global minima for the effective
    energy potential over product states. The correspondence is good,
    except at the avoided crossing where the quantum states are
    entangled.}
  \label{fig:magnetization}
\end{figure}

In a closed system, the quantum adiabatic evolution algorithm allows the system to tunnel from the old global minimum to the new one in the vicinity of $s_c$  \cite{farhi_quantum_2002}. It tunnels under a barrier whose maximum approximately corresponds to zero z-magnetization $q_1=0$.
The tunneling  corresponds to an avoided crossing between the two lowest  energy levels of the Hamiltonian $H_0(s)$ of Eq.~\eqref{eq:h_annealing} shown in Fig.~\ref{fig:gaps}. During tunneling, the total  spin of the left  cluster switches its direction. The  switching manifests itself  in Fig.~\ref{fig:magnetization} as a steep change in $s$-dependence  of the quantum mechanical average of the left cluster polarization  $\langle q_1\rangle$ in the instantaneous ground state. In contrast, the right cluster, which does not tunnel, displays  a smooth change in its average polarization  $\langle q_2\rangle$.

\begin{figure}[h]
  \centering
  \includegraphics[width=\columnwidth]{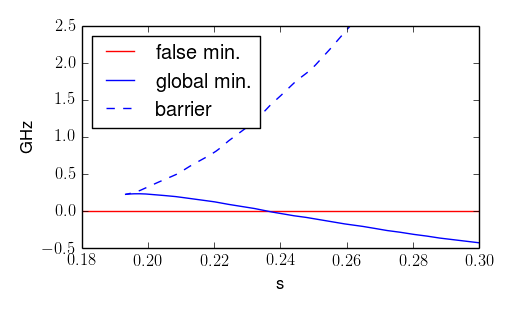}
  \caption{Bifurcation for $h_1=0.44$. The energies of the paths of
    the double-well minima and the height of the effective energy
    barrier are plotted in GHz and we have subtracted off the instantaneous
    energy of the final false minimum. The path corresponding to the
    global minimum appears as a bifurcation with higher energy. When
    this path crosses the path of the final false minimum, the
    height of the energy barrier is substantial.}
  \label{fig:bifurcation}
\end{figure}

In classical dynamical evolutions, when tunneling is not possible,
the system will continue to reside in the initial global minimum
emanating from the point $(0,0)$.  Fig.~\ref{fig:bifurcation} shows
the energies of the two classical paths corresponding to the two
minima of the energy potential $U(q_1, q_2, s)$ for
$h_1=0.44$.  Classical dynamical evolutions will get trapped in the
false minimum path due to the bifurcation seen in this figure. Classical trajectories can only reach
the global minimum through thermal excitations. In
Figure~\ref{fig:magnetization} we also show the value of the parameter
$q_1$ corresponding to the local minima of 
$U(q_1, q_2, s)$ as a function of the annealing parameter $s$.  The
parameter $q_1$ for the classical path connecting to the final false
minimum first aligns with the quantum ground state and later aligns
with the quantum first-excited state. This path starts at the
point $(q_1, q_2)=(0,0)$. The bifurcation path corresponding to the global
minimum first appears with a definite but small $q_1$, and later
aligns with the quantum ground state after the avoided crossing.

\section{Modeling the annealing dynamics}

\subsection{Characterization of noise and dissipation} \label{sec:noise_character}

Under realistic conditions, the performance of a quantum annealer as
an optimizer can be strongly influenced by the coupling to the
environment. In order to capture this effect  we present a {\it phenomenological} open
quantum system model by incorporating the experimental
characterization of the noise that was performed to date on D-Wave devices (the experimental platform for the present investigation).

We shall assume that each flux qubit is coupled to its own environment
with an independent noise source; this assumption is consistent with
experimental data ~\cite{lanting_cotunneling_2010}. We separate the
bath excitations into two parts. Excitations with frequencies lower
than the annealing rate in our experiments (5 KHz) will be treated as
a ``static noise" whose effect can be included by an appropriate
averaging of the success rate over the local field errors ($\sim 5\%$
for the D-Wave Two chip).  The excitations with higher frequencies
will be modeled as a bath of harmonic oscillators. This approach is
quite general and independent of the true physical source of the
noise. Its validity rests on the assumptions that the free bath
(decoupled from the qubits), is in thermal equilibrium, that it can be
treated within the conditions of linear response theory, and that it
has Gaussian fluctuations~\cite{Wilhelm:07}.  The corresponding
system-bath Hamiltonian is
\begin{equation}
H(s)=H_0(s)+\frac{1}{2}\sum_{\mu=1}^{2n} \sigma_\mu^z Q_\mu(s)+H_B,\quad s=t/t_{qa}.\label{eq:Hcont}
\end{equation}
Hereafter we will use for brevity a single index $\mu$ for
single-qubit Pauli matrices instead of the double indexation employed
in Sec.~\ref{sec:probe}.  In the equation above, $t_{qa}$ is the
duration of the quantum annealing process, $H_0(t)$ is the Hamiltonian
for the $2n$-qubit system (as in Eq.~\eqref{eq:h_annealing}), $H_B$ is
the standard Hamiltonian of the bosonic bath, and $Q_\mu(s)$ is a
bosonic noise operator that couples the $\mu^{\rm th}$ qubit to its
environment. The coupling parameters of bosonic bath operator
$Q_\mu(s)$ depend on the annealing parameter $s$ through the
persistent current (see App.~\ref{app:single_qubit_hamiltonian}). In
what follows we will often omit  the argument $s$  in  $Q_\mu$  for brevity.

The properties of the system's noise are determined by the noise spectral density $S(\omega)$ 
\begin{equation}
\int_0^\infty dt e^{i\omega t} \langle
e^{i H_Bt}Q_\mu e^{-iH_Bt} Q_\nu \rangle=S(\omega)\delta_{\mu\nu}\;,\label{eq:S}
\end{equation}
 where the inclusion of the Kronecker delta function
 $\delta_{\mu\nu}$  is a consequence of the
assumption of independent baths. 

The effect of the noise on multiqubit quantum annealing was studied
numerically in~\cite{Albash:2012qamme} in the case of Ohmic spectral
density~\cite{Leggett:1987}
\begin{equation}
S_{{\rm Ohmic}}(\omega)=\hbar^2 \frac{\eta\, \omega \exp(-\omega \tau_c)}{1- \exp(-\hbar\omega/k_B T)},\label{eq:S-ohmic}
\end{equation}
where $\eta$ is the Ohmic damping coefficient and $\omega_c=1/\tau_c$ is a
high-frequency cutoff.  That work assumed weak system-environment
coupling and utilized the Redfield formalism to derive the quantum
Markovian master equation in the basis of the (instantaneous)
adiabatic eigenstates of the qubit Hamiltonian. It was built on
earlier studies of open-system quantum annealing where similar
assumptions were
made~\cite{Lidar:05,Lidar:05pra,Love:08,Vega:10,Averin:09}.

In addition to Ohmic noise, an important role is played by a
low-frequency noise of the $1/f$
type~\cite{Martinis:2003,Harlingen:2004} produced by the spins in the
amorphous parts of the qubit
device~\cite{Martinis:2005,McDermott:2008,Paladino:2014}.  In current
D-Wave chips this noise is coupled to the flux qubit relatively
strongly as was shown in recent experiments~\cite{PhysRevB.83.180502}.
Additionally, our analysis shows that noise effects are significantly
enhanced by collective effects associated with multiqubit
tunneling. While future generations of quantum annealer chips will
hopefully have reduced levels of flux qubit noise, it will still
produce a highly nonlinear effect for sufficiently large number of
tunneling qubits.

In recent years, the noise spectrum of flux qubits was studied using a
variety of approaches that includes dynamical decoupling
schemes~\cite{Oliver:2011}, free-induction Ramsey
interference~\cite{Oliver:2012}, coherent spectroscopy with strong
microwave driving~\cite{Oliver:2014}, and macroscopic resonant
tunneling (MRT)
techniques~\cite{harris_experimental_MRT_2008,PhysRevB.83.180502}.

In MRT experiments performed on the D-Wave One chip the qubit state is
probed in a way that is most similar (compared with other methods) to
the quantum annealing process itself, with no use of high-frequency
drive and with just slow tunneling for each qubit within its
group-state manifold. While the exact microscopic models of
low-frequency noise are not well understood
\cite{Martinis:2005,Paladino:2014} its effect on the system evolution
in MRT experiments appeared to be well-described by phenomenological
models~\cite{harris_experimental_MRT_2008,PhysRevB.83.180502}.  There,
the quantity of interest is the incoherent-tunneling rate between the
``up" and ``down" eigenstates of the single flux qubit Hamiltonian $-
\frac{1}{2} (\epsilon_{1} \sigma^z + \Delta_{1} \sigma^x)$ as a
function of the bias $\epsilon_{1}$. In~\cite{amin_macroscopic_2008},
the Gaussian form of the MRT line is described with a noise model
whose spectral density is sharply peaked at low frequency.  In
\cite{PhysRevB.83.180502}, this model is extended by attributing the
linear form of the tails in the MRT line shape to the high-frequency
(Ohmic) part of the noise spectral density. The MRT data collected for
small tunneling amplitudes ( $\Delta_{1}/(2\pi\hbar) < $ 1 MHz) and in
a broad range of biases (0.4 MHz$-$4 GHz) and temperatures (21~mK $-$
38~mK) is surprisingly well-described by a phenomenological ``hybrid"
noise model
 \begin{equation}
  S(\omega)=S_{\rm lf}(\omega)+S_{\rm Ohmic}(\omega),\label{eq:hybrid}
  \end{equation}
  \noindent
 where the  high-frequency part of the spectral density  has Ohmic form~\eqref{eq:S-ohmic}  and the low-frequency part  $S_{\rm lf}(\omega)$ is described only  by the two parameters  related to the  width $W$ and shift $\epsilon_p$ of the MRT line~\cite{amin_macroscopic_2008}:
\begin{align}
W^2&=\int_{-\infty}^{\infty}\frac{d\omega}{2\pi} \frac{S_{\rm lf}(\omega)}{\hbar^2},\label{eq:W1}\\
\epsilon_p &= \int_{-\infty}^{\infty}\frac{d\omega}{2\pi} \frac{S_{\rm lf}(\omega)}{\hbar^2}  \frac{1}{\omega(1+\coth(\hbar\omega/(2 k_B T))}\label{eq:W}\;.
\end{align}
These parameters are also well known  in physics in the context of the Pekar-Huang-Rhys theory \cite{Pekar:1950,Huang:1950} of phonon broadening of the electron optical transitions in  impurity centers in solids. The shift $\epsilon_p$ is often called the reorganization energy, i.e. the energy change of the bath degrees of freedom during an
incoherent tunneling process (it is conventionally called  a Stokes shift  in the theory of optical transitions). The width $W$ and shift $\epsilon_p$
measured in~\cite{PhysRevB.83.180502} satisfy with high accuracy the
thermodynamic relation
\begin{equation}
\epsilon_p=\frac{\hbar W^2}{2 k_B T}\label{eq:fdt}\;.
\end{equation}
This requires  that the integrals in Eqs.~\eqref{eq:W1} and~\eqref{eq:W}  are dominated  by the range of frequencies   $\omega\lesssim \omega_{\rm lf}   \ll k_B T/\hbar$ where $\omega_{\rm lf}$ is a characteristic cutoff of the low-frequency noise. The above  hybrid model uses no information about the noise spectrum at frequencies below $W$ except the assumption that   $\omega_{\rm lf} \ll W$ which is well    justified because   in the   experiments $\hbar W$ is of the same order as $ k_B T$~\cite{PhysRevB.83.180502}.

The hybrid noise model parameters were measured on D-Wave One
~\cite{PhysRevB.83.180502} and D-Wave Two chips near the end of the
quantum annealing schedule, $s\simeq 1$. The values of the noise
parameters at a point during the annealing can be related to the
measured ones (see App.~\ref{sec:coupling_to_bath})
\begin{align}
\frac{\eta(s)}{\eta_{\rm MRT}}=\left(\frac{W(s)}{W_{\rm MRT}}\right)^2=\frac{B(s)}{B(1)}\label{eq:MRT}
\end{align}
\begin{equation}
W_{\rm MRT}/(2\pi \hbar)=0.4\,{\rm GHz},\quad  \eta_{\rm MRT}=0.24 \label{eq:MRT1}
\end{equation}
where the above values correspond to the measurements done with the
D-Wave Two chip.
 
\begin{figure}[h]
  \centering
  \includegraphics[width=\columnwidth]{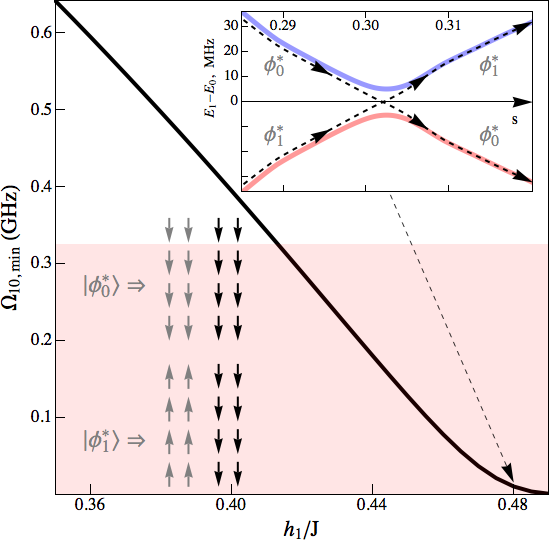}
  \caption{The minimum gap between the ground- and first excited-state levels  $\Omega_{10,\min}=\min_{s\in(0,1)} [E_{1}(s)-E_0(s)]/\hbar$  \eqref{eq:H0-eiegn} during quantum annealing as a function of the rescaled bias $h_1/J$ is shown by the solid green line. The horizontal boundary of the red-filled area  at 324MHz corresponds to 15.5mK, which is the lowest temperature in our experiments. The bias value $h_1/J=0.5$ corresponds to zero energy gap, achieved at the end of quantum annealing when the eigenstates of $H_0(t_{qa})=H_P$ with parallel and anti-parallel  cluster orientations are degenerate.  The upper inset shows the avoided crossing between the energy levels $\{E_{1}(s),\,E_0(s)\}$  in the weak-strong cluster problem at $h_1/J=0.48$. Dashed lines show the  energy levels corresponding to the diabatic basis of states $\{|\phi_{1}^{*}(s)\rangle, |\phi_{0}^{*}(s)\rangle\}$ \eqref{eq:rot-b} formed by the rotation of the adiabatic eigenstates \eqref{eq:H0-eiegn} that maximizes the average Hamming distance \eqref{eq:h}, \eqref{eq:hmax} between the spin configurations. The lower inset shows the spin configurations in both clusters that dominate the characteristics of the eigenstates before and after the avoided crossing.  While transversing the avoided crossing,  spins in the left cluster (shown in grey) reverse their orientations. }
  \label{fig:mingaps}
\end{figure}

\subsection{Three stages of quantum annealing process \label{sec:phases}}

During quantum annealing the system  follows  several stages   depending on the magnitude  of  the instantaneous  energy gap  between  the two lowest energy eigenstates  of the control Hamiltonian $H_0(s)$. The instantaneous energy spectrum is
 \begin{equation}
H_0(s)|\psi_\gamma(s)\rangle=E_\gamma(s) |\psi_\gamma(s)\rangle.\label{eq:H0-eiegn}
\end{equation}
At the beginning of quantum annealing, the qubit Hamiltonian $H_0$   is dominated by the driver Hamiltonian  term $H^D$   and the energy gap between  the ground state and  the first excited state ( 2$ A(s)$~$ \sim$ {\rm 5 GHz})  is very large compared to the temperature.  At that stage the system resides in a ground state $|\psi_0(s)\rangle$  with overwhelming probability. 

In our case the evolution of the system beyond the initial {\it
  gapped} phase is significantly different form the case of low
connectivity systems such as the linear chain considered in
\cite{Albash:2012qamme}.  Because the dominant interaction in the
problem Hamiltonian $H^P$ is intra-cell ferromagnetic coupling with
high degree of qubit connectivity the system dynamics during quantum
annealing is well described as an evolution of the coupled large spins
(spin value $n/2$) corresponding to the two ferromagnetic
clusters. The flipping of individual spins in each unit cell rises the
energy by a large amount and therefore, when $h_1$ is close to $J/2$,
the system evolution is well-described by the two lowest energy
eigenstates that corresponds to the clusters moving as a whole.
 
For local fields $h_1<J/2$, the system evolution goes through the
so-called ``avoided-crossing" region at intermediate times where the
two lowest eigenstates $E_1(s)$ and $E_0(s)$ approach closely to, and
then repel from, each other (see inset in Fig.~\ref{fig:mingaps}).
This level repulsion occurs due to the collective tunneling of qubits
in the left cluster between the opposite $z$-polarizations.  At the
point where the gap $\hbar \Omega_{10}(s)=E_1(s)-E_0(s)$ reaches its minimum
the adiabatic eigenstates $\{|\psi_0\rangle,|\psi_1\rangle\}$ are formed
by the, respectively, symmetric and anti-symmetric superpositions of
the cluster orientations.

In the minimum gap region the coupling of the qubit system to the
environment causes fast transitions between the states giving rise to
thermalization. Unlike the case of quantum annealing in a linear
qubit chain~\cite{Albash:2012qamme} only the two lowest levels will be
thermalized with the rest of the levels being unoccupied because the
energy splitting at the avoided crossing obeys $\hbar \Omega_{10} \lesssim k_B T$
and is much smaller than the separation to the next energy level
(which is in excess of 3GHz, see Fig.~\ref{fig:gaps}).

\begin{figure}[h]
  \centering
  \includegraphics[width=\columnwidth]{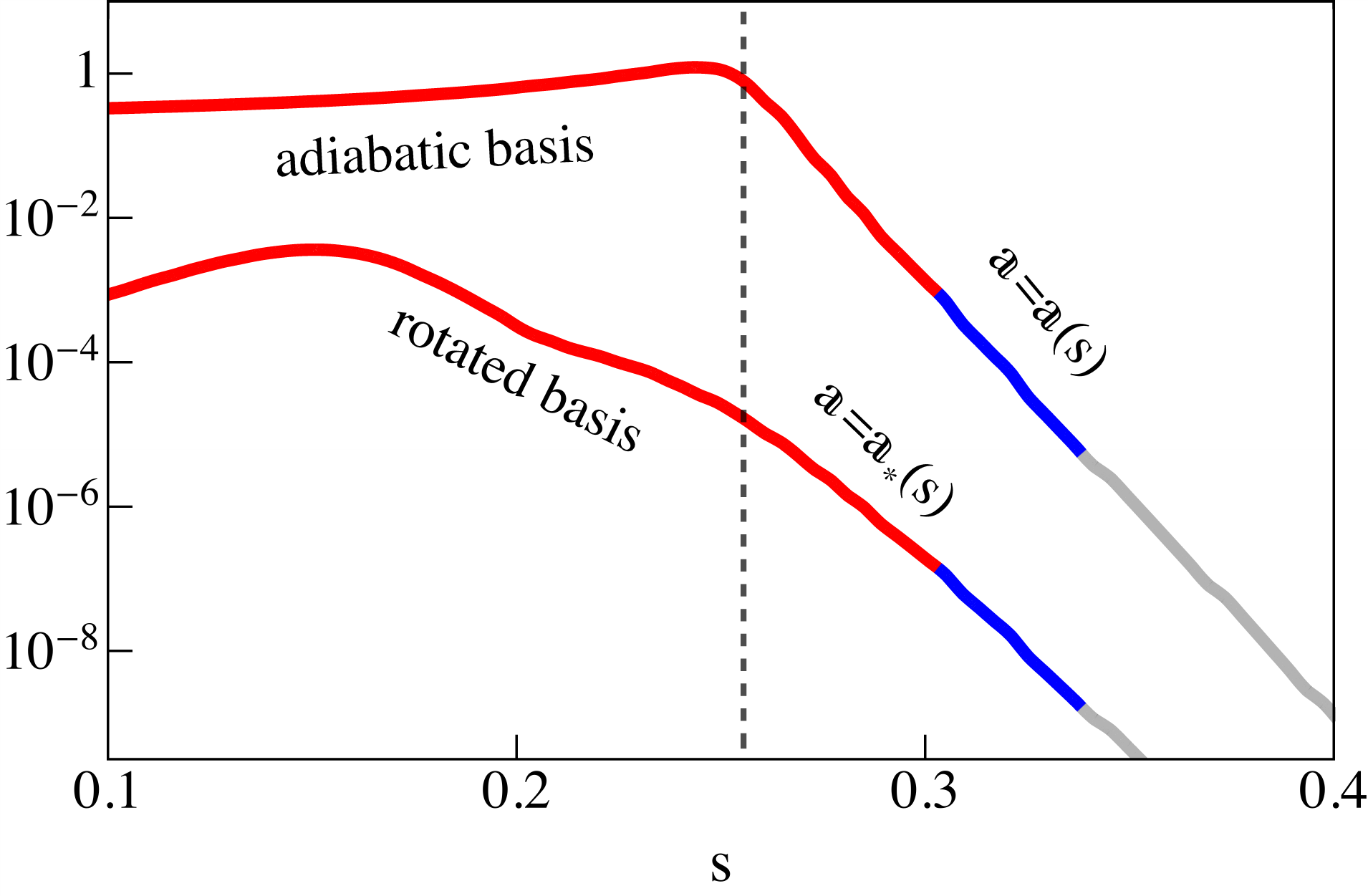}
  \caption{Dependence of the transition matrix element coefficient
    $\mathbb a$ on $s$ for $h_1=0.44 J$.  The two plots correspond to
    the coefficient in the adiabatic basis
    $\{\ket{\psi_0},\ket{\psi_1}\}$ and rotated basis (see
    \eqref{eq:rot-b}) of the pointer states.  Vertical dashed line
    indicates the point where the minimum energy gap is reached
    (avoided-crossing).  Red, blue, and gray colors indicate the
    different stages of quantum annealing (thermalization,
    slowdown, frozen) described in the text. The boundary between the
    thermalized and slowdown phases corresponds to $\Gamma_{1\rightarrow 0}
    t_{qa}=10$. The boundary between the slowdown and frozen phases
    corresponds to $\Gamma_{1\rightarrow 0} t_{qa}=0.1$ }
  \label{fig:a}
\end{figure}

\begin{figure}[h]
  \centering
  \includegraphics[width=\columnwidth]{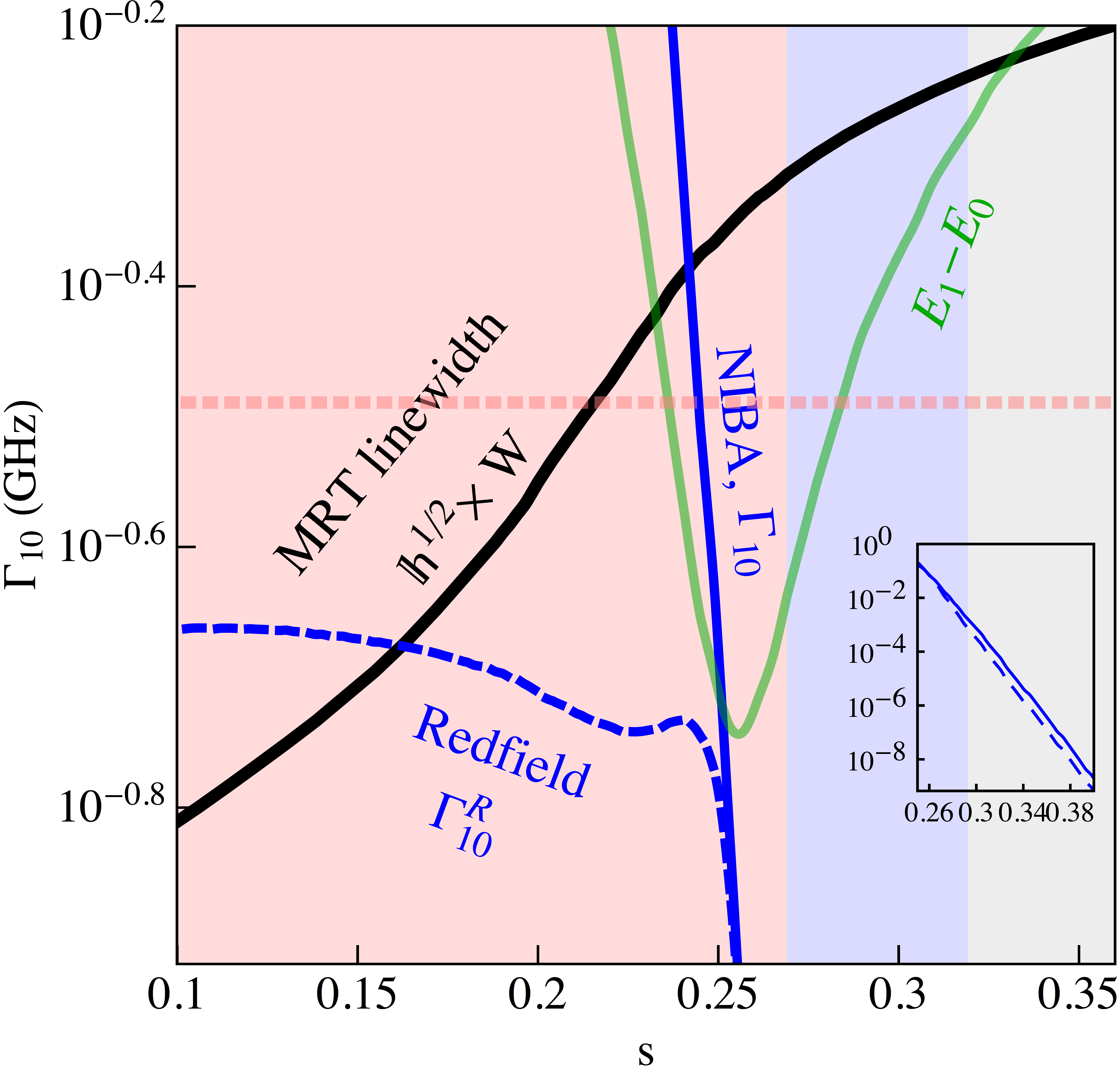}
  \caption{The solid black line shows the dependence of the effective MRT linewidth of the low-frequency noise   $\mathbb h^{1/2}(s) W(s)/(2 \pi)$ (GHz) on annealing  parameter $s$.  The dashed  blue line  is a plot of the single boson transition rate $\Gamma_{1\rightarrow 0}^{FGR}$ vs. $s$ calculated   using Fermi's golden rule, Eq.~\eqref{eq:W10fgr}, with Ohmic spectral density.  The solid blue is a plot of the transition rate $\Gamma_{1\rightarrow 0}$ calculated using the NIBA equation~\eqref{eq:G10}.  The green line shows the $s$-dependence of the energy gap $(E_{1}(s)-E_0(s))/(2 \pi \hbar)$ between the two lowest-energy levels during the annealing at 15.5 mK. The lowest temperature in our experiments is shown by the horizontal red dotted line.  All plots correspond to bias $h_1=0.44 J$.  At the early stage of quantum annealing $\mathbb h^{1/2}(s) W(s) \ll (E_{1}(s)-E_0(s))/\hbar$ and $\Gamma_{1\rightarrow 0}^{FGR}$ gives an accurate description of the dynamics. Red, blue and gray filling areas correspond to the phases  of quantum annealing (respectively, thermalization, slowdown and frozen) described in the main text.  The boundary between the thermalized  and slowdown phases corresponds to $\Gamma_{1\rightarrow 0} t_{qa}=10$. The boundary between the slowdown and frozen phases corresponds to $\Gamma_{1\rightarrow 0} t_{qa}=0.1$.
   The inset shows the transition rates $\Gamma_{1\rightarrow 0}^{FGR}$,  $\Gamma_{1\rightarrow 0}$ vs. $s$ for a greater range of values. One can see that even at the end of the thermalization phase  $\mathbb h^{1/2}(s) W(s) \gg \Gamma_{1\rightarrow 0}$,  justifying the NIBA approximation discussed in the main text.  }
 \label{fig:W}
\end{figure}

After the avoided-crossing region the ground state
$|\psi_{0}(s)\rangle $ and first excited state $|\psi_{1}(s)\rangle $ gradually
evolve into product states with the same and opposite cluster
orientations, respectively.  There the evolution of the system is
dominated by the spontaneous symmetry breaking signature of the
quantum phase transition: a steep increase of the $z$-magnetizations
of the clusters (see Fig.~\ref{fig:magnetization}). The transitions
between the states $ |\psi_{0}(s)\rangle $ and $ |\psi_{1}(s)\rangle $
involve progressively larger numbers of tunneling qubits leading to an
exponential slow-down of the transitions as shown in Fig.~\ref{fig:W}
(see also Fig.~\ref{fig:a}).

The {\it slowdown} phase of quantum annealing is followed by the
{\it frozen} phase where the transition rates are much slower than the
quantum annealing rate so that the population of the levels do not
change during this phase.  Part of the system population remains
trapped in the excited state $|\psi_{1}(s)\rangle $ until the end of
the quantum annealling process.  The success probability of quantum
annealing is (roughly) determined by the Boltzmann factor for the
relative thermal populations of the first two states during the
slow-down phase.

In the analysis of the transitions between the states we start from the initial (gapped) stage when the instantaneous energy gap $\hbar \Omega_{10}(s)=E_1(s)-E_0(s)$ is sufficiently large compared to $\hbar W(s)$  (see  Fig.~\ref{fig:W}) and  the coupling to the  environment   can be treated as a perturbation.  Then    the  transition rate    from   the  first excited state to the ground state  is given by Fermi's golden rule for a single-boson process:
\begin{equation}
\Gamma_{1\rightarrow 0}^{FGR}(s)=\frac{\mathbb{a}(s)  }{\hbar^2}S(\Omega_{10}(s))  ,\label{eq:W10fgr}
\end{equation}
where 
\begin{equation}
\mathbb a(s)=\frac{1}{4}\sum_{\mu=1}^{2n}| Z_{\mu}^{10}(s)|^2. \label{eq:a}
\end{equation}
Here and below we use the following notation  for the matrix elements:
\begin{equation}
Z_{\mu}^{\gamma\gamma^\prime}(s)=\langle \psi_\gamma (s)|\sigma_{\mu}^{z}|\psi_{\gamma^\prime}(s)\rangle,\quad \gamma,\gamma^\prime=\{0,1\}.\label{eq:Z10}
\end{equation}
This is an overlap factor  that determines how strongly the transition $1\leftrightarrow 0$ is coupled to the environment.

\begin{figure}[h]
   \centering
 \includegraphics[width=\columnwidth]{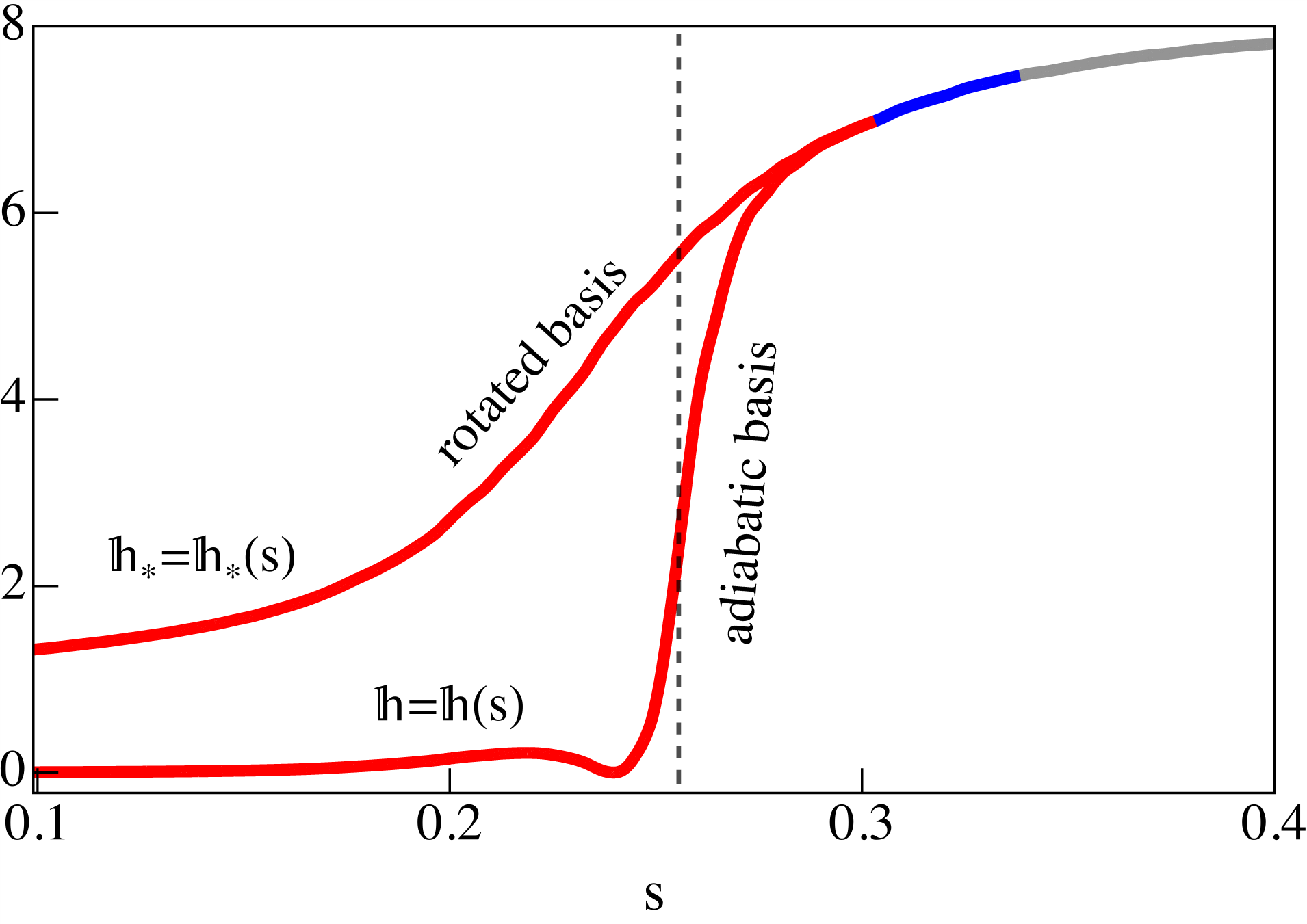}
  \caption{Dependence of the average Hamming distance between the two lowest-energy eigenstates   on $s$ for the adiabatic basis $\{\ket{\psi_0},\ket{\psi_1}\}$ and the rotated diabatic  basis  \eqref{eq:rot-b}  for $h_1=0.44J$. The vertical dashed line indicates the point  where the minimum energy gap is reached (avoided-crossing).  Red, blue, and gray colors indicate the different stages of quantum annealing described in the text. The boundary between the thermalized  and slowdown phases corresponds to $\Gamma_{1\rightarrow 0} t_{qa}=10$. The boundary between the slowdown and frozen phases corresponds to $\Gamma_{1\rightarrow 0} t_{qa}=0.1$}
  \label{fig:h}
\end{figure}

Fig.~\ref{fig:a} shows the dependence of $\mathbb a(s)$ on  the annealing parameter $s$.
We observe a steep exponential fall-off of this coefficient after the avoided crossing. This happens because, in accord with the   discussion above, starting from the avoided crossing region,
intra-cell ferromagnetic interaction plays a substantial role by causing the  spins in each unit cell  to move {\it in unison}, forming two clusters with  total spin value $n/2$ each. As can be seen from Fig.~\ref{fig:magnetization},  {\it after} the avoided crossing the first two eigenstates $\ket{\psi_0}$ and $\ket{\psi_1}$  correspond to opposite total spin $z$- projections  of the left cluster.  When $s$  increases, the  average ``Hamming distance" between the eigenstates,
\begin{equation}
\mathbb h(s)=\frac{1}{4}\sum_{\mu=1}^{2n}|Z_{\mu}^{11}(s)-Z_{\mu}^{00}(s)|^2,  \label{eq:h}
\end{equation}
also increases very steeply  as shown in Fig.~\ref{fig:h}. (We note that the maximum value of $\mathbb h(s)$ is proportional to $n$.)  In that region   the transition between the eigenstates requires  multiqubit tunneling of a progressively  higher order, leading to an exponential decay of the overlap coefficient  $\mathbb a(s)$ with $s$ and a steep deceleration of the
environment-induced transitions between the two states, as can be seen from the plot of $\Gamma_{1\rightarrow 0}(s)$ in Fig.~\ref{fig:W}.

The fact that qubits within each unit cell tend to move together,
forming large spins, amplifies the effect of the environment on their
quantum dynamics. In particular, we will show below that the effective
linewidth of the low-frequency noise as seen by the two-state system
$\{\ket{\psi_1(s)},\ket{\psi_0(s)}\}$ is $\mathbb h^{1/2}(s)
W(s)$ and that the effective Ohmic coefficient is $\mathbb h(s)
\eta(s)$. This amplification becomes important at the slowdown  stage of
quantum annealing when clusters increase their z-polarizations and
$\mathbb h \sim n \gg 1$ (see Fig.~\ref{fig:W}).  For sufficiently
large $\mathbb h^{1/2}(s) W(s)\gtrsim \Omega_{10}$, the description
of the system dynamics becomes substantially non-perturbative in the
spin-boson interaction. Equilibria of the environmental degrees of
freedom shift depending on the collective qubit-state, which in
turn affects the state itself causing the {\it polaronic} effect. In
this case, the adiabatic basis of instantaneous eigenstates
$\{\ket{\psi_1(s)},\ket{\psi_0(s)}\}$ formed by the superposition of
up and down cluster orientations loses its physical
significance. Instead, the dynamics occurs between the two states
$\{\ket{\phi_1(s)},\ket{\phi_0(s)}\}$ with the predominately opposite
cluster orientations corresponding (roughly) to the bottoms of the
wells of the classical potential separated by the barrier as shown in
Fig.~\ref{fig:V(q_{1,1},q_{2,1})}.

We introduce a unitary rotation on angle $\vartheta$ defining the
rotated basis of states $|\phi_{\gamma}(\vartheta,s)\rangle $:
 \begin{equation}
 |\phi_{\gamma}\rangle=\sum_{\gamma^\prime=0,1}(-1)^{\gamma \gamma^\prime+ \gamma^\prime+1}\,\cos\left(\frac{\vartheta}{2} -(-1)^{\gamma+\gamma^\prime}\frac{\pi}{4} \right) |\psi_{\gamma^\prime}\rangle,\label{eq:Ur}
 \end{equation}
\noindent  where $\gamma, \gamma^\prime=\{0,1\}$. The corresponding matrix elements are
\begin{equation}
Z_{\mu}^{\gamma\gamma^\prime}(\vartheta,s)=\langle \phi_\gamma (\vartheta,s)|\sigma_{\mu}^{z}|\phi_{\gamma^\prime}(\vartheta,s)\rangle,\quad \gamma,\gamma^\prime=0,1.\label{eq:Z10r}
\end{equation}
We find  the angle $\vartheta_{*}(s)$ that maximizes  the average Hamming distance between  the states
\begin{equation}
\mathbb h_{*}(s)=\max_{\vartheta 
}\frac{1}{4}\sum_{\mu=1}^{2n} |Z_{\mu}^{11}(\vartheta,s)-Z_{\mu}^{00}(\vartheta,s)|^2.
\label{eq:hmax}
\end{equation}
Then our new instantaneous basis will be  defined as
\begin{equation}
|\phi_{\gamma *}(s)\rangle = |\phi_\gamma(\vartheta_{*}(s),s)\rangle,\quad \gamma=\{0,1\} .\label{eq:rot-b}
\end{equation}
The system dynamics is mainly a hopping process between these states
associated with the incoherent tunneling of the spins in the left
cluster connecting states with predominantly ``up" ($\ket{\phi_{1
    *}}$) and ``down" ($\ket{\phi_{0 *}}$) qubit configurations. The
Hamming distance between these states is $\mathbb h_{*}(s)$ (shown in
Fig.~\ref{fig:h}). In fact the transition rate between the states in
the rotated basis (to be computed below) is minimized for the angle
of Eq.~\eqref{eq:hmax}.  In essence, this approach is related to the pointer
basis idea, that the system tends to be localized in states induced by
environmental coupling ~\cite{Zurek81}. In what follows a subscript $*$ will denote quantities computed in the rotated
basis.

\begin{figure}[h]
   \centering
 \includegraphics[width=\columnwidth]{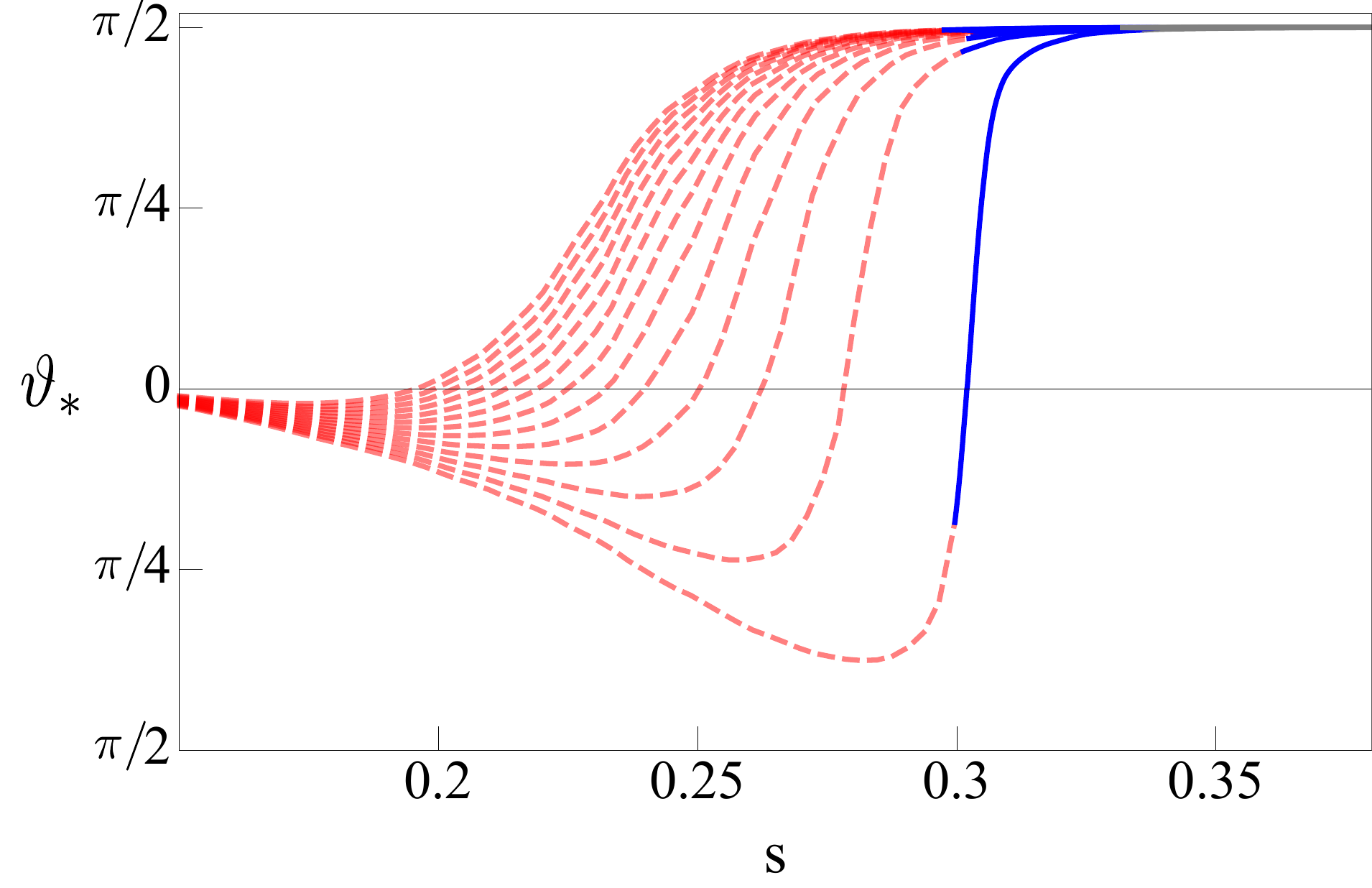}
  \caption{Optimal rotation angles $\vartheta_*$ vs. $s$  for  eigenstates in the basis of Eq.~\eqref{eq:rot-b} at different values of $h_1/J=0.35+0.01 \kappa$, with $\kappa=1,2,.\ldots, 13$ corresponding to the curves in the figure numbered from the bottom to the top.  Red corresponds to the thermalization phase,  blue corresponds to the loss of thermalization (slowdown phase) , and black corresponds to the frozen phase described in the main text. The boundary between the thermalized  and slowdown phases corresponds to $\Gamma_{1\rightarrow 0} t_{qa}=10$. The boundary between the slowdown and frozen phases corresponds to $\Gamma_{1\rightarrow 0} t_{qa}=0.1$. Optimal rotation angles only have physical meaning starting from about the end of the thermalization phase where the spins  in the clusters start to behave in a concerted manner. All angles approach $\pi/2$ in a frozen phase corresponding to the adiabatic eigenstates. }
  \label{fig:theta}
\end{figure}

The  rotation angle $\vartheta_*(s)$ during the annealing is shown in Fig.\ref{fig:theta} for different values of $h_1/J$.  In the later stages of the annealing  the angle $\vartheta$ always approaches the value  $\pi/2$ that corresponds to the  adiabatic (energy) basis with state $\ket{\phi_{1*}}$ being an excited state~\eqref{eq:Ur}.  In other words,  quantum  annealing along the pointer states arrives at the  encoded solution of the computational problem because, toward the end of the annealing, the rotated basis converges back to eigenstates of the problem Hamiltonian.

We will show below (cf. Eq.~\eqref{eq:rate}) that the non-perturbative treatment of the effects of noise and dissipation does not change the Markovian nature of the system dynamics but modifies the instantaneous  transition rate  $\Gamma_{1\rightarrow 0}(s)$ compared to its value $\Gamma_{1\rightarrow 0}^{FGR}(s)$ given in Eq.~\eqref{eq:W10fgr} for a single-boson process.
The non-perturbative analysis in the  rotated basis is   justified within the context of the theory of spin-boson interaction developed in~\cite{Leggett:1987}. The individual transitions  due to the coupling  to bosons are associated with so-called  ``blip-cojourn" pairs with blips forming   a dilute gas  if   the duration of the  blip         $\tau_b=1/\mathbb h^{1/2}(s) W(s)$  is much shorter than the characteristic inter-blip distance $\sim$ 1$/\Gamma_{1\rightarrow 0}(s)$, with
\begin{equation}
\Gamma_{1\rightarrow 0}(s)\ll \mathbb h^{1/2}(s) W(s).\label{eq:cond}
\end{equation}
In the adiabatic basis, $\mathbb h(s)$ can reach very small values
near the avoided crossing (see Fig.\ref{fig:h}) due to  
quantum  superpositions. In the optimally rotated 
basis, $\tau_{b}^{-1}=\mathbb h_{*}(s)^{1/2} W(s)$ is monotonically
increasing during the annealing to its maximum value ($n$=8 in the
problem of interest).  When the condition~\eqref{eq:cond} is satisfied
at the last stages of the annealing, the gas of blips is dilute and we
can apply the Noninteracting-blip Approximation (NIBA)
\cite{Leggett:1987}. While this method was developed to analyze a
Landau-Zener problem in a driven spin-1/2 system coupled to a finite
temperature bath in a number of papers
\cite{Ao:1989,Ao:1991,Kayanuma:1998}, its application in the present
context of strongly correlated qubit dynamics is novel and leads to 
qualitatively new features.

Before we proceed further with the NIBA analysis, we emphasize that
our theory will not provide an accurate treatment of the region of $s$
where $\mathbb h^{1/2}(s) W(s)\sim \Gamma_{1\rightarrow 0}$, representing a
crossover between the perturbative treatment based on single-boson
processes and Ohmic spectral density (with $\Gamma_{1\rightarrow 0}^{FGR}(s)$
given in Eq.~\eqref{eq:W10fgr}) and non-perturbative NIBA theory that
includes low-frequency noise.  In this intermediate region the
low-frequency noise cannot be described just by the two
characteristics of its spectral density treated here
(cf. Eqs.~\eqref{eq:W} and \eqref{eq:W1} ) and we are simply lacking the
experimental data for the theoretical analysis.

Our main observation  is that this region does not affect the  population of  the ground state at the end of quantum annealing.
Since the quantum annealing rate $1/t_{qa}$=50 kHz  is constant, the system stays  very close to instantaneous  thermal  equilibrium while
\begin{equation}
\Gamma_{1\rightarrow 0}(s) \gg 1/ t_{qa}\label{eq:COND1}
\end{equation}\noindent
(see Fig.~\ref{fig:W}). This is a ``thermalization phase'' of the annealing process. On the other hand, due to the strong effect of low-frequency noise  on  D-Wave qubits  the condition
\begin{equation}
1/t_{qa} \ll  \mathbb h^{1/2}(s) W(s) \label{eq:COND2}
\end{equation}
is held almost everywhere except for the very early stages (cf. Fig.~\ref{fig:W}). Therefore  the non-perturbative regime of condition~\eqref{eq:cond} is established well within  the thermalization phase where the (Gibbs) distribution  is not sensitive to the noise model.

 \begin{figure}[h]
   \centering
  \includegraphics[width=\columnwidth]{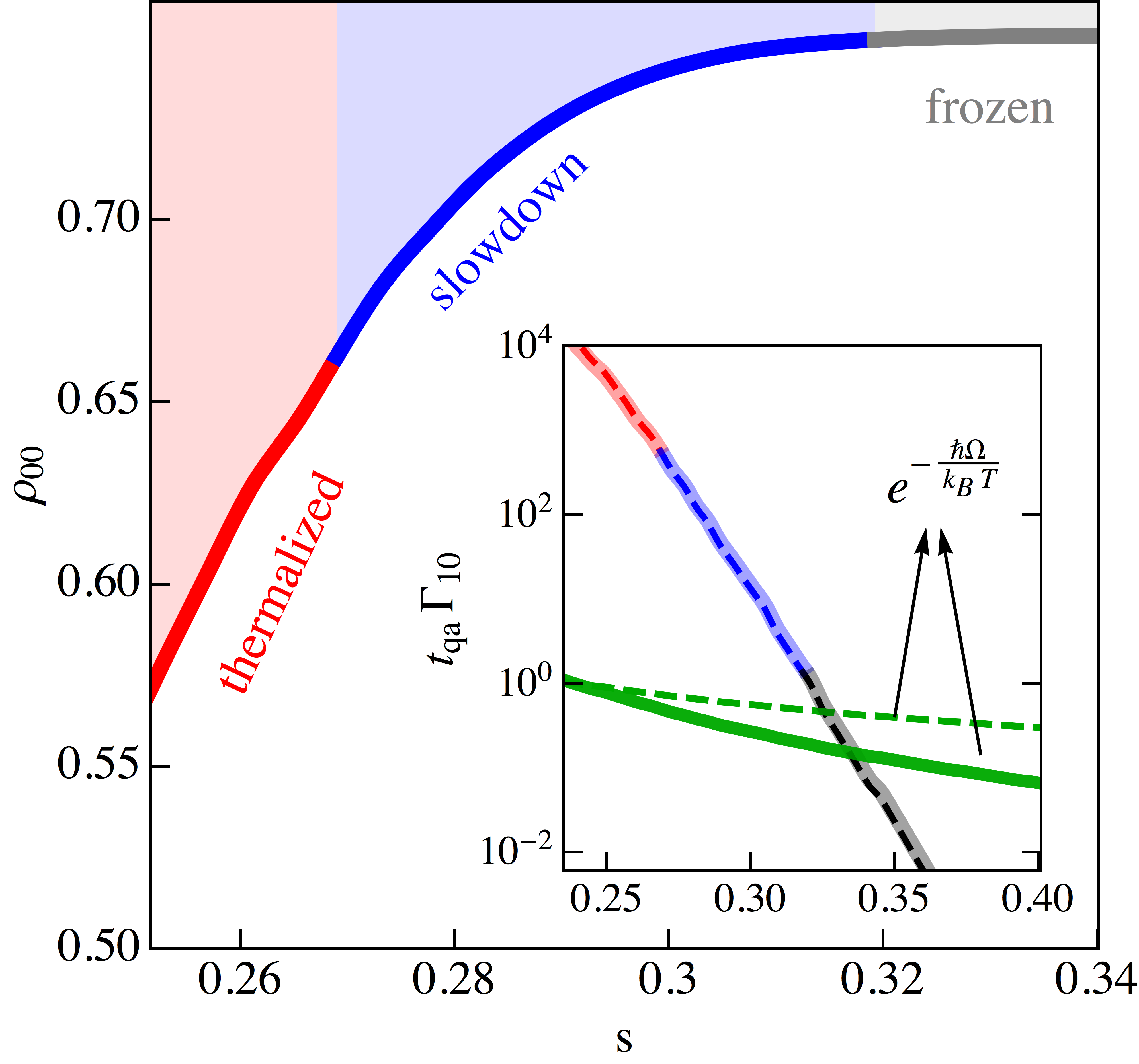}
  \caption{ Main plot shows the  evolution of the  population of the pointer state  $\rho_{00}=\langle\phi_0^*(s)|\rho|\phi_0^*\rangle $ as a function of the annealing parameter $s$.
Red, blue and gray filling areas correspond to the phases  of quantum annealing (respectively, thermalization, slowdown and frozen) described in the text.  The boundary between the thermalized  and slowdown phases corresponds to $\Gamma_{1\rightarrow 0} t_{qa}=10$. The boundary between the slowdown and frozen phases corresponds to $\Gamma_{1\rightarrow 0} t_{qa}=0.1$.
 In the inset  the solid line  shows the dependence of $t_{qa}\,\Gamma_{1\rightarrow 0}$ \eqref{eq:G10}  on the annealing  parameter  $s$. Different color portions correspond to the same annealing  phases  as in the main plot. 
 The solid green line is a plot of the Boltzmann factor for the instantaneous energy difference $\hbar\Omega(s)$  between the pointer  states given in \eqref{eq:Omega}. Both lines are plotted at 15.5mK. Dashed lines correspond to the temperature 35mK.  All data are for the value of the local field $h_1/J=0.44$.}
  \label{fig:freez}
\end{figure}

At a later stage the thermal distribution  can no longer be supported  due to the steep slowdown of the transition rates as shown in Fig.~(\ref{fig:freez}). Eventually  the system enters the final ``frozen" stage where the transitions are suppressed over the period of the annealing and part of the system population remains trapped in the excited state. As can be seen in the  Fig.~\ref{fig:freez}, the success probability is determined by the occupation of the ground state at the beginning of the frozen phase. This, in turn, is
given by the thermal equilibrium population $\frac 1 2 (1+\tanh
(\hbar \Omega(s)/2 k_B T))$ where $\hbar \Omega(s)$ is the difference of the energies of
the pointer states $|\phi_{1*}\rangle$ and $|\phi_{0*}\rangle$, as
given in Eq.~\eqref{eq:Omega}.  When the temperature grows, the
transition rates dominated by low-frequency noise change very little in the studied temperature range,
while the thermal population decreases appreciably. This reduces the success of quantum
annealing, as will be seen later, and is the origin of the observed thermal
reduction.

\subsection{Non Interacting Blip Approximation}

To implement  NIBA in our problem we need to explicitly represent the   boson operators for normal modes of the environmental  free-boson Hamiltonian
\begin{align}
  H_B&=\sum_{\mu=1}^{2n}\sum_{ u}\hbar \omega_{\mu u}(b_{\mu\, u}^{\dagger}b_{\mu u}+1/2)\\
  Q_\mu&=\sum_{ u}\lambda_{\mu u}(b_{\mu\, u}^{\dagger}+b_{\mu
    u}),\quad \mu=1,\ldots,2n.
\end{align}
 Here $\mu$ indexes qubits, $ u$ indexes the bath modes, and
 $\lambda_{\mu u}$ and $\omega_{\mu u}$ are the microscopic
 parameters that will not enter into any observable directly but will
 do so only via the spectral function of Eq.~\eqref{eq:hybrid}, which
 takes the form
\[S(\omega,s)=\frac{2\pi \sum_{ u}\lambda_{\mu u}^2 \delta(\omega-\omega_{\mu u})}{1-\exp(-\hbar\omega/k_B T)}\]
identical for each qubit. We explicitly note that   $S\equiv S(\omega, s)$ depends on the quantum annealing parameter $s$ (see Eqs.~\eqref{eq:MRT} and \eqref{eq:MRT1} as well as Sec.~\ref{app:single_qubit_hamiltonian}). 
We proceed by making  a small polaron transformation  of the original Hamiltonian with unitary operator
\[U(s)=\exp\left[\sum_{\gamma} -\Lambda_{\gamma}(s)|\phi_{\gamma *}(s)\rangle\langle \phi_{\gamma *} (s)| \right ] \;. \]
Here  we use the notation
 \begin{align}
 \Lambda_{\gamma }(s)&=\sum_{\mu u}\xi_{\mu u }^{\gamma\gamma}(s)(b_{\mu u}-b^{\dagger}_{\mu u}),\label{eq:L}\\
 \xi_{\mu u }^{\gamma\gamma^{\prime}}(s)&=\frac{\lambda_{\mu u}}{\hbar \omega_{\mu u}}Z_{\mu *}^{\gamma\gamma^{\prime}}(s),
\nonumber
 \end{align}
where $\gamma\in\{0,1\}$ and the matrix elements $Z_{\mu *}^{\gamma\gamma^{\prime}}$ are   defined in the  basis of optimally rotated states as
\begin{equation}
Z_{\mu *}^{\gamma\gamma^\prime}(s)=\langle \phi_{\gamma *} (s)|\sigma_{\mu}^{z}|\phi_{\gamma^\prime *}(s)\rangle. \label{eq:Z}\end{equation}
The  system-bath Hamiltonian \eqref{eq:Hcont} after the  polaron transformation $\mathbb H= U H U^{-1}$ is 
\begin{equation}
\mathbb H(s)=\mathbb H_0(s)+\mathbb H_{\rm int}(s)+H_B
\end{equation}
where 
\begin{equation}
\mathbb H_0(s)=\hbar \Omega (s) \left( |\phi_{1 *}(s)\rangle\langle\phi_{1 *}(s)|-|\phi_{0 *}(s)\rangle\langle\phi_{0 *}(s)|\right),
\end{equation}
\vspace{-0.2in}
\begin{align}
 \Omega(s)&=\Omega_{10}^{*}(s) -\epsilon_p(s) \mathbb d_*(s),\label{eq:Omega}\\
\Omega_{10}^{*}(s)&= \Omega_{10} (s) \sin\vartheta_*(s) \label{eq:Omega10}
\end{align}
\noindent
and $ \Omega_{10} (s)= (E_1(s)-E_0(s))/\hbar$.
Here  $\Omega_{10}^{*}(s)$  corresponds to the system  energy gap in a rotated  basis and the second term gives the polaronic shift due to the reorganization energy of the environment. Similar to the linewidth, it is renormalized with respect to its  single qubit MRT value by a coefficient   $\mathbb d_*(s)$ reflective of a collective qubit behavior:
\begin{equation}
\mathbb {d}_*(s)=\frac{1}{4}\sum_{\mu=1}^{2n}\left[(Z_{\mu *}^{11})^2(s)-(Z_{\mu *}^{00})^2(s)\right].\label{eq:38}
\end{equation}
The  Hamiltonian $\mathbb H_{\rm int}(s)$ is strictly non-diagonal in the rotated   basis and  explicitly involves boson operators  
\begin{widetext}
\begin{align}
\mathbb H_{\rm int}^{\gamma\gamma^{\prime}}(s)=\left(\frac{\hbar \Delta_*(s)}{2}+
\sum_{\mu, u} Z_{\mu *}^{\gamma\gamma^{\prime}}(s)\lambda_{\mu u}(b^{\dagger}_{\mu u}+b_{\mu u}-2\xi_{\mu u *}^{\gamma\gamma^{\prime}}(s))\right )\,e^{\Lambda_{\gamma^{\prime}\gamma}(s)},\label{eq:Hint}
\end{align}
\end{widetext}
where $\Lambda_{\gamma^{\prime}\gamma}=\Lambda_{\gamma^{\prime}}-\Lambda_{\gamma}$ and $\Delta_*$ represent off-diagonal elements of the system Hamiltonian in the rotated  basis
\begin{equation}
\Delta_*(s)=-\Omega_{10}(s) \cos\vartheta_*(s)\label{eq:Delta}\;.
\end{equation}

While the original NIBA calculation~\cite{Leggett:1987} was quite involved, a very simple prescription for how to apply it was given  in~\cite{Dekker:1987}, which we will follow below. We start from the quantum Liouville  equation for the full system-bath density operator $\varrho(s)$:
\begin{equation}
i \hbar \frac{d\varrho}{dt}=[\mathbb H_0(s(t))+\mathbb {\bar H}_{\rm int}(s(t),t),\varrho(t)]\label{eq:Lio},
\end{equation}
where  we employed the interaction picture for bosons 
\begin{equation}
\mathbb {\bar H}_{\rm int}(s,t)=\exp(i H_B t/\hbar) \mathbb H_{\rm int}(s)\exp(-i H_B t/\hbar).\label{eq:Hintt}
\end{equation}
We write  the density matrix in the  rotated basis $\{\ket{\phi_{\gamma *}}\}$
and express its non-diagonal  matrix elements  through the diagonal ones as
\begin{widetext}
  \begin{align}
    \varrho_{10}&(t) =\varrho_{10}(0)+\frac{1}{i\hbar}\int_{0}^{t}d\tau e^{\frac{i }{\hbar} \int_{\tau }^{t}d\tau'\Omega(s(\tau^\prime ))} \left[\mathbb {\bar
        H}_{\rm int}^{10}(s(\tau),\tau)\varrho_{00}(\tau)-\varrho_{11}(\tau)\mathbb
      {\bar H}_{\rm int}^{10}(s(\tau),\tau)\right],\label{eq:nd}
  \end{align}
\end{widetext}
where  $s(t)=t/t_{qa}$ is the quantum annealing parameter. Next we use the fact that the initial state for quantum annealing corresponds to $ \varrho_{10}(0)=0$. We then
plug the non-diagonal matrix elements from \eqref{eq:nd} back into \eqref{eq:Lio} and obtain the   equations for the diagonal matrix element
\begin{widetext}
\begin{align}
\frac{d \varrho_{11}}{dt}&=-\frac{1}{\hbar^2}\int_{0}^{t}d\tau e^{\frac{i }{\hbar} \int_{\tau }^{t}d\tau'\Omega(s(\tau^\prime ))} \left[\mathbb {\bar H}_{\rm int}^{10}(s(t),t)\mathbb {\bar H}_{\rm int}^{01}(s(\tau),\tau)\varrho_{11}(\tau)-\mathbb {\bar H}_{\rm int}^{10}(s(t),t)\varrho_{00}(\tau)\mathbb {\bar H}_{\rm int}^{01}(s(\tau),\tau)\right]+\text{h.c.}\label{eq:dd}
\end{align}
\end{widetext}
and a similar equation for $\varrho_{00}$. Here we  neglected the basis dragging terms of the type  $\langle \phi_{1}(s)|\dot\phi_0(s)\rangle$, because we focus on   the situation where Landau-Zener transitions are negligible ($t_{qa} \Omega_{10}^{\min} \gg 1$). Other than that the expression above is {\it exact}.

We introduce an approximation  and  insert the free-bath dynamics into the expressions for $H_{\rm int}^{\gamma\gamma^{\prime}}(t)$ by replacing  time-dependent boson operators $b_{\mu u}(t)$ with  $b_{\mu u}(t) e^{-i\omega_{\mu u}t}$. We then  introduce the  decoupling  anstatz for the full density matrix: $\varrho = \rho\otimes \rho_B$, where $\rho_B$ is the Gibbs density operator for the bath and
$ \rho =Tr_B[\varrho]$ is the reduced density matrix of the qubit system.
We then  average  Eq.~\eqref{eq:dd} with respect to the bath, which can be done in a tedious but straightforward manner since averaging involves only free-boson operators.  We write the resulting equation of the difference in populations  of  the rotated (pointer) states $z(t)=\rho_{11}(t)-\rho_{00}(t)$ as
\begin{equation}
\frac{d z(t)}{dt} = \int_{0}^{t}dt^{\prime} h(t-t^{\prime},s(t)) -z(t)  \int_{0}^{t}dt^{\prime} g(t-t^{\prime},s(t))\label{eq:dzdt},
\end{equation}
where the functions $h$ and $g$ are given below.  In the above equation we removed   the time delay for the population difference by replacing $z(t-t^{\prime})$ with   $z(t)$. This is possible to do   because the time scale for its variation (inverse transition rates    $1/\Gamma_{1\rightarrow 0}$) is assumed to be   much longer than the correlation time of the environment  $\tau_b$ (see equation \eqref{eq:cond} and discussion  there). The precise  condition for this will be discussed later. In Eq.~\eqref{eq:dzdt} we also neglected the variation of the annealing parameter $s(t')$ over the range of integration $t-t'$ because  $t_{qa} \gg \tau_b$ (see Eq.~\eqref{eq:COND2}).

The functions  $h$ and $g$ used in \eqref{eq:dzdt} are
\begin{align}
h(\tau,s) & =2 Re[e^{i\Omega(s)\tau}(C_{01}^{*}(\tau,s)-C_{10}(\tau,s))]\label{eq:hf}\\
g(\tau,s) & =2 Re[e^{i\Omega(s)\tau}(C_{01}^{*}(\tau,s)+C_{10}(\tau,s))]\label{eq:gf}\;,
\end{align}
\noindent
where   \begin{equation}
C_{\gamma\gamma'}(\tau,s)=\frac{1}{\hbar^2}Tr\left[\rho_B \,\mathbb {\bar H}_{\rm int}^{\gamma\gamma'}(s,\tau)  \mathbb {\bar H}_{\rm int}^{\gamma'\gamma}(s,0)\right ]
\end{equation}
 One can show that $C_{\gamma\gamma^\prime}(\tau,s)$ satisfies the Kubo-Martin-Schwinger (KMS) condition 
\begin{equation}
C_{\gamma \gamma'}(t)=C_{\gamma' \gamma}(-t-i \beta )
\end{equation}
as well as
\begin{equation}
C_{\gamma \gamma'}^{*}(t)=C_{\gamma \gamma'}(-t)\;.
\end{equation}
These conditions ensure that for a fixed  $s(t)={\rm const}$ the stationary solution for $\rho_{\gamma \gamma}(t)$ is a thermal equilibrium  distribution at temperature $T=\hbar/(k_B\beta)$.

The explicit form of $C_{\gamma\gamma'}(\tau,s)$   is the central result of our  analysis as it is distinct from the
conventional NIBA theory (cf. Eqs.~(7.5),(7.6) in~\cite{Leggett:1987}). It has the following form
\begin{equation}
C_{10}(\tau,s)=F_*(\tau,s) \,e^{-\mathbb h_{*}(s) f(\tau,s)},\label{eq:C}
\end{equation}
where
\begin{align}
F_*(\tau,s)=&\mathbb a_*(s) f_{\tau\tau}(\tau,s)\label{eq:F}\\
&+ (\epsilon_p \,\mathbb c_{+}^{*}(s)- i \mathbb c_{-}^{*}(s) f_{\tau}(\tau,s)-\Delta_*(s)/2)^2\;.\nonumber
\end{align}
Here, the off-diagonal matrix element $\Delta_*(s)$ is given in \eqref{eq:Delta} and we denoted $f_{\tau\tau}=\partial^2 f/\partial \tau^2$. We also  used the average Hamming distance \eqref{eq:hmax} and the transition matrix element coefficient $\mathbb a(s)$ in Eq.~\eqref{eq:a} calculated in the rotated basis using matrix elements from Eq.~\eqref{eq:Z}.  The function $f(\tau,s)$ is related to the spectral density $S(s,\omega)$  and appears in the context of the MRT theory of flux qubits~\cite{amin_macroscopic_2008,PhysRevB.83.180502}  and Marcus theory~\cite{Marcus1:1956}:
\begin{equation}
f(\tau,s)=\int_{-\infty}^{\infty}\frac{d\omega}{2\pi}S(\omega,s)\frac{1-e^{-i\omega t}}{(\hbar\omega)^2}.\label{eq:f}
\end{equation}
The coefficients $\mathbb c_{\pm}^{*}(s)$ are defined in the optimally rotated (pointer basis)
\begin{equation}
\mathbb c_{\pm}^{*}(s)=\frac{1}{4}\sum_{\mu=1}^{2n}Z_{\mu\,*}^{10}(s)(Z_{\mu}^{11}(s)\pm Z_{\mu\,*}^{00}(s)), \label{eq:cpm1}
\end{equation}
where the matrix elements $Z_{\mu\,*}^{\gamma\gamma^\prime}$ are given in  Eq.~\eqref{eq:Z}.
Using the hybrid model of noise in Eq.~\eqref{eq:hybrid} introduced in~\cite{PhysRevB.83.180502} the function $f(\tau,s)$ takes the  form
 \begin{equation}
 f(\tau,s)=i \epsilon_p(s) \tau  +\frac{1}{2} W^2(s)\tau^2 -\frac{\eta}{2\pi} \ln G(\tau). \label{eq:fhy}
 \end{equation}
 Here the function $G(\tau) $ is closely related to the well-known functions $Q_1(\tau), Q_2(\tau)$ discussed in Ref.~\cite{Leggett:1987} for the case of an Ohmic environment. Its explicit form is
 \begin{widetext}
   \begin{align}
     G(\tau)=e^{-i\tan^{-1}(\tau\omega_c)}\sqrt{1+(\omega_c\tau)^2}
     \frac{\Gamma((1-i\tau\omega_c)/\beta\omega_c)\Gamma((1+i\tau\omega_c)/\beta\omega_c)}{\Gamma^2(1/\beta\omega_c)}\label{eq:G}
   \end{align}
 \end{widetext}
where $\Gamma(x)$ is the Gamma function. We note that in the range of parameters relevant to the present discussion one can treat $1/\omega_c$  as a positive infinitesimally-small quantity that   serves to regularize the integral in \eqref{eq:dzdt} and whose exact value does not enter the final result.

If follows from \eqref{eq:C} and \eqref{eq:f} that the functions $g(\tau,s)$ and $h(\tau,s)$ decay exponentially with $\tau$ on the scale $\tau_b=1/(\mathbb h^{1/2}W(s))$ in correspondence with the discussion  above (cf. \eqref{eq:C}    and \eqref{eq:fhy}).  The equation \eqref{eq:dzdt} can be re-written as follows: 
\begin{equation}
\frac{dz}{dt}=-\Gamma(s) (z-\tan(\beta\Omega(s)/2))\;,\label{eq:rate}
\end{equation}
where $s=t/t_{qa}$ and 
\begin{equation}
 \Gamma(s)=\Gamma_{0\rightarrow 1}+\Gamma_{1\rightarrow 0},\quad \Gamma_{0\rightarrow 1}=\Gamma_{1\rightarrow 0}e^{-\beta\varepsilon(s)}\;.
\end{equation}
Here 
\begin{equation}
\Gamma_{1\rightarrow 0}(s)=2 Re\int_{0}^{\infty}d\tau\,e^{i\Omega(s)\tau-\mathbb h_*(s) f(\tau,s)}F_*(\tau,s)\;,\label{eq:G10}
\end{equation}
where the function $f$ is given in \eqref{eq:f} and \eqref{eq:fhy} and the function $F$ is given in \eqref{eq:F} and \eqref{eq:cpm1}.  
The above equation can be simplified by making use of the condition that the temperature is much smaller than the cutoff frequency for Ohmic noise 
\begin{equation}
\beta \omega_c\gg 1\;.
\end{equation}
Expanding the Gamma functions in Eq.~\eqref{eq:G} we get
\begin{widetext}
\begin{equation}
\Gamma_{1\rightarrow 0}(s)=\int_{-\infty}^{\infty}d\tau e^{i(\Omega_{10}^{*}-(\mathbb h_*+\mathbb d_* ) \epsilon_p)\tau -\frac{1}{2}\mathbb h_* W^2\tau^2}\left [\frac{i\beta}{\pi \tau_c} \sinh\frac{(\tau-i \tau_c)}{\beta/\pi}\right]^{-\frac{ \mathbb h_* \eta}{2\pi}} F_*(\tau,s),\label{eq:res}
\end{equation}
\end{widetext}
where the expression for $F_*$ is given in Eq.~\eqref{eq:F}. This is a transition rate from the state 1 to 0 that appears in  a rate equation \eqref{eq:rate} for the population difference. This is a  central result of our NIBA theory analysis. 
 
\begin{figure}[h]
  \centering
  \includegraphics[width=\columnwidth]{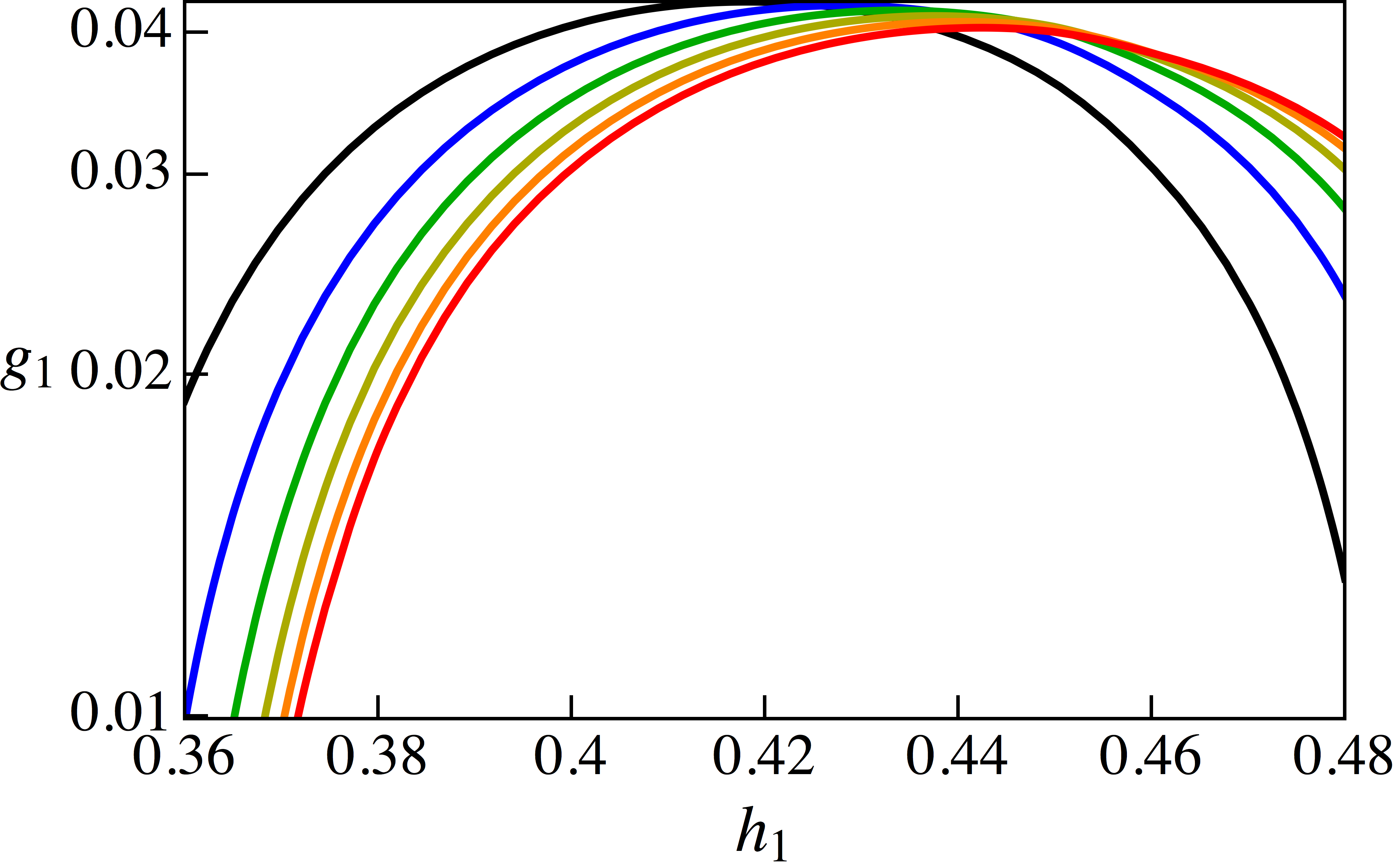}
  \caption{Value of the coefficient $g_1$ in the NIBA criterion \eqref{eq:g1c} as a function of the local field $h_1/J$ at different temperatures. Black, blue, green, yellow, orange, and red curves correspond to the temperature values of  15.5, 20, 25, 30, 35, and 40 mK, respectively.}
  \label{fig:g1}
\end{figure}

The precise condition for neglecting the time-delay for $z(t)$ in
\eqref{eq:dzdt} can be derived within the context of the
Keldysh-Schwinger non-equilibrium diagrammatic technique with respect
to spin boson interaction. This approach was utilized in a number of
studies of Landau-Zener phenomena with dissipation
\cite{Kayanuma:1998,Ao:1991,Pokrovsky:2008}.  In the limit of short
noise correlation time $\tau_b$ the {\it pairing-off theorem}
\cite{Kayanuma:1998} states that non-crossing and non-overlapping
diagrams on double-path Keldysh contour provide a dominant
contribution to the dynamics. This theorem is analogue to the theorem
proven by Abrikosov and Gorkov in their theory of impurities in a
metal \cite{AGD:1963} as was pointed out in \cite{Pokrovsky:2008}. It
also precisely corresponds to non-interacting-blip approximation
(NIBA) developed in Ref.~\cite{Leggett:1987}, where the criterion for
the validity of this approximation is given.  In our case this criterion amounts
to the condition
\begin{equation}
g_1(s)=2 Re\int_{0}^{\infty}d\tau\,\tau\,e^{i\Omega(s)\tau-\mathbb h_*(s) f(\tau,s)}F(\tau,s)\ll 1\;.\label{eq:g1c}
\end{equation}
We note that the expressions \eqref{eq:G10} and \eqref{eq:g1c} differ only by the presence of the $\tau$ in the integrand. This condition corresponds to $\Gamma_{1\rightarrow 0} \,\tau_b\ll 1$ because $\tau_b=1/\mathbb h^{1/2}(s) W(s)$ is the size of the interval beyond which the integrand in \eqref{eq:G10} decays exponentially. As it was discussed in the previous section, we only need to inspect the condition \eqref{eq:g1c} at the end of the thermalization region. For the sake of numerical investigation, we define this by the value of $s\approx s_{eq}$ where the deviation of the system state from the instantaneous Gibbs distribution is 1$\%$. We then compute $g_1$ at various values of parameters with the results given in Fig.~\ref{fig:g1}. It can be seen that $g_1\ll 1$ in the entire parameter range under study. 

We note in passing that the choice of the optimal basis rotation angle
$\vartheta_*(s)$ could be done differently from that given in
Eq.~\eqref{eq:hmax} that maximizes the Hamming distance between
clusters.  Instead, one can calculate the transition rate
$\Gamma_{1\rightarrow 0}=\Gamma_{1\rightarrow 0}(\vartheta,s)$ as a function of the angle of
the basis rotation and choose the angle so that it minimizes the
transition rate between the states.  In fact the optional angle we
obtained in this way is extremely close to that obtained by maximizing
the Hamming distance.

\begin{figure}[h]
  \centering
  \includegraphics[width=\columnwidth]{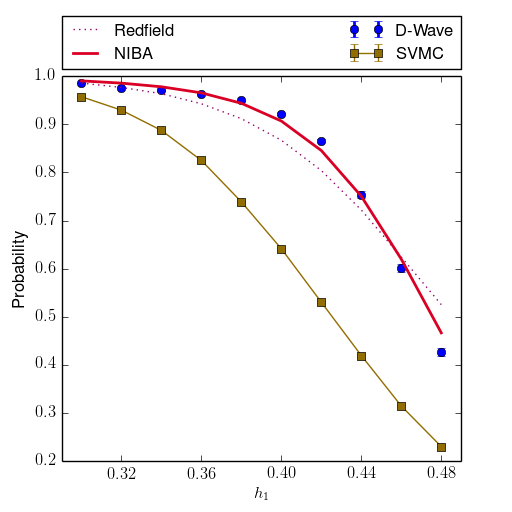}
  \caption{Probability of success versus $h_1$ for D-Wave (purple
    $\circ$ marker), open system quantum numerics and the classical-path model (SVMC, brown $\square$ marker). The open system
    numerics are obtained from the Redfield quantum master equation based on the transition rate \eqref{eq:W10fgr} (dotted line) and from the NIBA Quantum
    Master Equation (continuous red line). The NIBA Quantum Master
    Equation is a surprisingly good fit to the data. Error bars are
    smaller than markers. }
  \label{fig:p_vs_h1}
\end{figure}

We observe a very close correspondence between the results of the
analysis with the NIBA Quantum Master Equation and the D-Wave data
displayed in Fig.~\ref{fig:p_vs_h1}. This figure shows the probability
of success versus $h_1 = [0.3, .., 0.48]$. Tunneling can only be
present for $h_1 < 0.5$ when there is an energy barrier between the
local and global minimum (seen as a bifurcation in the semi-classical
energy landscape analysis, see Fig~\ref{fig:bifurcation}).  We
emphasize that for NIBA (and Redfield) we do not have any parameter
fitting. We use parameters obtained experimentally from MRT
measurements on the device (see
Eqs.~\eqref{eq:MRT}~\eqref{eq:MRT1}~\eqref{eq:fhy} and
App.~\ref{app:single_qubit_hamiltonian}).

\subsection{Numerical simulation of classical paths }\label{sec:svmc}

Our main purpose is to study multiqubit tunneling under experimental
conditions with current technology for programmable quantum annealing,
such as the D-Wave Two chip. One important component of this study
is the comparison to the detailed open quantum system theory outlined
in the previous section. In addition, we will compare the experimental
data with semi-classical numerics that simulate the evolution under the
effective potential $U(q_1, q_2, s)$, as introduced in
Sec.~\ref{sec:effective_potential}. We are interested in 
numerical methods that fulfill the following conditions:
\begin{itemize}
\item They must be constrained to quantum product states, quantum correlations are disallowed.
\item They do not include collective state or cluster updates. This prevents
  quantum tunneling to be included in the simulation. The dynamical
  equations of the numerical method must specify only
  equations of motion for each individual qubit in the product state
\item The simulation must be capable of including the effective
  potential $U(q_1, q_2, s)$ of Sec.~\ref{sec:effective_potential}.
\end{itemize}

One such method was introduced recently in Ref.~\cite{shin2014quantum}
and studied in related
works~\cite{vinci2014distinguishing,albash2014reexamining}. For these methods,
dynamics are constrained to spin-vector product states, with one spin
vector per qubit. For a given product state, we denote by $\theta_\mu$ the
angle of the spin vector for qubit $j$ with the $x$ quantization
axis. For a given Hamiltonian $H_0(s)$, we denote the corresponding
energy by $E_s(\theta_1,\ldots,\theta_{n_q})$, where $n_q$ is the number of qubits. The evolution consists of a
sequence of sweeps along the Hamiltonian path $\{H_0(s)\}$. In each
sweep, a Monte Carlo update is proposed for each qubit in the
following manner:
\begin{itemize}
\item A new angle $\theta'_\mu$ is drawn from the uniform distribution
  in $[0,2\pi]$.
\item The spin vector for qubit $j$ is updated $\theta_\mu \leftarrow
  \theta'_\mu$ according to the Metropolis rule for the energy difference
  \begin{align*}
    \quad D = E_s(\ldots,\theta'_\mu,\ldots)-
  E_s(\ldots,\theta_\mu,\ldots)\;.
  \end{align*}
  That is, the move is
  always accepted if $D$ is negative, and with probability given by the
  Boltzmann factor $\exp(-D/k_B T)$ if $D$ is positive. 
\end{itemize}

We call this method Spin Vector Monte Carlo (SVMC). The initial state
is chosen to be the global minimum of the transverse field. When the
spin vectors of each cluster are aligned with the parameters
$\{q_1,q_2\}$ we obtain $E_s(q_1,\ldots,q_2)=U(q_1, q_2, s)$. For low $T$ and sufficient sweeps,
the evolution proceeds along the false minima path of
Fig.~\ref{fig:magnetization}. That is, the numerical method at low
temperature simulates the classical-path model outlined in
Sec.~\ref{sec:effective_potential}.

\begin{figure}[h]
  \centering
  \includegraphics[width=\columnwidth]{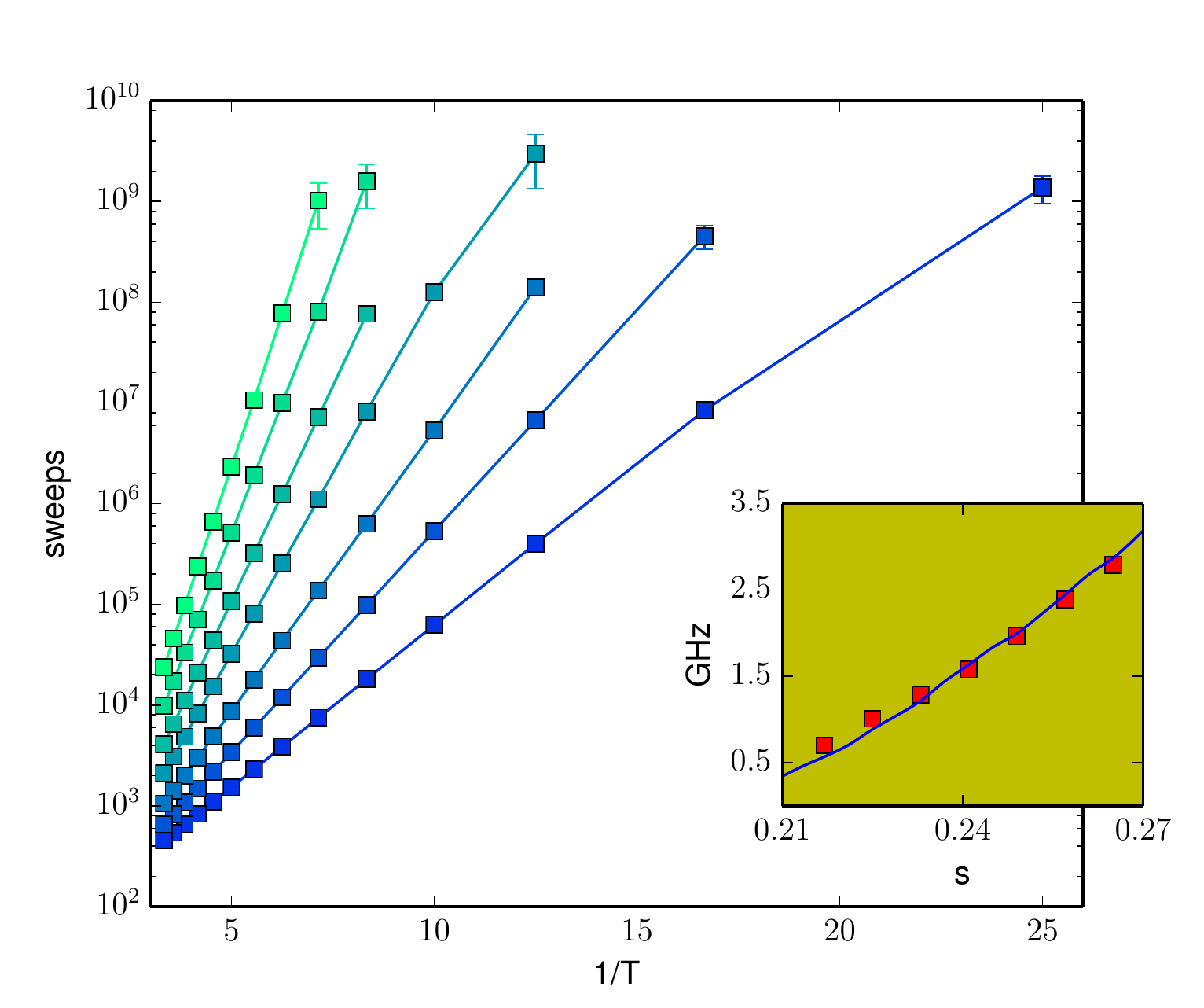}
  \caption{Analysis of the activation energy for Kramer's scape  for SVMC. The main figure
    shows, in a semilog scale, the average number of sweeps as a
    function of temperature. We plot lines for different points in the
    annealing schedule, from $s=0.217$ (dark blue) to $s=0.265$
    (green). The embedded figure shows the activation energy (red dots) and the
    semi-classical energy barrier (blue). There is a good
    correspondence between SVMC and the effective energy potential. }
  \label{fig:svmc_kramers}
\end{figure}

This numerical method allows us to study thermal hopping between the
minima of the effective potential $U(q_1, q_2, s)$. To check this
correspondence, we studied the height of the energy barrier obtained
from Kramer's theory applied to SVMC. For the potential $U(q_1, q_2,
s)$ at a fixed value of $s$, we initialized the spin vector state at a
local minima, and watch for Kramer events. A Kramer event
corresponds to the arrival at the other minima under thermal
activation. According to Kramer's theory, the dependence on
temperature for the expected number of sweeps necessary for a Kramer
event follows the formula $\exp(\Delta U/T)$, where $\Delta U$ is the
height of the energy barrier. We extract the energy barrier by fitting
the curve of the average number of sweeps for different $T$. We find
that this matches almost exactly the energy barrier height from
$U(q_1, q_2, s)$ in Fig.~\ref{fig:bifurcation} for different values of
$s$, see Fig.~\ref{fig:svmc_kramers}. We also studied other
semi-classical methods, such as a mean-field Redfield model similar to
Forster's theory, and a Landau-Lifshitz-Gilbert model related to the
one studied in Ref.~\cite{crowley2014quantum}. As we could not recover
the barrier height of the effective energy potential $U(q_1, q_2, s)$
from the Kramer events with these other numerical
methods, we will use SVMC in what follows.

\begin{figure*}[t]
  \centering
  \includegraphics[width=\textwidth]{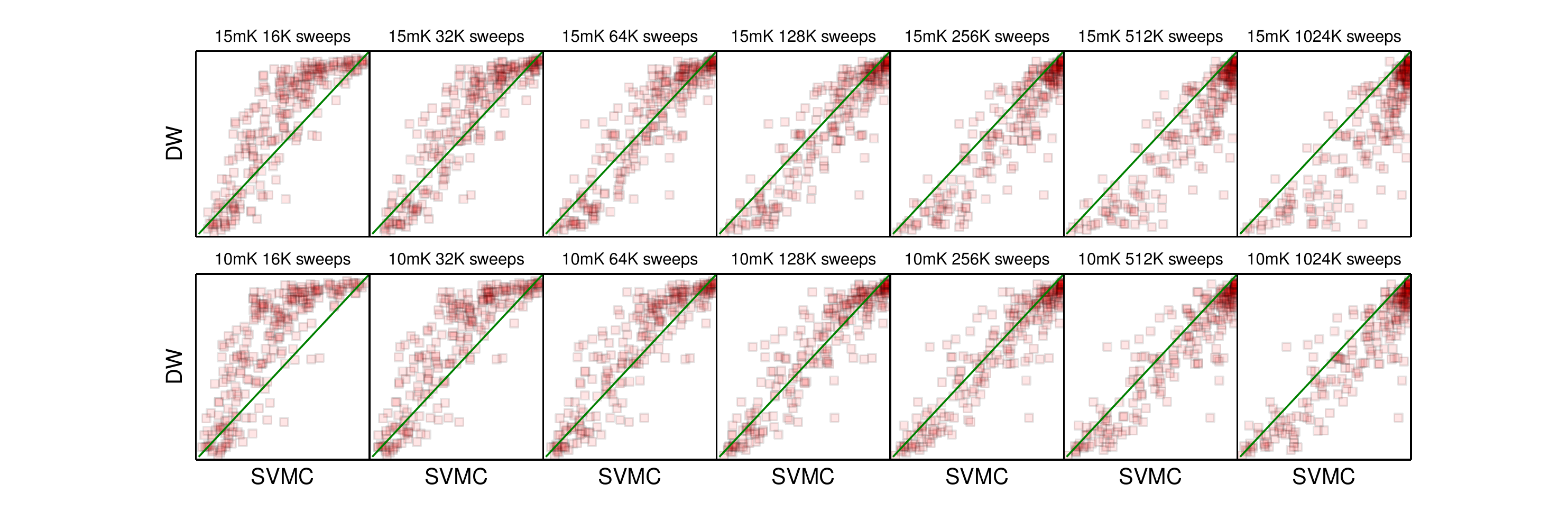}
  \caption{Scatter plots showing the correlation of D-Wave Two data and SVMC
    for the random Ising benchmark for different algorithmic
    temperatures (in mK) and number of sweeps. We will use the
    parameters T = 15 mK, and sweeps = 128,000 in the rest of the
    paper. }
  \label{fig:dw_vs_svmc}
\end{figure*}

A disadvantage of SVMC as outlined above and introduced in
Ref.~\cite{shin2014quantum} is that there is no natural choice to
relate the number of sweeps to the physical evolution time. As in
other works, we will choose the number of sweeps in order to obtain a
good correlation with the probability of success of the D-Wave chip
for a benchmark of random Ising models with binary couplings $J_{\mu\nu}
\in \{1,
-1\}$~\cite{boixo_evidence_2014,shin2014quantum,ronnow_defining_2014,albash2014reexamining}.
This will allow us to phenomenologically correlate the number of sweeps
to physical time.  We set the algorithmic temperature of SVMC to be
the same as the physical temperature because we are interested in the
dependence of the success probability with temperature. There are no
important differences for the correlation with other temperature
choices. The correlation with the random Ising benchmark for 128,000
sweeps (see Fig.~\ref{fig:dw_vs_svmc}) is 0.92, and the residual
probabilities $p_{\rm SVMC} - p_{\rm D-Wave}$ have a mean of 0.05 and
a standard deviation of 0.12. This is consistent with the best values
found over a wide range of parameters. We will use 128,000 sweeps at 15
mK as our reference rate for the rest of the paper.

Another parameter, the so-called qubit background susceptibility
$\chi$, has been introduced in the literature to improve the
correlations between numerical simulations and the D-Wave
data~\cite{vinci2014distinguishing,albash2014reexamining}. While the
physical motivation for this parameter is well understood, it is also
treated in those works as a free parameter together with the number of
sweeps. Increasing values of $\chi$ have the effect of decreasing the
barrier height for $h_1 < J/2$. We have designed a specific problem to
bound the range of choices of $\chi$ for SVMC compatible with
experimental data. We find $\chi=0.0025$ to be the value most
consistent with the data for the device used in our paper, as seen in
App.~\ref{app:chi_probe}. Plots that include this choice of $\chi$ for
SVMC are also presented in the Appendix.

\section{Experimental results from the D-Wave Two processor and fit to
  theory}

\subsection{Double-well potential with two clusters}\label{sec:experimental_results}

One of the most distinctive signatures of quantum tunneling when
compared to thermal hopping is the response to
temperature variations at low temperatures. Consider first the quantum
tunneling situation. For low temperatures compared to the gap, and when
the tunneling rate is fast compared to the evolution time, the final
success probability is close to unity. As we increase the temperature in
the range of the gap, we expect to see thermal excitations and a lower
probability of success. Therefore, the expected tendency at low
temperatures is a decrease of probability of success with increasing
temperature.

Consider now the situation with thermal hopping. At very
low temperatures, the state follows the classical path along the local
minimum through the evolution. If this path does not connect to the
global minimum, the probability of success is close
to zero. As we increase the temperature, we also increase the
probability of a thermal excitation over the energy
barrier, and therefore increase the probability of success. Consequently, the
expected tendency at low temperatures is an increase in the
probability of success for models that follow the classical paths.

\begin{figure}[h]
  \centering
  \includegraphics[width=\columnwidth]{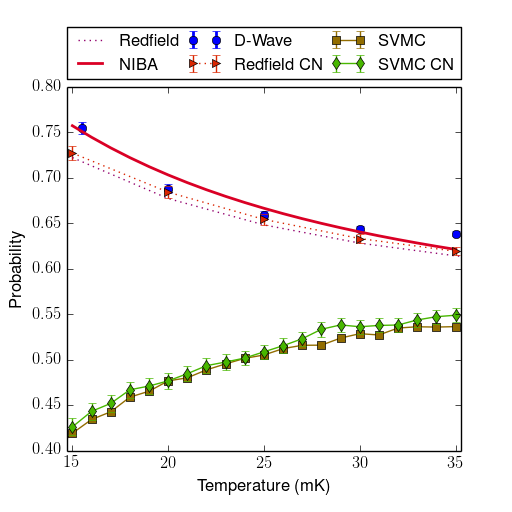}
  \caption{Probability of success versus temperature at $h_1 = 0.44$
    for D-Wave (purple $\circ$ marker), open system quantum numerics
    and the classical-path model (SVMC). The open system numerics are
    Redfield (doted line), Redfield with physically estimated control
    noise of $\sigma_h = 0.05$ for the local fields and
    $\sigma_J=0.035$ for the couplings (doted line with
    $\vartriangleright$ marker) and the NIBA Quantum Master Equation
    (continuous red line). The two SVMC curves correspond to SVMC
    (brown $\square$ marker) and SVMC with the physically estimated
    control noise of $\sigma_h = 0.05$ for the local fields and
    $\sigma_J=0.035$ for the couplings (SVMC-CN, green $\lozenge$
    marker). Error bars are smaller than markers when not seen. D-Wave
    data fits well the quantum models. The temperature dependence of
    SVMC is the opposite. It is important to emphasize that in this
    temperature range the lowest two energy states (the ones that
    participate in the double well potential) account for all the
    probability (0.9998 in the experimental D-Wave data, 0.99998 for SVMC).}
  \label{fig:p_vs_t}
\end{figure}

Figure~\ref{fig:p_vs_t} shows the success probability as a function of
temperature for D-Wave, open system quantum numerics and the
classical-path model (SVMC) for $h_1 = 0.44$ in the Hamiltonian of
Eq.~\eqref{eq:Hp_total}. There is a clear tendency towards lower
probability of success as temperature increases in the experimental
D-Wave data. The same is true of the various open system quantum
master equations models. This is a consequence of quantum
tunneling. Interestingly, SVMC shows a positive correlation between
success probability and temperature. This is a consequence of thermal
hopping above the energy barrier. The probability of success obtained
with the Redfield quantum master equation matches well the D-Wave
data, and it is not affected by the control noise of the D-Wave
chip. SVMC is run at an algorithmic temperature equal to the physical
temperature indicated in the horizontal axis, and with 128,000 sweeps,
as explained in Sec.~\ref{sec:svmc}. We plot SVMC without control
noise and SVMC with the physically estimated control noise. Averaging
over control noise does not have a significant effect on the
probability of success for SVMC. As seen in Figure~\ref{fig:p_vs_h1},
the NIBA Quantum Master Equation model is a better match to the data
than the Redfield model. In the region $h_1 < 0.5$ there is an energy
barrier between the local and global minimum. In this region, we see
that the probability of success for SVMC is significantly lower than
the probability of success for D-Wave and open system quantum models.

\begin{figure}[h]
  \centering
  \includegraphics[width=\columnwidth]{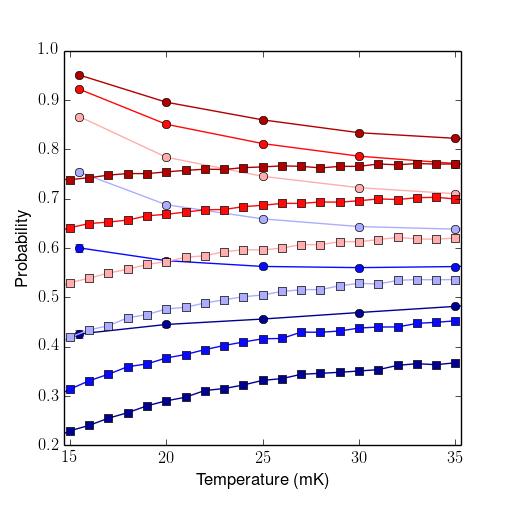}
  \caption{Probability of success versus temperature for D-Wave data
    ($\circ$ markers) and SVMC numerics ($\square$ markers). We plot
    (from top to bottom, and red to blue) $h_1 = [0.38, 0.4, 0.42,
    0.44, 0.46, 0.48]$. Error bars are smaller than markers. We use
    SVMC with 15 mK algorithmic temperature and 128,000 sweeps, as
    explained in the text.}
  \label{fig:dwave_p_vs_t}
\end{figure}

\begin{figure}[t]
  \includegraphics[width=\columnwidth]{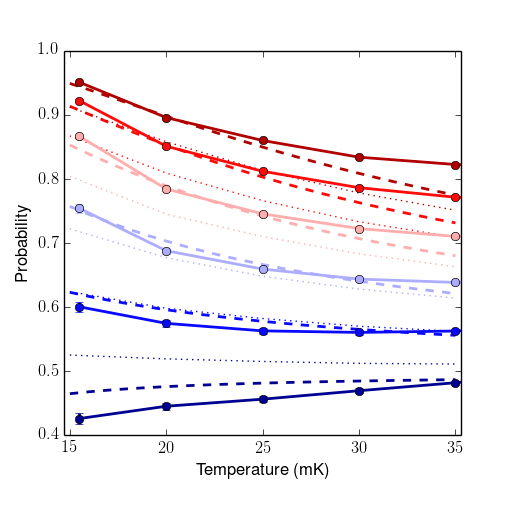}
  \caption{Probability of success as a function of temperature for
    $h_1 = [0.38, 0.4, 0.42, 0.44, 0.46, 0.48]$. We plot D-Wave data
    ($\circ$ markers), Redfield (dotted line) and the NIBA Quantum
    Master Equation (dashed line). The main qualitative difference is
    that for $h_1 = 0.48$ the NIBA Quantum Master Equation predicts a
    much lower probability of success, which increases with
    temperature. We see the same feature in D-Wave's experimental
    data. In the NIBA Quantum Master Equation, this is due to the
    suppression of the tunneling rate by the low-frequency noise. The
    gap at the avoided crossing for $h_1=0.48$ is 10 MHz. The standard
    Redfield model does not include low-frequency noise.}
  \label{fig:redfield_polaron}
\end{figure}

\begin{figure}[t!]
  \centering  
  \includegraphics[width=.5\textwidth]{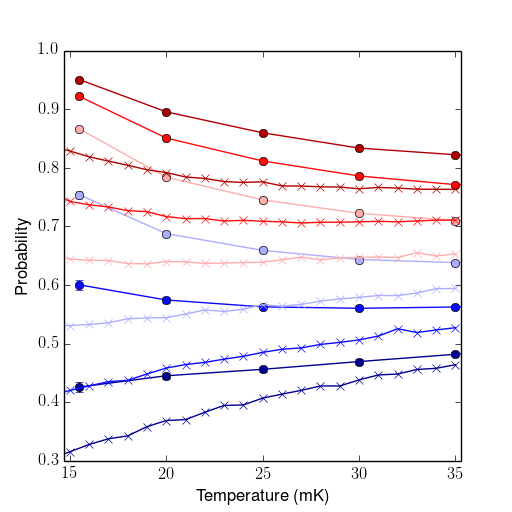}
  \caption{Probability of success versus temperature for D-Wave data
    ($\circ$ markers) and Quantum Annealing Path Integral Monte Carlo Quantum Annealing
    (PIMC-QA) numerics ($\times$ markers). We plot
    (from top to bottom, and red to blue) $h_1 = [0.38, 0.4, 0.42,
    0.44, 0.46, 0.48]$. Error bars are smaller than markers. PIMC-QA is
    not a good fit to D-Wave's data.}
  \label{fig:p_vs_t_sqa}
\end{figure}

Figure~\ref{fig:dwave_p_vs_t} shows the probability of success versus
temperature for D-Wave data and SVMC numerics for $h_1 = [0.38, 0.4,
0.42, 0.44, 0.46, 0.48]$. The probability of success decreases with
temperature for D-Wave in instances with a significant coherent
tunneling contribution to the dynamics.  For SVMC the probability of
success increases with temperature in all cases. As noted before, the
probability of success form SVMC is lower than the probability of
success of D-Wave. Figure~\ref{fig:redfield_polaron} shows the
probability of success versus temperature for D-Wave data and open
quantum systems numerics for the same values of $h_1$.  The D-Wave
data reproduces the reduction in probability of success predicted by
the quantum models. The probability of success does increase with
temperature for D-Wave for the instance with $h_1 = 0.48$, where the
minimum gap is 10 MHz. The limitation in this case is strong coupling
to low frequency noise. This behavior is not captured by standard
Redfield theory. To explain it, we must take into account the
reorganization energy induced by low frequency noise, as in standard
Marcus theory. The NIBA Quantum Master Equation does capture this
effect correctly. For this gap size, coherent quantum tunneling is
suppressed.


\subsection{Larger problems that contain the weak-strong cluster ``motives'' as subproblems}\label{sec:locally_correct_main}

\begin{figure}[h]
     \centering     
     \subfloat[Stacked clusters]{
       \includegraphics[width=0.45\columnwidth]{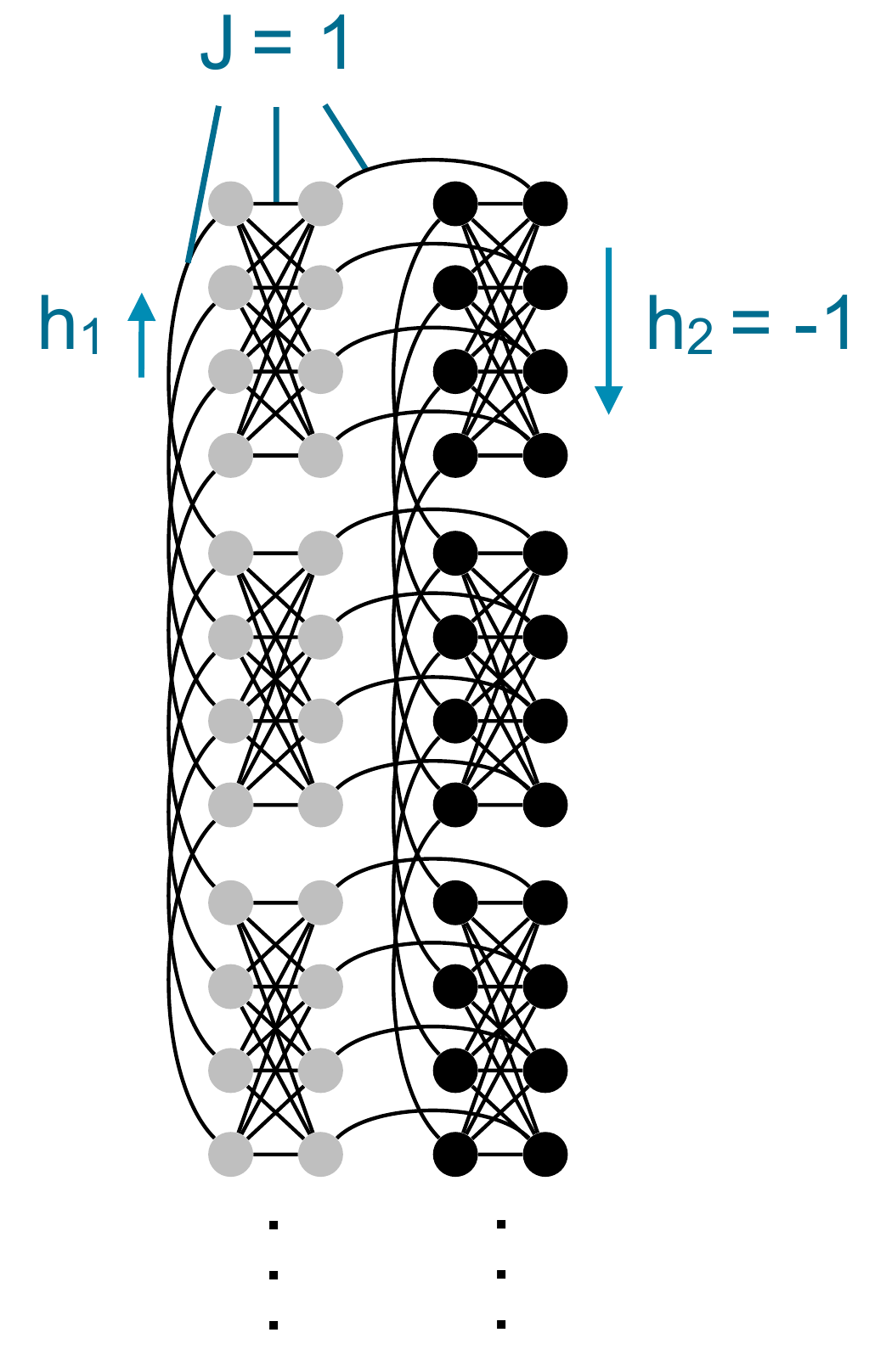}
       \label{fig:stacked_cells_graph}
     }
    \subfloat[Glass of weak-strong cluster pairs]{
      \includegraphics[width=0.45\columnwidth]{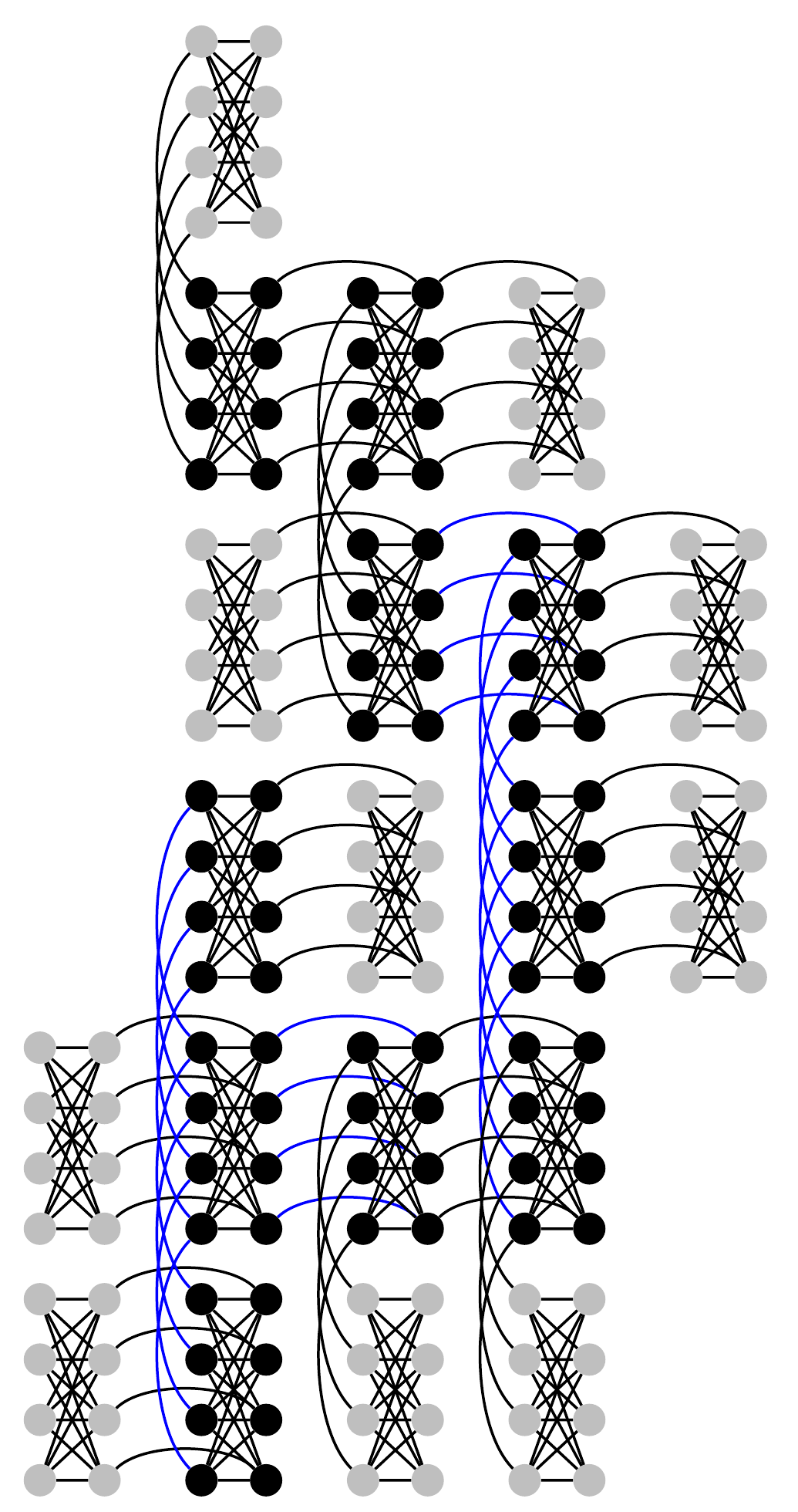}
      \label{fig:locally_correct_graph}
    }
    \caption{Larger problems that contain the weak-strong cluster ``motives'' as subproblems. (a) A stack of weak-strong cluster pairs. (b) Weak-strong cluster pairs connected in a glassy fashion by setting all connections between any two neighboring strong clusters randomly to either $-1$ or $+1$ The $-1$ anti-ferromagnetic connections are depicted in blue. }
\end{figure}

In the previous sections we established that quantum tunneling assists the D-Wave Two processor in finding the global minimum of the weak-strong cluster probe problem. The 16 qubit problem we considered was specifically designed to be suitable for studying the role of tunneling by analytical, numerical and experimental means. A generalization to a larger number of qubits is achieved by studying problems that contain the weak-strong cluster ``motive'' multiple times within the connectivity graph.

\begin{figure}[h]
  \centering
  \includegraphics[width=\columnwidth]{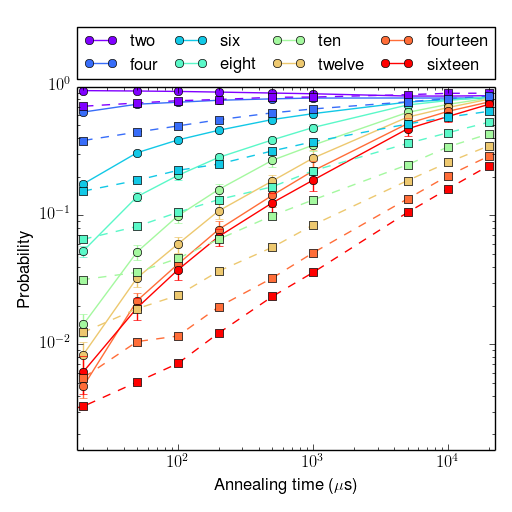}
  \caption{Success probabilities for varying numbers of stacked
    weak-strong cluster pairs as a function of annealing time for a
    weak local field $h_1=0.4$ and $J=1$. The different colors show
    'two' clusters, one 'weak' and one 'strong', as in
    Sec.~\ref{sec:experimental_results}, 'four' clusters (two weak,
    two strong), 'six' clusters (as in
    Fig.~\ref{fig:stacked_cells_graph}), etc... We show D-Wave data
    ($\circ$ marker, continuous lines) and SVMC ($\square$ marker,
    dashed lines). Note that for larger number of qubits the success
    probability for the D-Wave increases faster with annealing time
    than for SVMC with a proportional increase in the number of
    sweeps.}
  \label{fig:clusters_stacked}
\end{figure}

The first generalized configuration we studied is a stack of weak-strong
cluster pairs with $h_1 = 0.4$ setting all connections between the left columns of the
unit cells in the Chimera graph to ferromagnetic 1 (see
Fig.~\ref{fig:stacked_cells_graph}). As the number of stacked cluster
pairs grows, the success probability decreases for the annealing time
of 20 $\mu$s that was used in the previous sections. This behavior is
expected since the minimum gap also decreases. When we increase the
annealing time to 20 ms the success probability grows significantly. The
increase of the success probability for SVMC is much slower with a
proportional increase in the number of sweeps (note the logarithmic
scale) even for instances with 128 qubits, see
Fig.~\ref{fig:clusters_stacked}.

\begin{figure}[h]
  \centering
  \includegraphics[width=\columnwidth]{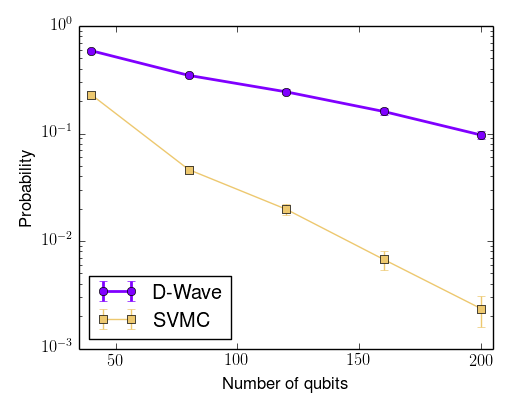}
  \caption{
    Success for a glass of weak-strong clusters as a function of the
    number of qubits involved. D-Wave data is plotted with purple
    $\circ$ markers. SVMC is plotted with $\square$ 
    markers. The fitting exponent for the D-Wave data is $-(1.1 \pm
    0.05)\cdot 10^{-2}$, while the fitting exponent for the SVMC numerics is $-(2.8
    \pm 0.17)\cdot 10^{-2}$. The error estimates for the exponents are obtained by
    bootstrapping.}
  \label{fig:clusters_glass}
\end{figure}

In a second experiment we again placed a number of weak-slow cluster
pairs across the Chimera graph. Then we connected the strong clusters
in a glass like structure by randomly setting all Chimera connections
between neighboring pairs of strong clusters to +1 or
-1. Fig.~\ref{fig:locally_correct_graph} depicts a problem instance
constructed this way.  The success probabilities are shown in
Fig.~\ref{fig:clusters_glass}. We fit the average probability $p(n_q)$ as $p(n_q) \propto \exp(-
\alpha n_q)$, where $n_q$ is the number of qubits. The fitting exponent $\alpha$
for the D-Wave data is $-(1.1 \pm 0.05)\cdot 10^{-2}$, while the
fitting exponent for the SVMC numerics is $-(2.8 \pm 0.17)\cdot
10^{-2}$. For additional data including problems for which the strong
fields have been set to zero please refer to Appendix
\ref{sec:locally_correct}.

\section{Conclusion}

There has been a great deal of debate as to whether quantum resources in
the D-Wave Two processor are employed in a manner that enhances the
probability for the device to return low energy solutions of encoded optimization problems.
To address this question we programmed optimization problems corresponding to  two weakly coupled   ferromagnetic clusters with strong intra-cluster interactions and local fields acting on two  clusters   in the opposite  directions.  This construction encodes  the simplest non-convex
optimization problem that only exhibits one false and one global
minimum in a time-dependent effective potential. The time evolution is
such that a path in the potential over product states connects the
initial global minimum with the final false minimum. The final global
minimum can only be reached by traversing an energy
barrier. Experimentally, we found that for this situation the D-Wave Two
quantum annealer returns the solution that minimizes the energy with
consistently higher probability than physically plausible models of the hardware
that only employ product states which do not allow for multiqubit
tunneling transitions. On the contrary open system quantum mechanical
models are in a very close correspondence with the  hardware data without using any fitting parameters.  We developed a multiqubit Quantum Master
Equation using the Non Interacting Blip approximation (NIBA) which takes high and low frequency noise into account.  It
continuously rotates the basis in multiqubit Hilbert space to coincide with the basis
that minimizes the transition rate between the first two levels. In
this way we find the most robust states under decoherence. One can
think of this as working in the instantaneous pointer
basis~\cite{Zurek81,Zurek93}. The polaron transform was used since the
interaction of the qubits with their oscillator baths forms
polaron-like quasi particles. To increase our confidence that quantum
mechanical models are indeed required to describe the D-Wave annealing
dynamics properly, we performed a series of experiments in which we varied
the temperature of the chip. Regardless of specific parametrizations of
quantum and classical models, the apparent trend between temperature and
success probability revealed by these experiments is consistent only with quantum models.

Quantum correlations are strongest during the annealing near the
avoided crossing. In our problem fast collective tunneling processes involving
many qubits near the avoided crossing give rise to adiabatic
eigenstates where the states of the 8 spins forming a tunneling
cluster contain quantum correlations. Environmental effects modify the
ideal system (adiabatic) states into pointer states that still retain
quantum superpositions (non-product nature). In this our study is
different from the previous study of incoherent qubit tunneling near
the minimum point observed using the D-Wave-I chip
\cite{dickson2013thermally}.

In contrast to a simple 2-level system tunneling, the overlap between
the states is relatively large at the point of minimum gap, and
decreases exponentially after that.  The initial overlap gives rise to
fast transitions and thermalization which is maintained throughout the
avoided crossing region.  After the avoided crossing the energy
splitting increases, leading to thermal de-population of the excited
state. Transitions slow down because of the steep decrease of the
collective tunneling matrix elements. Eventually, the population of
the excited state freezes at the point where the transition rate is
reduced to the level of the quantum annealing rate. It should be noted
that the freezing described here is a multiqubit effect that is to be
distinguished from the single qubit freezing that occurs at the end of the
annealing evolution due to the raising of the barrier of the individual
qubits. Higher temperature will lead to greater thermal population of
the excited state at this freezing point, and to the decrease of the
success probability. This results in the opposite temperature effect
compared to the limit of incoherent tunneling, where the level
population immediately after avoided crossing is inverted (the excited
state has a larger population). In this case an increase of
temperature would lead to an increase of the success probability.

In our studies we employed single qubit noise model parameters
reconstructed from the MRT studies of D-Wave One
\cite{PhysRevB.83.180502} and D-Wave Two systems.  Results for the 16
qubit quantum annealing success rates are in remarkably close
correspondence with experimental data over the range of local fields
and temperatures studied here (see Fig.~\ref{fig:p_vs_h1}).

The correlation between D-Wave's experimental data and Path Integral Monte Carlo along the Quantum Annealing schedule (PIMC-QA) has been studied in recent
works~\cite{boixo_evidence_2014,ronnow_defining_2014,albash2014reexamining}. Unfortunately,
there is no known formal connection between Monte Carlo updates in PIMC-QA and open system quantum
dynamics. The relationship between PIMC-QA and quantum tunneling is also not well
understood. We show in Fig.~\ref{fig:p_vs_t_sqa} the
probabilities for PIMC-QA as a function of temperature for different
values of $h_1$. We use similar parameters for PIMC-QA as in
Ref.~\cite{boixo_evidence_2014}. The probability of success for PIMC-QA
is lower than the probabilities observed for D-Wave. On the one hand, there is an inverse relationship between temperature and success probability for small $h_1/J$ ratios (big
minimum gaps in the QA spectrum). On the other hand, this dependence
is opposite to D-Wave's data for $h_1 = 0.44$, the main case studied in Sec.~\ref{sec:experimental_results}.

Beyond the original 16 qubit probe problem we also explore larger
problems of up to 200 qubits that contain multiple weak-strong cluster
pairs. We found that classical-path models that only operate on
product states do not explain the hardware performance. The difference
in the fitting exponents is given in
Sec.~\ref{sec:locally_correct_main}. Appendix~\ref{sec:locally_correct}
contains the fitting exponents for different choices of parameters of
SVMC and PIMC-QA.

A way to think of multiqubit tunneling as a computational resource is
to regard it as a form of large neighborhood search. Collective
tunneling transitions involving K qubits explore a K variable
neighborhood. We find that the current generation D-Wave Two annealer
enables tunneling transitions involving at least 8 qubits. It will be
an important future task to determine the maximal K for the current
hardware and how large it can be made in next generation hardware.
The larger K, the easier it should be to translate the quantum
resource ``K-qubit tunneling'' into a computational speedup. We want
to emphasize that we do not claim to have established a quantum
speedup in this work. To this end one would have to demonstrate that
no known classical algorithm finds the optimal solution as fast as the
quantum process.  To establish such an advantage it will be important
to study to what degree collective tunneling can be emulated in
classical algorithms by employing cluster update methods. However the
collective tunneling phenomena demonstrated here present an important
step towards what we would like to call a {\it physical speedup}: a
speedup relative to a hypothetical version of the hardware operated
under the laws of classical physics.

To summarize, in this work, we demonstrate that a noticeable
computational role of collective tunneling can already be observed in existing
quantum annealing hardware, such as the D-Wave Two processor. This is
despite substantial environmental noise and control errors. Our study
can thus inspire the design and development of future generations of
powerful medium- or large-scale quantum information processors that
utilize collective tunneling as a useful quantum-computational resource. Future
designs might operate under longer coherence times, reduced control
errors, higher graph connectivity, and benefit from advanced quantum
error protection or correction procedures. Such improvements might be
necessary to convert the collective tunneling resource into a computational
advantage.

\vspace{0.5in} \textbf{Acknowledgements --} We would like to thank
Edward Farhi and John Martinis for reviewing and
discussing earlier versions of the paper. We also thank Ryan Babbush and Bryan O'Gorman for reviewing the manuscript, and Damian Steiger, Daniel Lidar and Tameem Albash for comments about the temperature experiment. The work of V.N.S. was
supported in part by the Office of the Director of National
Intelligence (ODNI), Intelligence Advanced Research Projects Activity
(IARPA), via IAA 145483 and by the  AFRL Information Directorate under grant F4HBKC4162G001.

\newpage
\onecolumngrid
\appendix

\section{Single Flux Qubit Hamiltonian}\label{app:single_qubit_hamiltonian}

\subsection{Full Flux Qubit Hamiltonian}

\begin{figure}
  \centering
  \includegraphics[width=.7\textwidth]{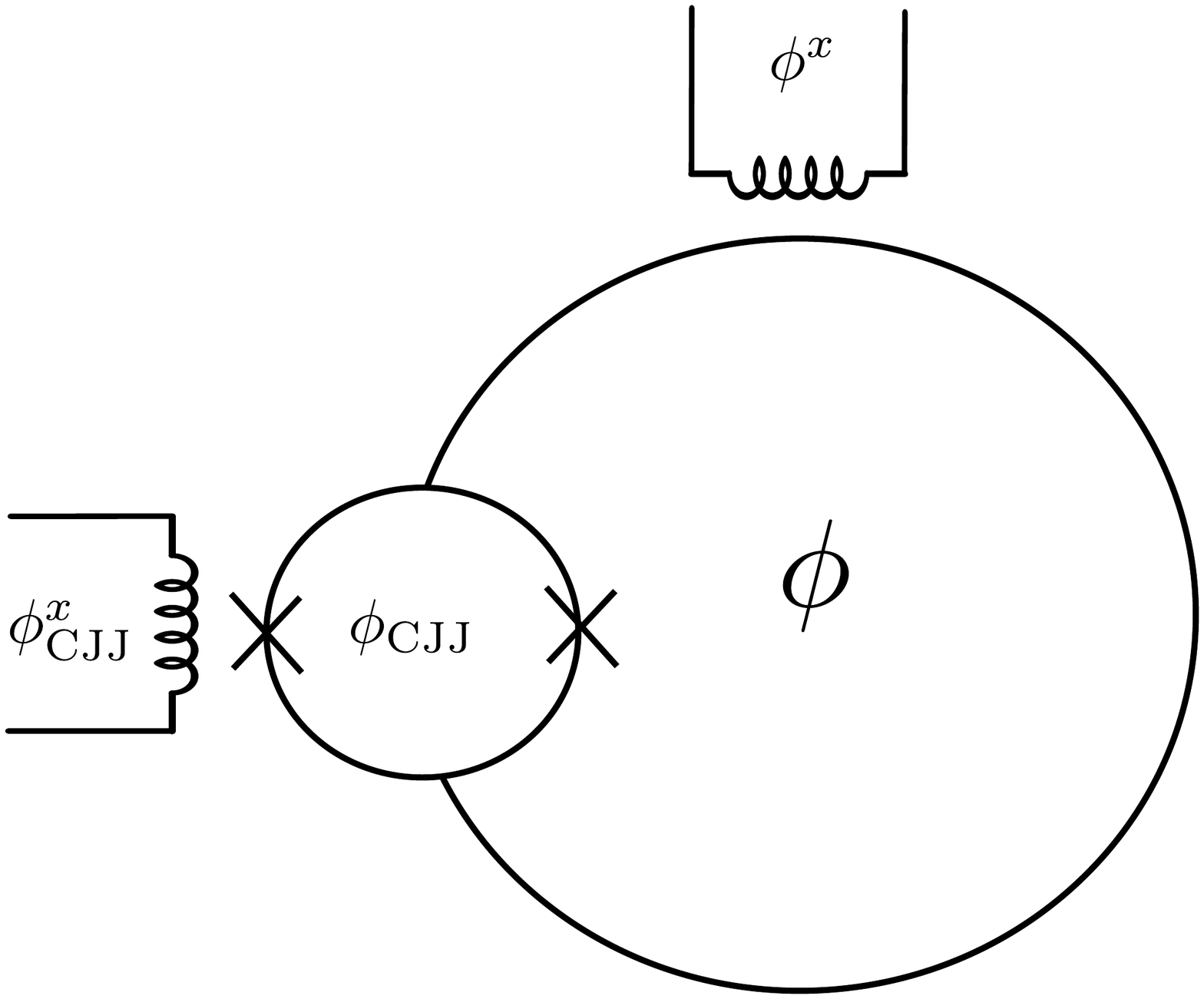}
  \caption{Schematic of a Compound Josephson Junction (CJJ) qubit. }
  \label{fig:cjj_qubit}
\end{figure}

The full Compound Josephson Junction (CJJ) flux qubit Hamiltonian is better expressed
in terms of flux phases, defined as renormalized fluxes $\phi = 2 \pi
\Phi / \Phi_0$ for any flux $\Phi$. The Hamiltonian
is~\cite{harris_experimental_2010}
\begin{align}
  - E_\C \partial_\phi^2  -E_\CCJJ \partial_{\phi_\CJJ}^2 + E_\J \cos(\phi)\cos(\phi_\CJJ/2) + E_\L \frac {(\phi-\phi^x)^2} 2 + E_\LCJJ \frac {(\phi_\CJJ - \phi^x_\CJJ)^2} 2\;,
\end{align}
where $\phi$ is the body flux phase to be quantized, $\phi^x$ is the
external flux phase, $\phi_\CJJ$ is the flux phase of the CJJ and
$\phi^x_\CJJ$ is the external flux of the CJJ (see
Fig.~\ref{fig:cjj_qubit}). The energies of the different terms are
given by
\begin{align*}
  E_\C &= \frac {(2e)^2}{2 C}  & E_\CCJJ &= \frac {(2e)^2}{2 (C/2)} & E_\J &= \frac {I_c \Phi_0}{2\pi} \\
  E_\L &= \(\frac {\Phi_0} {2 \pi}\)^2 \frac 1 {L + L_\CJJ/4}  &  E_\LCJJ &= \(\frac {\Phi_0} {2 \pi}\)^2 \frac 1 {L_\CJJ}& &
\end{align*}
The parameters are the capacitance $C$, the body inductance of the
main flux loop $L$ and of the Compound Josephson Josephson $L_\CJJ$,
and the effective critical current of the Compound Josephson Junction
$I_c$.

The median values for D-Wave's CJJ flux qubits in GHz are
\begin{align}\nonumber
  \frac {E_\C} {2 \pi \hbar} &= 0.67 \GHz &
  \frac {E_\CCJJ} {2 \pi \hbar} &= 1.35 \GHz &
  \frac {E_\J} {2 \pi \hbar} &= 1071 \GHz \\
  \frac {E_\L} {2 \pi \hbar} &= 537 \GHz &
  \frac {E_\LCJJ} {2 \pi \hbar} &= 11680 \GHz && \nonumber\;.
\end{align}

\begin{figure}
  \centering
  \includegraphics[width=10cm]{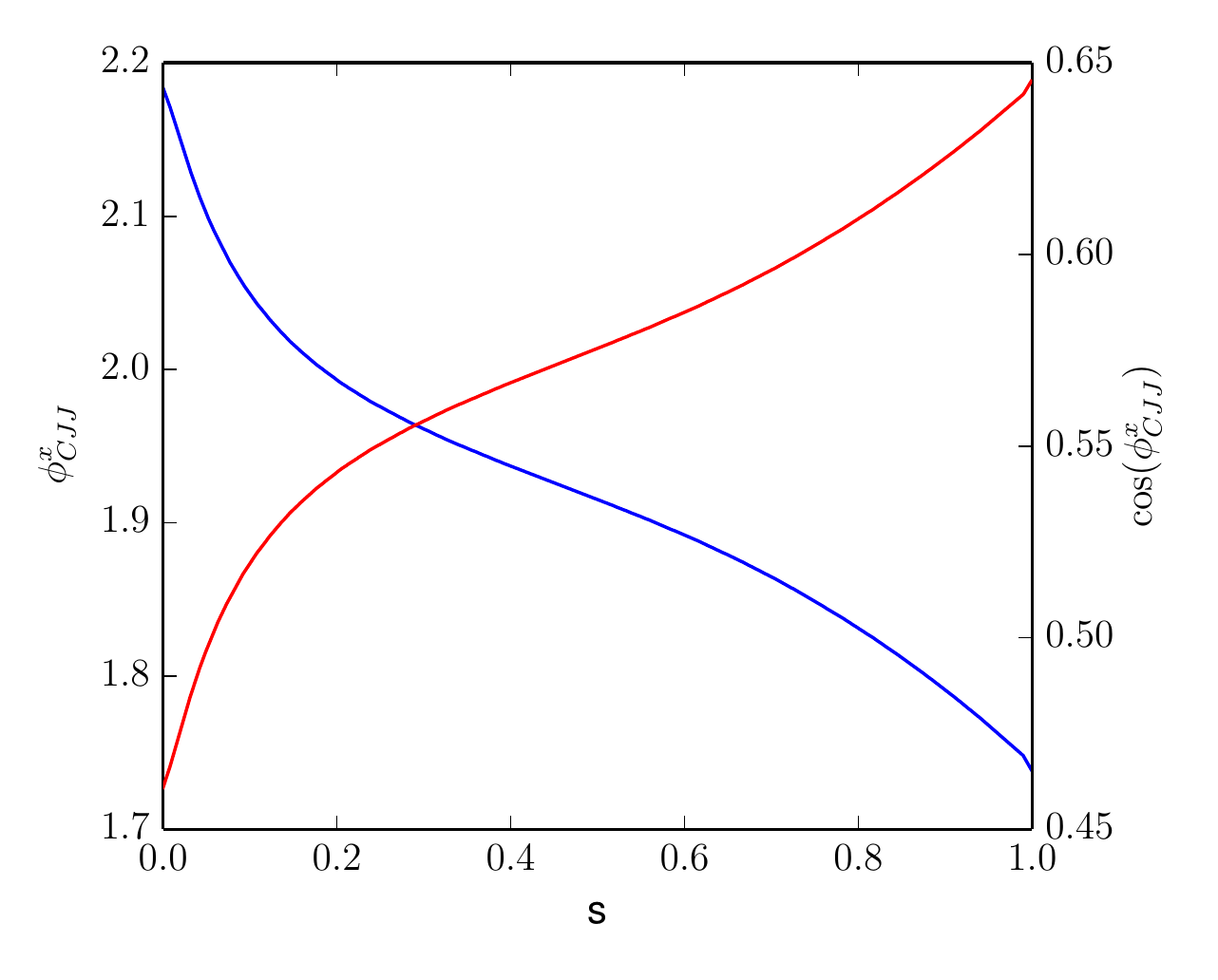}
  \caption{The CJJ external phase $\phi^x_\CJJ$ (blue) and
    $\cos(\phi^x_\CJJ)$ (red). The dependence of $\phi^x_\CJJ$ on the
    parameter $s$ is chosen so that the persistent current $I_p(s)$
    scales linearly with $s$, see Eq.~\eqref{eq:lin_ip}.}
  \label{fig:cjj_phase}
\end{figure}

The CJJ flux phase $\phi^x_\CJJ$ controls the quantum annealing
evolution. The function $\phi^x_\CJJ(s)$ as a function of the
annealing parameter for the quantum annealing schedule employed in
this paper is plotted in Fig.~\ref{fig:cjj_phase}.

\begin{figure}
  \centering
  \subfloat[Simplified 1D potential.]{\includegraphics[width=8cm]{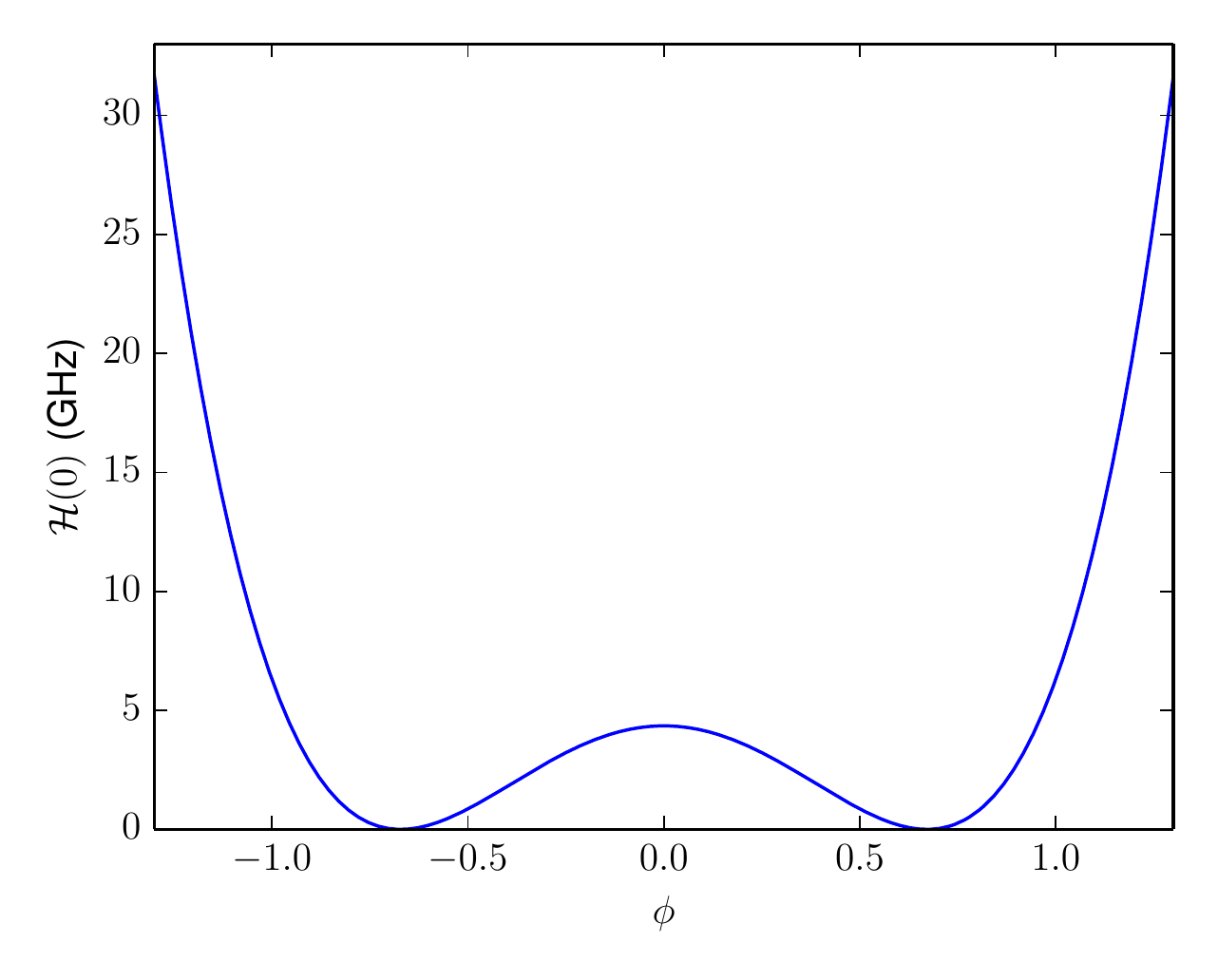}\label{fig:simplified_potential}}
  \subfloat[First two eigenstates.]{\includegraphics[width=8cm]{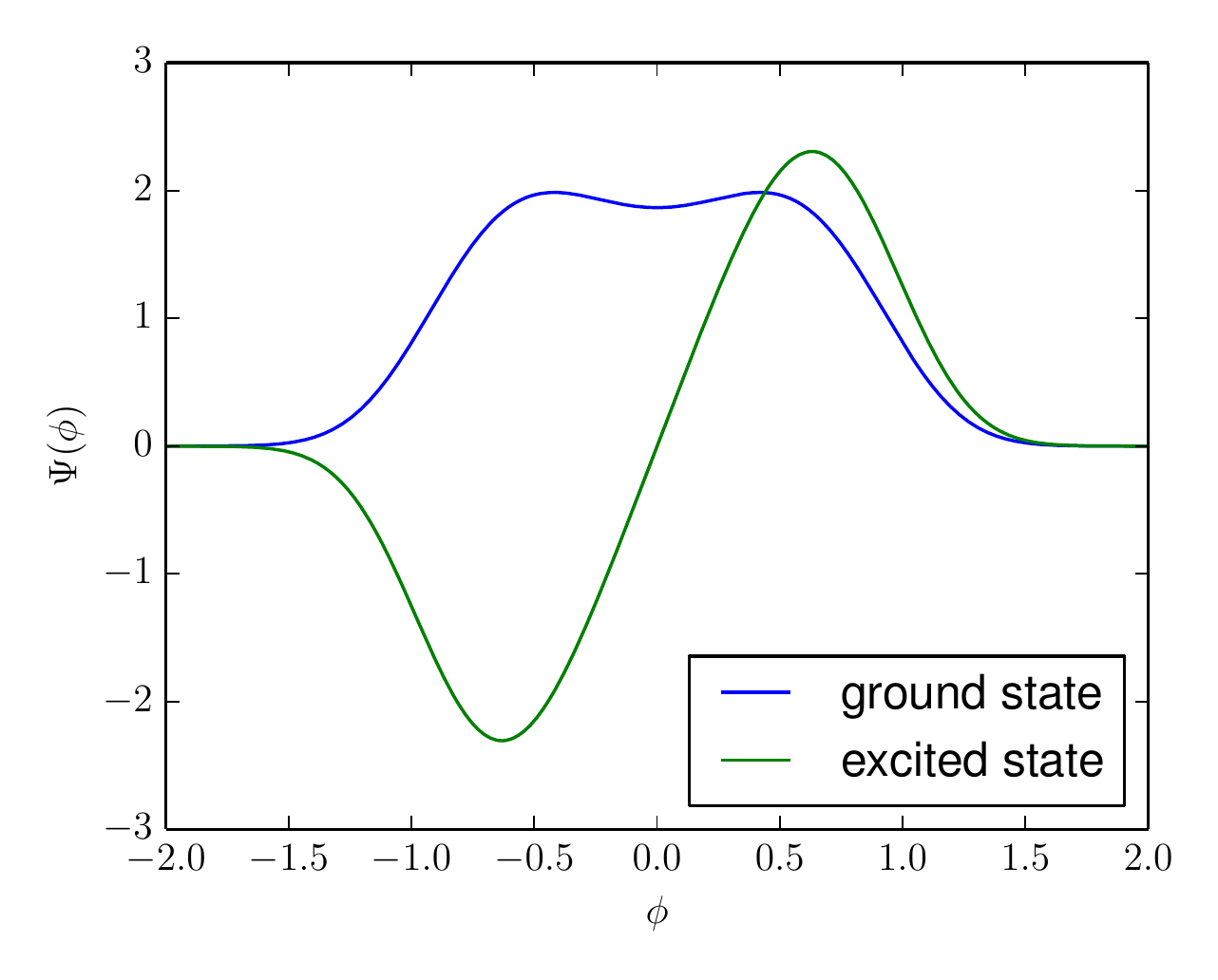}\label{fig:flux_eigenvectors}}
  \caption{(a) The simplified 1D potential $\mathcal{H}(\phi^x)$
    of Eq.~\eqref{eq:simplified_hamiltonian} for annealing parameter
    $s=0.278$. (b) The first two eigenvectors
    of the flux qubit potential for $s=0.278$.}
  \label{fig:flux_qubit_potential}
\end{figure}

Because $E_\LCJJ \gg
E_\L$ the phase $\phi_\CJJ$ can be assumed to be centered at the
value given by $\phi^x_\CJJ$, as a first approximation. The approximated
flux qubit Hamiltonian is then 
\begin{align}\label{eq:simplified_hamiltonian}
  \mathcal{H}_s(\phi^x) = - E_\C \partial_\phi^2 + E_\J \cos(\phi)\cos(\phi^x_\CJJ(s)/2) + E_\L \frac {(\phi-\phi^x)^2} 2 \;.
\end{align}
This potential is plotted in Fig.~\ref{fig:simplified_potential}.

\subsection{Effective qubit Hamiltonian}\label{sec:effective_qubit}
The effective qubit Hamiltonian is the simplified Hamiltonian
$\mathcal{H}_s(\phi^x)$ of Eq.~\eqref{eq:simplified_hamiltonian}
projected into the two lowest energy energy levels $\{\ket {{\mathfrak
    g}(s)}, \ket {{\mathfrak e}(s)}\}$ of $\mathcal{H}_s(0)$
\begin{align}
   \mathcal{H}_s(\phi_x) \Big|_{\{\ket {{\mathfrak g}(s)},
     \ket {{\mathfrak e}(s)}\}} =  \mathcal{H}_s(0) + \phi^x \frac{\partial \mathcal{H}_s(0)}{\partial \phi^x}  \Big|_{\{\ket  {{\mathfrak g}(s)},
     \ket {{\mathfrak e}(s)}\}} \;.
\end{align}The eigenvectors of $\mathcal{H}_s(0)$ are symmetric and
anti-symmetric superpositions of the flux up and down state in the
double-well potential (see Fig.~\ref{fig:flux_eigenvectors})
\begin{align}
  \ket {{\mathfrak g}(s)} &= \frac 1 2 (\ket{\uparrow(s)} + \ket{\downarrow(s)}) \\
  \ket {{\mathfrak e}(s)} &= \frac 1 2 (\ket{\uparrow(s)} - \ket{\downarrow(s)}) \;.
\end{align}
The gap between the ground state and the third excited state,
depending on the annealing parameter $s$, goes between 10 and 8 GHz in
the region of interest. This justifies the projection into the two
lowest energy levels as long as the linear term in $\phi_x$ remains
well below this energy. 

Note that
\begin{align}
    \phi^x \frac{\partial \mathcal{H}_s(0)}{\partial \phi^x} = \Phi^x \frac {\Phi}  {L + L_\CJJ/4}\;,
\end{align}
where $\Phi^x$ is the external flux and $\Phi/( L + L_\CJJ/4 )$ is the
persistent current operator. The eigenvectors of this operator are the
flux up and down states $\ket{\uparrow(s)}, \ket{\downarrow(s)}$, with
eigenvalues $\pm I_p(s)$. This defines the persistent current $I_p(s)$. 
We denote the gap between these states by $\Delta_1(s)$. In the basis of
the up and down flux states we write
\begin{align}\label{eq:single_qubit}
  \mathcal{H}_s(0) + \phi^x \frac{\partial \mathcal{H}_s(0)}{\partial \phi^x}  \Big|_{\{\ket {{\mathfrak g}(s)},
     \ket {{\mathfrak e}(s)}\}} = - \frac 1 2 (\Delta_1(s) \sigma^x+ \epsilon_1(\phi^x) \sigma^z)\;,
\end{align}
where
\begin{align}
\epsilon_1(\phi^x) &= 2 I_p(s) \Phi^x\label{eq:epsilon_1}\\
\Phi/( L + L_\CJJ/4 ) &= I_p(s) \sigma^z\label{eq:ip_sigmaz}\;,
\end{align}
in this basis.

\subsection{Coupling between qubits}
The coupling between qubits has the form~\cite{PhysRevB.80.052506}
\begin{align}
  - J_{\mu\nu} E_\M (\phi_\mu -
  \phi^x_\mu)(\phi_\nu-\phi^x_\nu) \approx - J_{\mu\nu}E_\M \phi_\mu\phi_\nu\;,
\end{align}
where $J_{\mu\nu} \in [-1,1]$ is the dimensionless coupling.
The corresponding energy is
\begin{align}
    \frac {E_\M} {2 \pi \hbar} = \frac 1 {2 \pi \hbar} \(\frac {\Phi_0} {2\pi}\)^2 \frac
    {M_\AFM}{(L+L_\CJJ/4)^2} = 2.44 \GHz\;,
\end{align}
where in our case $M_\AFM$ is measured to be $1.41$ pico henries.

In the two level qubit Hamiltonian approximation we use $\Phi/( L + L_\CJJ/4 ) \equiv I_p(s) \sigma^z$ to write
\begin{align}
    -J_{\mu\nu}E_\M \phi_\mu\phi_\nu \approx -J_{\mu\nu} M_\AFM I_p^2(s) \sigma^z_\mu \sigma^z_\nu = -B(s) J_{\mu\nu} \sigma^z_\mu \sigma^z_\nu\;,
\end{align}
with the annealing function $B(s)$ defined as $B(s) = M_\AFM I_p^2(s)$. The
superconducting flux qubits are calibrated so that $I_p(s)$ is the
same for each of them.

\subsection{External flux phase $\phi^x$}
The value of the external flux phase $\phi^x$ controls the strength of
the local field in the single qubit Hamiltonian of
Eq.~\eqref{eq:single_qubit}. Note that $B(s)$, as defined above, scales with the persistent
current squared. This is the reason why the external field flux is
chosen (using our sign convention) as $\Phi^x = h M_\AFM I_p(s)$ so
then (see Eq.~\eqref{eq:epsilon_1})
\begin{align}
  \epsilon_1(\phi^x) = 2 h M_\AFM   I_p^2(s) = 2 h B(s)\;.
\end{align}
Here $h \in [-1,1]$ is the dimensionless value of the local field as in Eq.~\eqref{eq:h_annealing}
With this choice we write the annealing Hamiltonian as in Eq.~\eqref{eq:h_annealing}:
\begin{align}\label{eq:h_annealing_explained}
  H_0(s) &= - \frac 1 2 \Delta_1(s) \sum_\mu \sigma^x_\mu - \frac 1 2 \epsilon_1 (\phi^x) \sigma^z_\mu - \sum_{\mu\nu} J_{\mu\nu}E_\M \phi_\mu\phi_\nu\\
  &= - A(s) \sum_\mu \sigma_\mu^x   - B(s) \(\sum_\mu h_\mu \sigma_\mu^z + \sum_{\mu\nu} J_{\mu\nu} \sigma_\mu^z \sigma_\nu^z\)\;.
\end{align}

As mentioned above, the CJJ flux phase $\phi^x_\CJJ$ controls the quantum annealing
evolution. Its value $\phi^x_\CJJ(s)$ plotted in
Fig.~\ref{fig:cjj_phase} was chosen so that $I_p(s)$ scales linearly. In our case we have, for the Google-NASA D-Wave Two chip, 
\begin{align}
    M_\AFM I_p(s) \frac {2 \pi} {\Phi_0} \approx 10^{-3}(4.11 s + 1.21)\;.\label{eq:lin_ip}
\end{align}
The energy functions $A(s)$ and $B(s)$ are shown in
Fig.~\ref{fig:annealing_functions}.

\subsection{Coupling to the bath}\label{sec:coupling_to_bath}
The interaction Hamiltonian of a single qubit with the bath is
dominated by fluctuations on the flux body bias. The dimensional
interaction Hamiltonian is
\begin{align}
  \mathcal{H}_\SB = \hat I \delta \Phi_x = \frac {\hat\Phi - \Phi_x}{L} \delta \Phi_x\;.
\end{align}
Projecting into the subspace $\{\ket {{\mathfrak g}(s)}, \ket
{{\mathfrak e}(s)}\}$ as before we write (see Eq.~\eqref{eq:ip_sigmaz})
\begin{align}\label{eq:bath1}
  \mathcal{H}_\SB(s) = I_p(s) \sigma^z \delta \Phi_x = \frac 1 2 \sigma^z Q(s)
\end{align}
where
\begin{align}\label{eq:qs}
  Q(s) &= 2 I_p(s) \delta \Phi_x\;.
\end{align}
The flux bias fluctuations are measured using microscopic resonant
tunneling (MRT), as mentioned in the text.  In particular MRT is
performed at a point $s$ with small tunneling amplitude $\Delta < 1$
MHz. Under these conditions we obtain the parameters for the noise
spectral density $S_\MRT(\omega)$ which is defined in terms of a
correlation function of the bath operators $Q(s)$ through the equation
\begin{equation}
  S(\omega)_{\mu\nu} = \int_0^\infty dt e^{i\omega t} \langle
  e^{i H_Bt}Q_\mu e^{-iH_Bt} Q_\nu \rangle\;,
\end{equation}
where $\mu$ and $\nu$ are qubit's indexes.  From Eq.~\eqref{eq:qs}
\begin{align}
  \delta \Phi_x = \frac {Q_\MRT}{2 I_p(\MRT)}\;,
\end{align}
which implies
\begin{align}
  Q(s) = \frac {I_p(s)} {I_p(\MRT)} Q_\MRT \approx \frac {I_p(s)}  {I_p(1)} Q_\MRT\;.
\end{align}

This is the source of the dependence of the noise parameters on the
annealing parameter, as mentioned in the text. Note that $S(\omega)$ in Eq.~\eqref{eq:qs} is quadratic in $Q(s)$. The open system parameters $W(s)$ and $\eta(s)$ enter the theoretical model through the function 
\begin{equation}
f(\tau,s)=i \epsilon_p(s) \tau  +\frac{1}{2} W^2(s)\tau^2 -\frac{\eta}{2\pi} \ln G(\tau)\;. 
\end{equation}
as given in Eq.~\eqref{eq:fhy}. 
The relation to $S(\omega)$ is given by Eq.~\eqref{eq:f}
\begin{equation}
f(\tau,s)=\int_{-\infty}^{\infty}\frac{d\omega}{2\pi}S(\omega,s)\frac{1-e^{-i\omega t}}{(\hbar\omega)^2}\;.
\end{equation}
Consequently, we obtain Eq.~\eqref{eq:MRT}
\begin{align}
\frac{\eta(s)}{\eta_{\rm MRT}}=\left(\frac{W(s)}{W_{\rm MRT}}\right)^2=\frac{B(s)}{B(1)}\;,
\end{align}
where the last equality follows from $B(s) \propto I_p^2(s)$, and the fact that the MRT measurements are done at a point with very small $\Delta_1$, very close to $s=1$.
The values corresponding to measurements done at the D-Wave Two chip are
\begin{equation}
W_{\rm MRT}/(2\pi \hbar)=0.4\,{\rm GHz},\quad  \eta_{\rm MRT}=0.24 \;.
\end{equation}

\section{Villain representation}\label{app:villain}
In the spin basis $\ket{M,S}$ for total spin $S$, we introduce scaled
spin operators $\ss^\alpha = S^\alpha/S$ for $\alpha = x,y,z$, and $q =
M/S$ an scaled quantum number. Denote $\epsilon = 1/S$ and
\begin{align}
  \ss^z \ket q &= q \ket q \\
  \ss^{\pm} \ket q &= \sqrt{q+\epsilon - q(q \pm \epsilon)} \ket{q \pm \epsilon}\;.
\end{align}
We further introduce the canonically conjugated momentum
operator $p=-i\,\epsilon\, \frac{\partial}{\partial q}$. The Villain representation in the limit of small $\epsilon$ (big $n$) is~\cite{enz_spin_1986,boulatov_quantum_2003}
\begin{align}\label{eq:villain1}
  \ss^+ &= e^{-i p} \sqrt{1+\epsilon - q (q + \epsilon)} \\
  \ss^- &= \sqrt{1+\epsilon - q (q + \epsilon)}e^{i p} \;.\label{eq:villain2}
\end{align}
These operators are Hermitian conjugates in this representation, and we will see that they have the correct action in the coordinate representations of the wave form
\begin{align}
  \ket \Psi = \int dq \;\Psi(q) \ket q\;.
\end{align}
We will use the property
\begin{align}
  e^{- \epsilon \frac {\partial}{\partial q}} F(q) = \sum_{a = 0}^\infty \frac {(-\epsilon)^n}{n!} \frac {\partial^n}{\partial q^n} F(q) = F(q - \epsilon)\;.
\end{align}
We get
\begin{align}
  \ss^+ \ket \Psi &= \int dq\; e^{-\epsilon \frac {\partial}{\partial q}} \sqrt{1 + \epsilon - q(q+\epsilon)} \;\Psi(q) \ket q 
  = \int dq\; \sqrt{1 + \epsilon -(q-\epsilon) q} \;\Psi(q - \epsilon) \ket q \\
  &= \int dq\; \sqrt{1 + \epsilon -q(q+\epsilon)}\; \Psi(q) \ket{q+\epsilon}\;,
\end{align}
and also 
\begin{align}
  \ss^- \ket \Psi &= \int dq\; \sqrt{1 + \epsilon - q(q+\epsilon)}  e^{\epsilon \frac {\partial}{\partial q}}\;\Psi(q) \ket q\
  = \int dq\;  \sqrt{1 + \epsilon -q(q+\epsilon)}\;\Psi(q + \epsilon) \ket q \\
  &= \int dq\; \sqrt{1 + \epsilon -q(q-\epsilon)}\; \Psi(q) \ket{q-\epsilon}\;
\end{align}
Ignoring factors of order $\epsilon$ in Eqs.~\eqref{eq:villain1} and~\eqref{eq:villain2} we approximate
\begin{align}
  \ss^x = \frac 1 2 (\ss^+ + \ss^-) \approx \sqrt{1-q^2} \cos p\;.
\end{align}

We can check the adequacy of the semiclassical Hamiltonian above even
for $n=8$ from the qualitative agreement of the gap of the original
Hamiltonian and that obtained from the semiclassical Hamiltonian by
the standard instanton
method~\cite{enz_spin_1986,farhi_quantum_2002,boulatov_quantum_2003}. From
the definition of the momentum operator $p=-i\,\epsilon\,
\frac{\partial}{\partial q}$ we see that $\epsilon = 1/S$ plays the
role of $\hbar$ in the WKB approximation, and we write the WKB ansatz
for the semiclassical eigenstates $\Psi \propto \exp(W/\epsilon)$. The
semiclassical Hamiltonian is
\begin{align}
  -m(q,t)(\cos p-1) + V(q,t) \;,\label{eq:semiclassical_H}
\end{align}
where the effective $q$ dependent mass is
\begin{align}
  m(q,t) = 
A(t) \sqrt{1-q^2} / \epsilon\;.
\end{align}

Using the instanton technique, the gap can be estimated as
\begin{align}\label{eq:instanton}
  R \exp\(- \frac 1 \epsilon \int_{q_a}^{q_b} dq\,p(q)\)\;,
\end{align}
where the exponent is the Euclidean action and $p(q)$ is the instanton
trajectory between the double-well minima $q_a$ and $q_b$. The
instanton trajectory is obtained by going to imaginary time (mapping
$p\rightarrow- i p$) and solving $p(q)$ in Eq.~\eqref{eq:semiclassical_H} 
to obtain
\begin{align}
  m(q,t) \(1-\cosh p \) + V(q,t) = V(q_a,t)\;.
\end{align}
We obtain
\begin{align}
  p(q) = \cosh^{-1} \( \frac{V(q,t) - V(q_a,t)}{m(q,t)} + 1\)\;.
\end{align}
For the WKB attempt rate $R$ we use the separation between the first
and third eigenstates of the quantum Hamiltonian, $R \approx 3$ GHz,
as a proxy for the gap of the possible single well bound
states. Plugging into Eq.~\eqref{eq:instanton}, we obtain a sufficient
qualitative agreement with the exact gaps 
\begin{align}
  \begin{tabular}{c|c|c}
    $h_1$ &{\textrm exact gap} & {\textrm instanton gap}\\
    \hline  \hline
    0.48 & 10 MHz & 5 MHz \\
    0.47 & 36 MHz & 33 MHz \\
    0.46 & 78 MHz & 85 MHz \\
    \hline
  \end{tabular}
\end{align}
The agreement improves for increasing $n$~\cite{enz_spin_1986,farhi_quantum_2002,boulatov_quantum_2003}

\section{Further comparisons of larger problems that contain the weak-strong cluster ``motives'' as subproblems.}\label{sec:locally_correct}
\subsection{SVMC and PIMC-QA results compared against D-Wave results}
\begin{figure}[h!]
\centering
\includegraphics[scale=0.4]{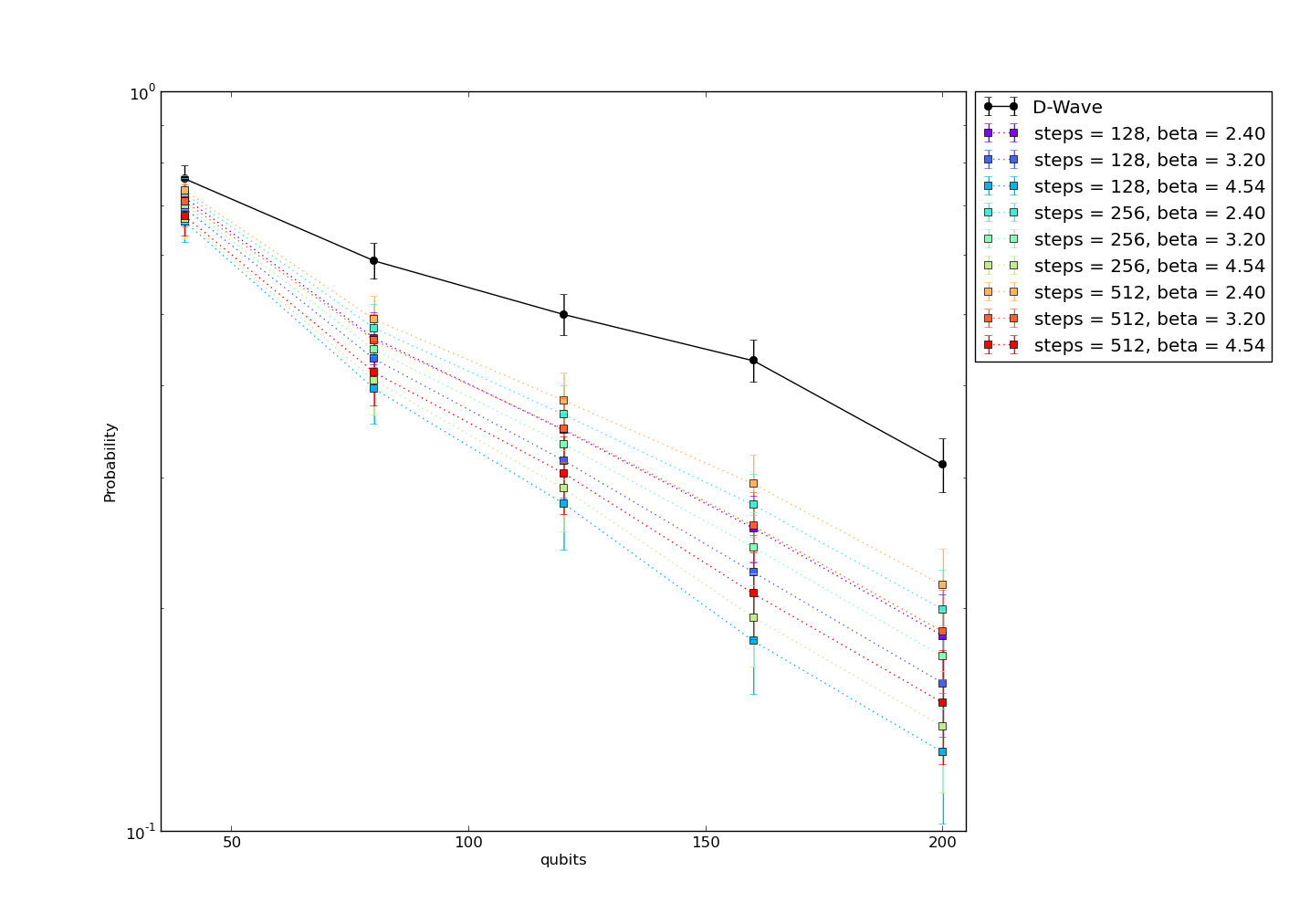}
\caption{D-Wave and SVMC (without $\chi$-correction) results for instances without strong fields.}
\label{svmc_ff0}
\end{figure}

\begin{figure}[h!]
\centering
\includegraphics[scale=0.4]{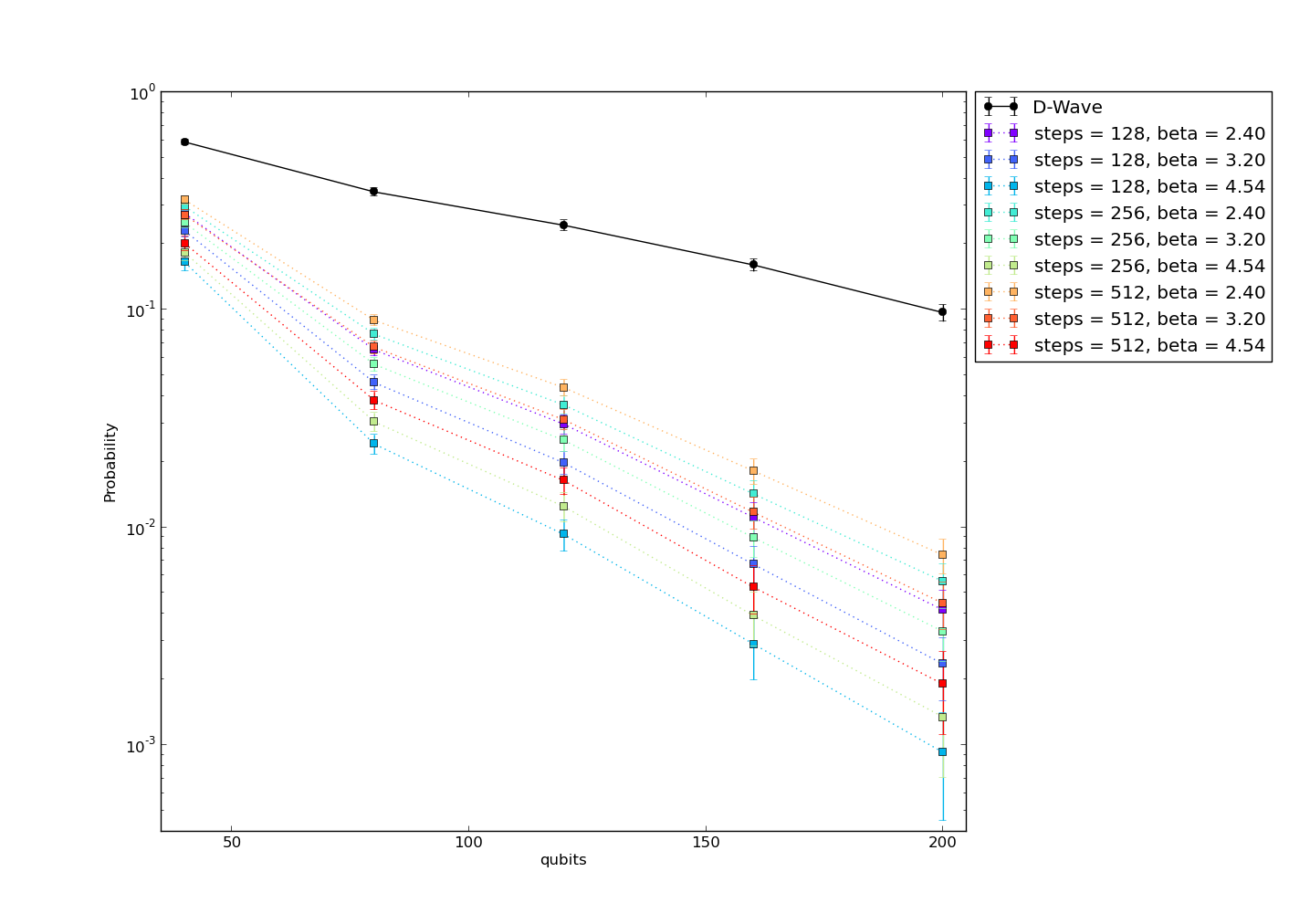}
\caption{D-Wave and SVMC (without $\chi$-correction) results for instances with strong fields.}
\label{svmc_ff1}
\end{figure}

\begin{figure}[h!]
\centering
\includegraphics[scale=0.4]{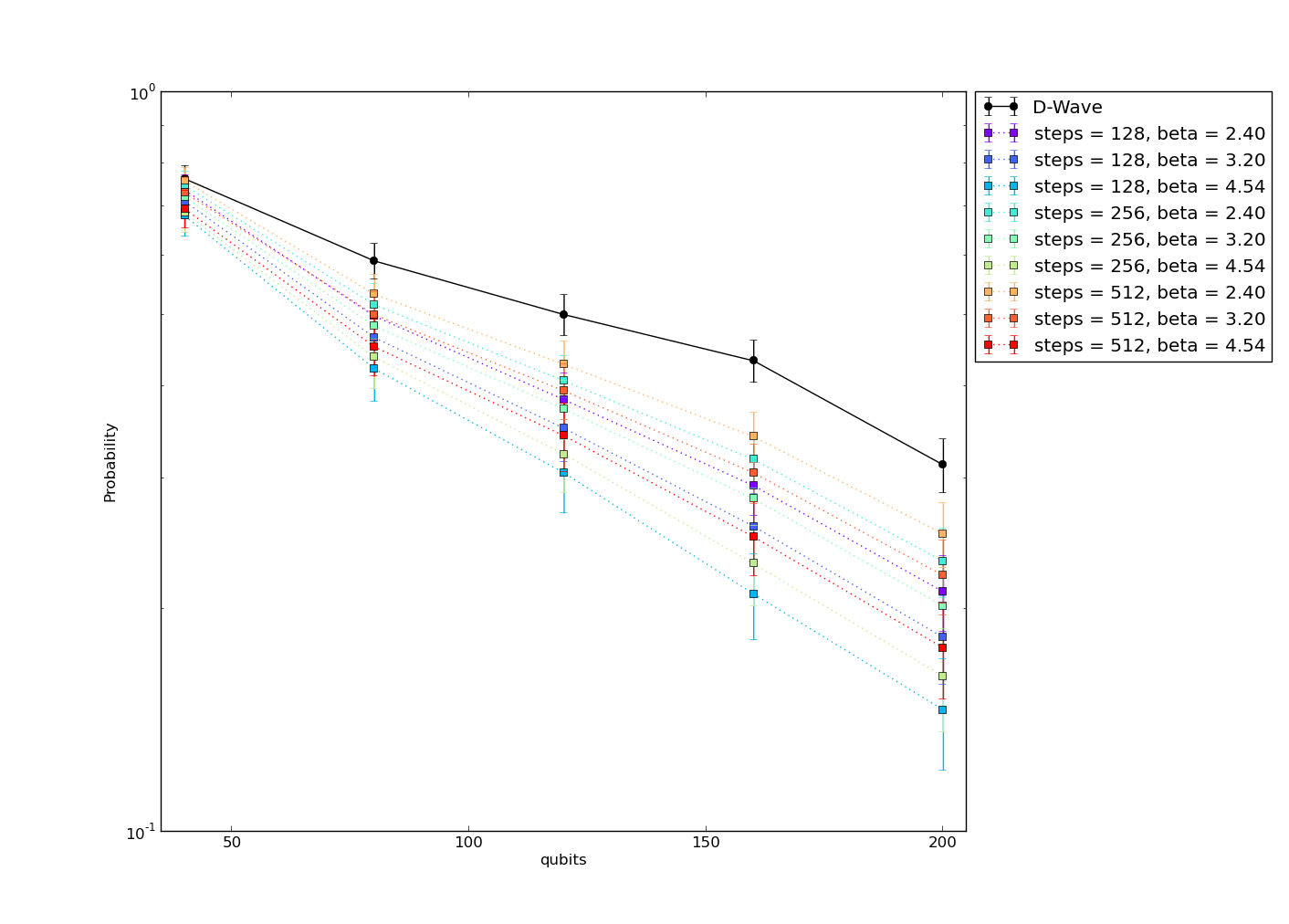}
\caption{D-Wave and SVMC (with $\chi$-correction) results for instances without strong fields.}
\label{svmc_ff0_chi}
\end{figure}

\begin{figure}[h!]
\centering
\includegraphics[scale=0.4]{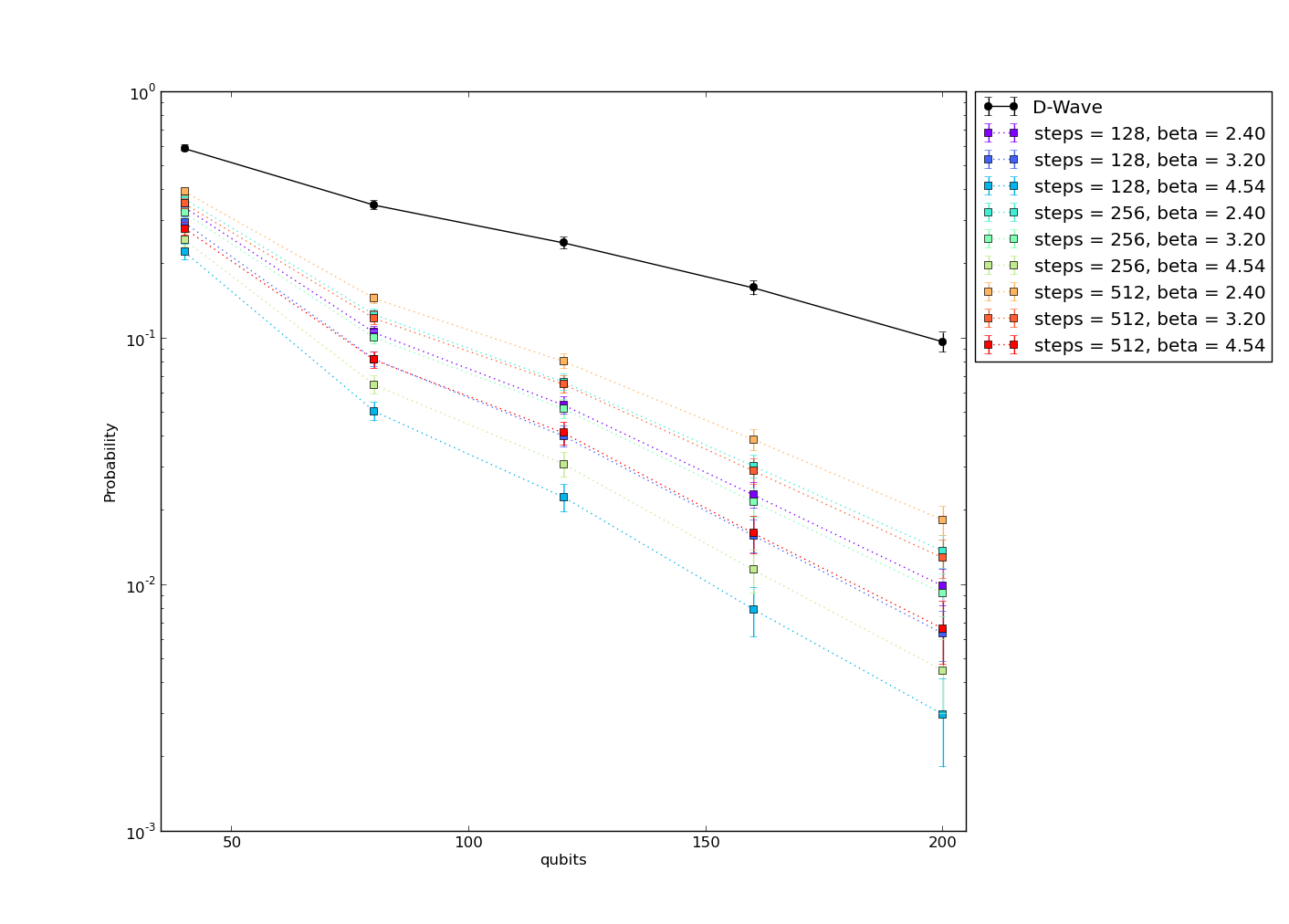}
\caption{D-Wave and SVMC (with $\chi$-correction) results for instances with strong fields.}
\label{svmc_ff1_chi}
\end{figure}

\begin{figure}[h!]
\centering
\includegraphics[scale=0.4]{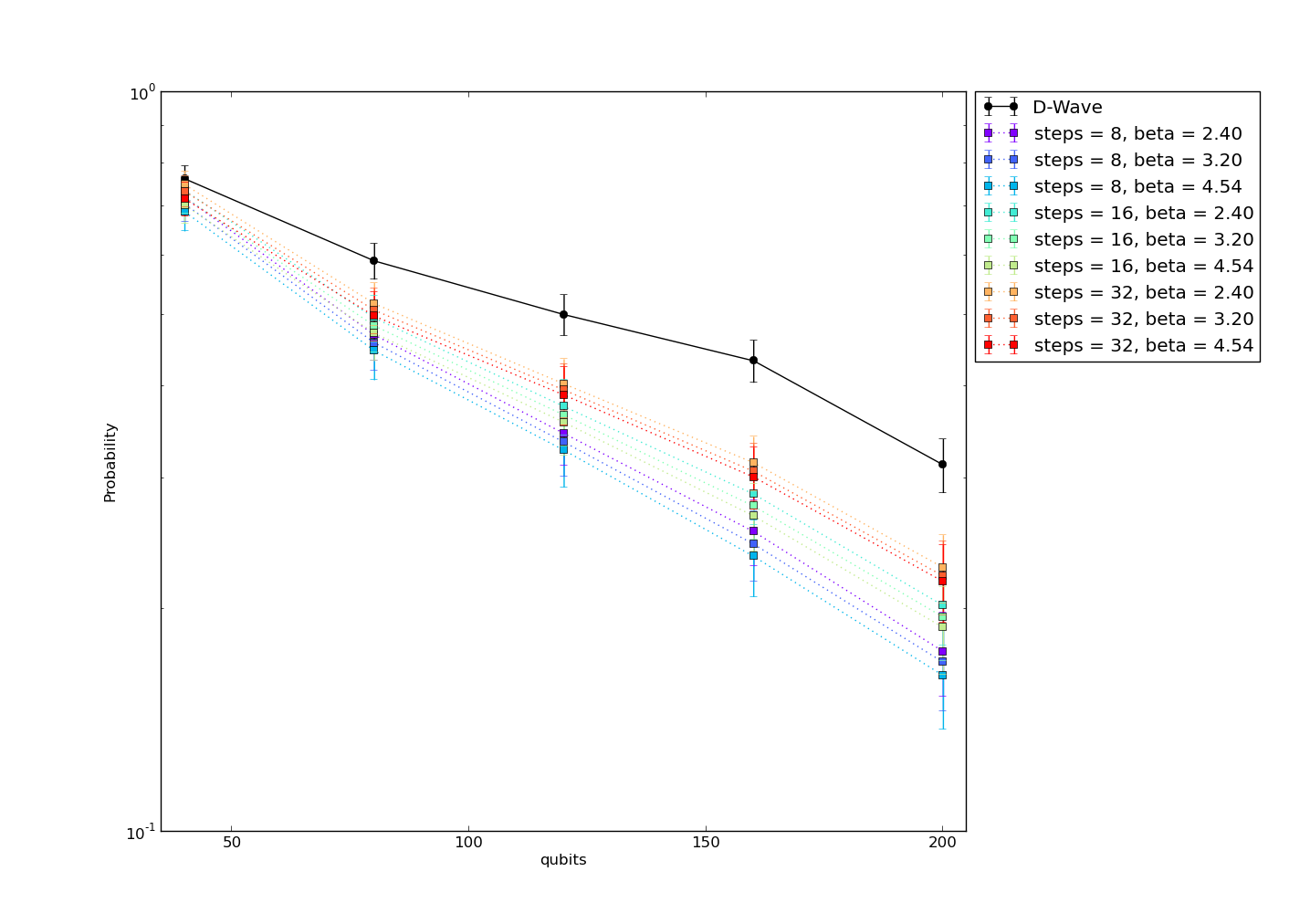}
\caption{D-Wave and PIMC-QA results for instances without strong fields.}
\label{pimc_ff0}
\end{figure}

\begin{figure}[h!]
\centering
\includegraphics[scale=0.4]{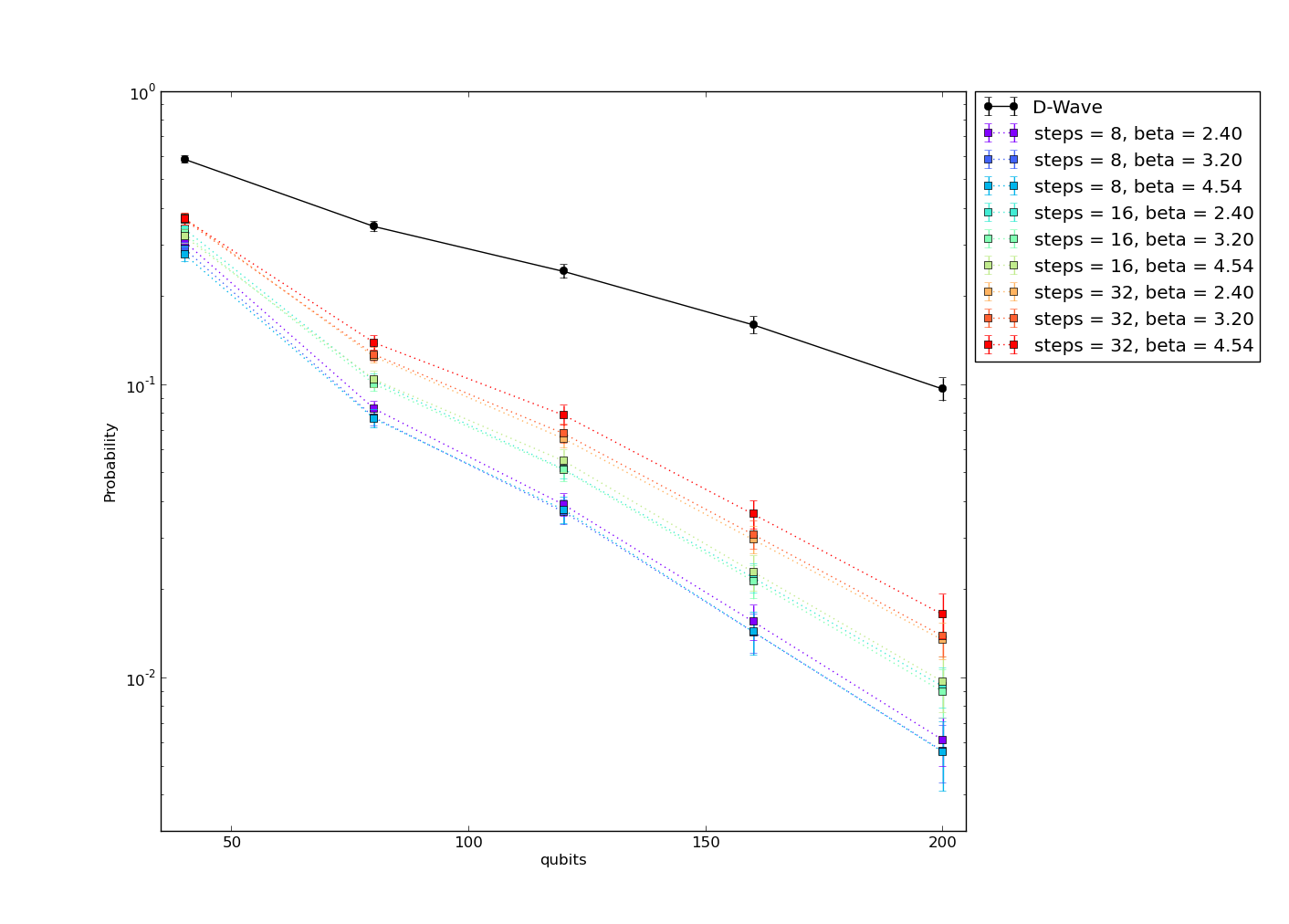}
\caption{D-Wave and PIMC-QA results for instances with strong fields.}
\label{pimc_ff1}
\end{figure}

Figs.~\ref{svmc_ff0}-\ref{pimc_ff1} show comparisons of SVMC (with and
without $\chi$-correction) and PIMC-QA against D-Wave results on the
larger problems that contain the weak-strong cluster ``motives'' as
subproblems. At each problem size (40, 80, 120, 160, and 200 spins) we
tested 100 random instances. D-Wave was executed with 16 gauges at
each instance. SVMC and PIMC-QA were each executed with 9 different
parameters settings: 3 values for $steps$ and 3 values for
$\beta$. The total number of sweeps is $1000 \times steps$. Each sweep
attempts to update all qubits or spins in the problem instance. The
plotted results were obtained by bootstrapping over the success
probabilities obtained from individual instances and the error bars
represent the bootstrapped estimate of standard error. The standard
Student's T-test ($\alpha = 0.05$) was applied to verify the
statistical significance of the difference in means between
SVMC/PIMC-QA and D-Wave results. The null hypothesis was rejected at
all parameter settings except for SVMC with $\chi$-correction and
$steps = 512$, $beta = 2.4$, $size = 5$ on the problem without strong
fields.

\subsection{Fitting curves}
In order to obtain a rough estimate of scaling behavior, we performed bootstrapped linear fits on the logs of success probabilities obtained from the same instances whose results are shown in Figs.~\ref{svmc_ff0}-\ref{pimc_ff1}.

The resulting exponential fits in linear probability space for D-Wave are: 
\begin{itemize}
\item Without strong fields: $y(x) = e^{-0.0710 \pm 0.0573} e^{(-0.0052 \pm 0.0005) x}$
\item With strong fields: $y(x) = e^{-0.1174 \pm 0.0498} e^{(-0.0110 \pm 0.0005) x}$
\end{itemize}

Additionally, Tables~\ref{table_svmc_ff0}-\ref{table_pimc_ff1} show the corresponding fitting coefficients (in log probability space) done also on the SVMC and PIMC-QA results.

\begin{table}[h!]
\centering
{\footnotesize
\begin{tabular}{| c | c | c | c |}
\hline
steps\textbackslash $\beta$ & 2.4 & 3.2 & 4.54 \\ \hline
128 & $(-0.0083 \pm 0.0008) * x - 0.0430 \pm 0.0823$ & $(-0.0091 \pm 0.0009) * x - 0.0480 \pm 0.0948$ & $(-0.0103 \pm 0.0012) * x - 0.0440 \pm 0.1146$ \\ \hline
256 & $(-0.0079 \pm 0.0008) * x - 0.0491 \pm 0.0783$ & $(-0.0086 \pm 0.0009) * x - 0.0561 \pm 0.0903$ & $(-0.0098 \pm 0.0011) * x - 0.0552 \pm 0.1102$ \\ \hline
512 & $(-0.0075 \pm 0.0007) * x - 0.0533 \pm 0.0744$ & $(-0.0082 \pm 0.0008) * x - 0.0617 \pm 0.0856$ & $(-0.0094 \pm 0.0010) * x - 0.0633 \pm 0.1050$ \\ \hline
\end{tabular}
}
\caption{SVMC without $\chi$-correction and without strong fields}
\label{table_svmc_ff0}
\end{table}

\begin{table}[h!]
\centering
{\footnotesize
\begin{tabular}{| c | c | c | c |}
\hline
steps\textbackslash $\beta$ & 2.4 & 3.2 & 4.54 \\ \hline
128 & $(-0.0255 \pm 0.0012) * x - 0.4461 \pm 0.1065$ & $(-0.0280 \pm 0.0017) * x - 0.5607 \pm 0.1471$ & $(-0.0320 \pm 0.0029) * x - 0.8125 \pm 0.2404$ \\ \hline
256 & $(-0.0242 \pm 0.0011) * x - 0.4118 \pm 0.0961$ & $(-0.0265 \pm 0.0015) * x - 0.5150 \pm 0.1300$ & $(-0.0304 \pm 0.0026) * x - 0.7377 \pm 0.2167$ \\ \hline
512 & $(-0.0229 \pm 0.0010) * x - 0.3810 \pm 0.0880$ & $(-0.0251 \pm 0.0013) * x - 0.4689 \pm 0.1157$ & $(-0.0287 \pm 0.0023) * x - 0.6755 \pm 0.1907$ \\ \hline
\end{tabular}
}
\caption{SVMC without $\chi$-correction and with strong fields}
\label{table_svmc_ff1}
\end{table}

\begin{table}[h!]
\centering
{\footnotesize
\begin{tabular}{| c | c | c | c |}
\hline
steps\textbackslash $\beta$ & 2.4 & 3.2 & 4.54 \\ \hline
128 & $(-0.0076 \pm 0.0007) * x - 0.0402 \pm 0.0729$ & $(-0.0083 \pm 0.0008) * x - 0.0495 \pm 0.0843$ & $(-0.0095 \pm 0.0010) * x - 0.0519 \pm 0.1026$ \\ \hline
256 & $(-0.0071 \pm 0.0006) * x - 0.0451 \pm 0.0682$ & $(-0.0077 \pm 0.0008) * x - 0.0567 \pm 0.0790$ & $(-0.0089 \pm 0.0009) * x - 0.0640 \pm 0.0972$ \\ \hline
512 & $(-0.0066 \pm 0.0006) * x - 0.0483 \pm 0.0639$ & $(-0.0072 \pm 0.0007) * x - 0.0618 \pm 0.0742$ & $(-0.0083 \pm 0.0009) * x - 0.0726 \pm 0.0916$ \\ \hline
\end{tabular}
}
\caption{SVMC with $\chi$-correction and without strong fields}
\label{table_svmc_ff0_chi}
\end{table}

\begin{table}[h!]
\centering
{\footnotesize
\begin{tabular}{| c | c | c | c |}
\hline
steps\textbackslash $\beta$ & 2.4 & 3.2 & 4.54 \\ \hline
128 & $(-0.0215 \pm 0.0009) * x - 0.3469 \pm 0.0830$ & $(-0.0234 \pm 0.0013) * x - 0.4257 \pm 0.1095$ & $(-0.0267 \pm 0.0021) * x - 0.6079 \pm 0.1764$ \\ \hline
256 & $(-0.0200 \pm 0.0008) * x - 0.3166 \pm 0.0744$ & $(-0.0217 \pm 0.0011) * x - 0.3830 \pm 0.0962$ & $(-0.0248 \pm 0.0018) * x - 0.5419 \pm 0.1535$ \\ \hline
512 & $(-0.0187 \pm 0.0007) * x - 0.2868 \pm 0.0672$ & $(-0.0202 \pm 0.0010) * x - 0.3438 \pm 0.0857$ & $(-0.0229 \pm 0.0015) * x - 0.4808 \pm 0.1327$ \\ \hline
\end{tabular}
}
\caption{SVMC with $\chi$-correction and with strong fields}
\label{table_svmc_ff1_chi}
\end{table}

\begin{table}[h!]
\centering
{\footnotesize
\begin{tabular}{| c | c | c | c |}
\hline
steps\textbackslash $\beta$ & 2.4 & 3.2 & 4.54 \\ \hline
8 & $(-0.0086 \pm 0.0008) * x - 0.0186 \pm 0.0795$ & $(-0.0088 \pm 0.0008) * x - 0.0346 \pm 0.0860$ & $(-0.0089 \pm 0.0009) * x - 0.0522 \pm 0.0934$ \\ \hline
16 & $(-0.0079 \pm 0.0007) * x - 0.0284 \pm 0.0730$ & $(-0.0080 \pm 0.0008) * x - 0.0451 \pm 0.0791$ & $(-0.0080 \pm 0.0008) * x - 0.0645 \pm 0.0859$ \\ \hline
32 & $(-0.0072 \pm 0.0006) * x - 0.0353 \pm 0.0674$ & $(-0.0073 \pm 0.0007) * x - 0.0520 \pm 0.0724$ & $(-0.0072 \pm 0.0007) * x - 0.0746 \pm 0.0778$ \\ \hline
\end{tabular}
}
\caption{PIMC-QA without strong fields}
\label{table_pimc_ff0}
\end{table}

\begin{table}[h!]
\centering
{\footnotesize
\begin{tabular}{| c | c | c | c |}
\hline
steps\textbackslash $\beta$ & 2.4 & 3.2 & 4.54 \\ \hline
8 & $(-0.0238 \pm 0.0010) * x - 0.3787 \pm 0.0906$ & $(-0.0241 \pm 0.0012) * x - 0.4207 \pm 0.1063$ & $(-0.0239 \pm 0.0014) * x - 0.4530 \pm 0.1232$ \\ \hline
16 & $(-0.0219 \pm 0.0009) * x - 0.3369 \pm 0.0792$ & $(-0.0220 \pm 0.0010) * x - 0.3587 \pm 0.0904$ & $(-0.0214 \pm 0.0012) * x - 0.3822 \pm 0.1036$ \\ \hline
32 & $(-0.0202 \pm 0.0008) * x - 0.3012 \pm 0.0705$ & $(-0.0200 \pm 0.0009) * x - 0.3122 \pm 0.0772$ & $(-0.0190 \pm 0.0009) * x - 0.3147 \pm 0.0850$ \\ \hline
\end{tabular}
}
\caption{PIMC-QA with strong fields}
\label{table_pimc_ff1}
\end{table}

\newpage
Based on these fits, the ratios between scalings of SVMC/PIMC-QA and corresponding scalings of D-Wave were computed and are summarized as follows:

\begin{itemize}
\item Without strong fields
\begin{itemize}
\item Ratios between SVMC without $\chi$-correction and D-Wave
\begin{itemize}
\item Min: 1.4423 (at $steps = 512$, $\beta = 2.4$)
\item Max: 1.9808 (at $steps = 128$, $\beta = 4.54$)
\end{itemize}
\item Ratios between SVMC with $\chi$-correction and D-Wave
\begin{itemize}
\item Min: 1.2692 (at $steps = 512$, $\beta = 2.4$)
\item Max: 1.8269 (at $steps = 128$, $\beta = 4.54$)
\end{itemize}
\item Ratios between PIMC-QA and D-Wave
\begin{itemize}
\item Min: 1.3846 (at $steps = 32$, $\beta = 2.4$)
\item Max: 1.7115 (at $steps = 8$, $\beta = 4.54$)
\end{itemize}
\end{itemize}
\item With strong fields
\begin{itemize}
\item Ratios between SVMC without $\chi$-correction and D-Wave
\begin{itemize}
\item Min: 2.0818 (at $steps = 512$, $\beta = 2.4$)
\item Max: 2.9091 (at $steps = 128$, $\beta = 4.54$)
\end{itemize}
\item Ratios between SVMC with $\chi$-correction and D-Wave
\begin{itemize}
\item Min: 1.7000 (at $steps = 512$, $\beta = 2.4$)
\item Max: 2.4273 (at $steps = 128$, $\beta = 4.54$)
\end{itemize}
\item Ratios between PIMC-QA and D-Wave
\begin{itemize}
\item Min: 1.7273 (at $steps = 32$, $\beta = 4.54$)
\item Max: 2.1909 (at $steps = 8$, $\beta = 3.2$)
\end{itemize}
\end{itemize}
\end{itemize}

\section{Chi probe for SVMC}\label{app:chi_probe}

The single qubit Hilbert space modeled as a spin vector in the SVMC numerics is obtained from the two lowest energy wave functions of the continuous flux qubit Hamiltonian with zero flux body bias (see App.~\ref{app:single_qubit_hamiltonian}). For sufficiently high flux body bias, these two wavefunctions mix with higher energy wavefunctions of the continuous flux qubit Hamiltonian. We have checked that, up to the freezing point, the flux bias remains low for the problems that we study in this paper. Nevertheless, a model introduced by D-Wave to deal with this error has been treated as a fitting parameter for SVMC numerics in previous works~\cite{vinci2014distinguishing,albash2014reexamining}. Our own derivation of this model gives the following equations modifying the couplings and local fields of the original Hamiltonian 
\begin{align}
  h'_i &= h_i - \chi \sum_j J_{ij} h_j \\
  J_{ij}' &= J_{ij} - 2 \chi \sum_{k} J_{ik} J_{kj}\label{eq:cross_talk}\;.
\end{align}
While these equations are slightly different from those used in other
works~\cite{vinci2014distinguishing,albash2014reexamining}, their
effect is the same for the problems under study here. Explicitly, the
problem becomes more ferromagnetic, and this has the effect of
decreasing the barrier height for $h_1 < J/2$. 

\begin{figure}[ht]
  \centering
  \subfloat[Chi-probe problem]{\includegraphics[width=.3\textwidth]{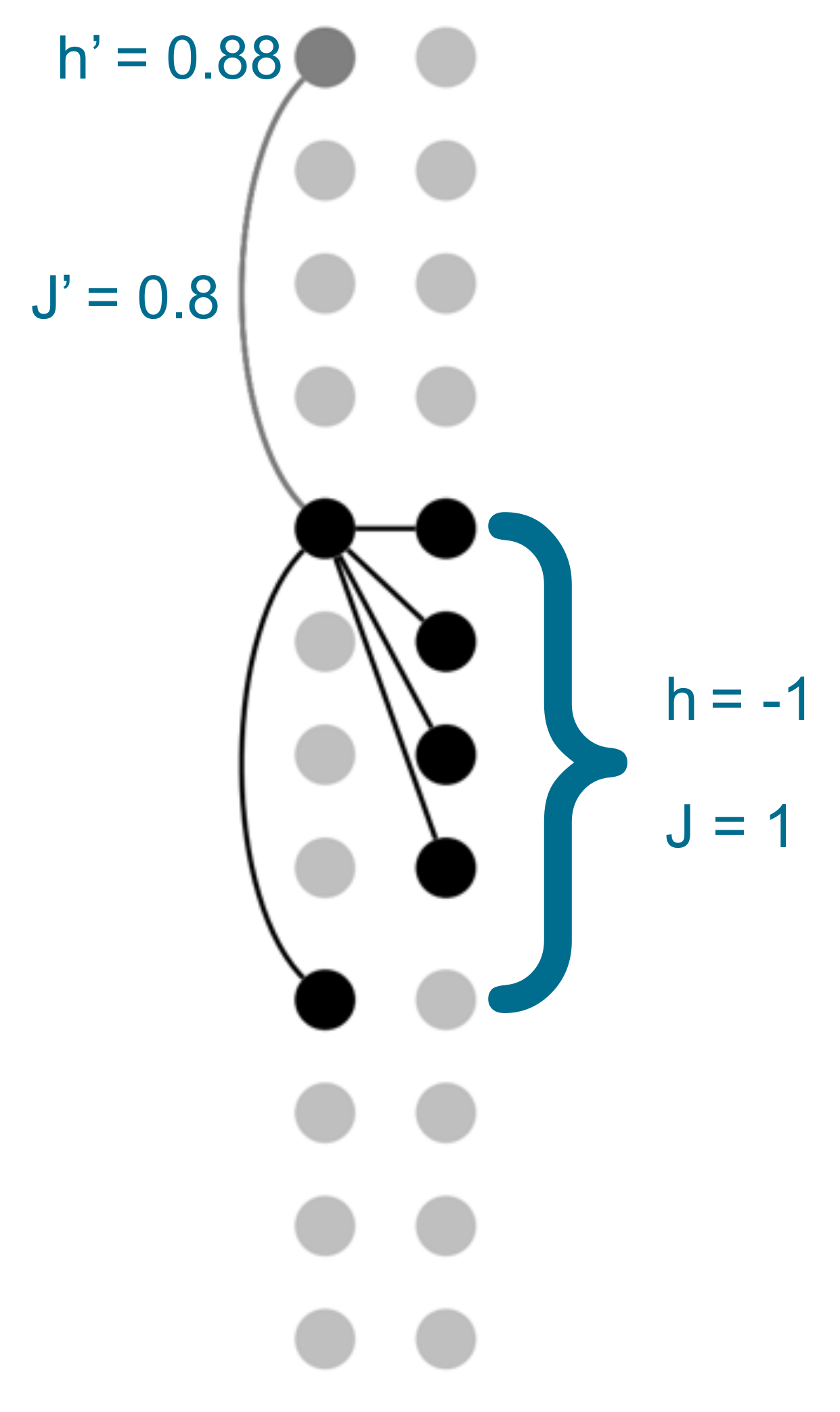}\label{fig:chi_probe_problem}}
  \subfloat[Chi-probe probabilities]{\includegraphics[width=.7\textwidth]{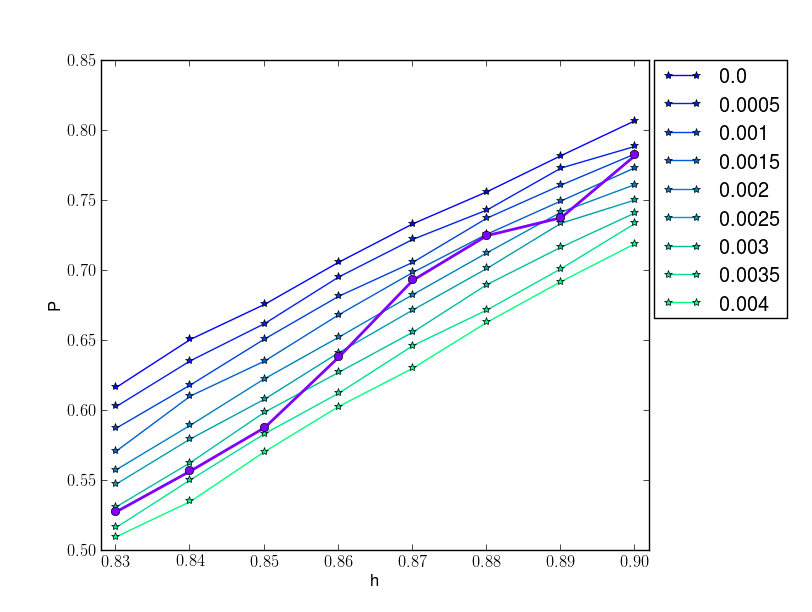}\label{fig:chi_probe_probabilities}}
  \caption{Problem to constraint values of $\chi$ compatible with SVMC. (a) Layout of the problem. (b) D-Wave data success probabilities (thicker line with $\circ$ markers) and SVMC success probability for the ``chi-probe'' problems. The lines with $\star$ markers correspond to SVMC with different values of $\chi$, from 0 to 0.004. The minimum residual error is found for $\chi=0.0025$. We use SVMC with 125K sweeps and T=15 mK, as explained in the main text.}
  \label{fig:chi_probe}
\end{figure}

We want to constrain the possible values of $\chi$ to be consistent with
SVMC when $\chi$ is treated as a fitting parameter.  To this effect we
introduce a ``chi-probe'' problem related to the weak-strong
cluster motif, but without a multi-spin energy barrier. We also
introduce many extra nearest-neighbor ferromagnetic couplings to
increase the sensitivity cross-talk of
Eq.~\eqref{eq:cross_talk}. Figure~\ref{fig:chi_probe_problem} shows
the layout of this problem. Figure~\ref{fig:chi_probe_probabilities}
shows the D-Wave data success probabilities and SVMC success
probabilities for the ``chi-probe'' problems. The minimum residual error
is found for $\chi=0.0025$, and we use this value of $\chi$ in the
main text when appropriate. Because our factor of 2 in
Eq.~\eqref{eq:cross_talk} this value of $\chi$ is roughly equivalent
to a value of $0.05$ for the equations used in
Ref.~\cite{vinci2014distinguishing}. 

\begin{figure}[ht]
  \centering  
  \includegraphics[width=.5\textwidth]{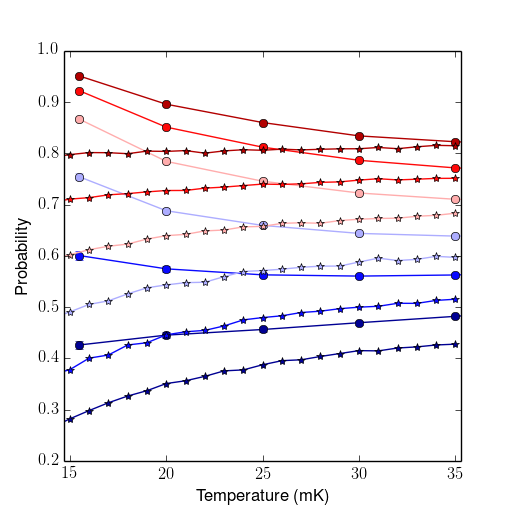}
  \caption{Probability of success versus temperature for D-Wave data
    ($\circ$ markers) and SVMC numerics ($\lozenge$ markers). We plot
    (from top to bottom, and red to blue) $h_1 = [0.38, 0.4, 0.42,
    0.44, 0.46, 0.48]$. Error bars are smaller than markers. We use
    SVMC with $\chi=0.0025$ and 15 mK algorithmic temperature for 128,000 sweeps, as
    explained in the text.}
  \label{fig:dwave_p_vs_t_chi}
\end{figure}

\begin{figure}[ht]
  \centering
  \subfloat[Probability vs. T for $h_1 = 0.44$.]{\includegraphics[width=.5\textwidth]{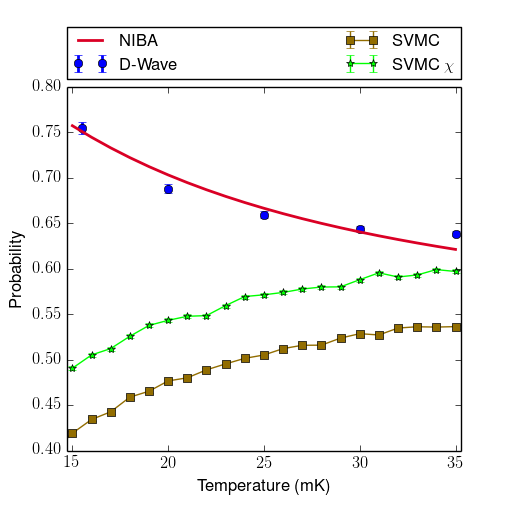}\label{fig:p_vs_t_h1_0p44_all_chi}}
  \subfloat[Probability vs. $h_1$.]{\includegraphics[width=.5\textwidth]{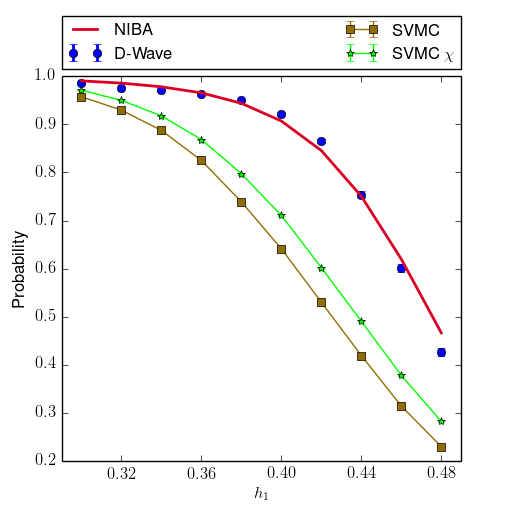}\label{fig:p_vs_h1_all_chi}}
  \caption{Plots including SVMC with $\chi=0.0025$ for the double-well
    potential problem introduced in the main
    text. (a) Probability of success
    versus temperature at $h_1 = 0.44$ for D-Wave (purple $\circ$
    marker), the NIBA Quantum Master Equations (continuous red line)
    and the classical-path model (SVMC). The two SVMC curves
    correspond to SVMC (brown $\square$ marker) and SVMC with
    $\chi=0.0025$ (green $\star$ marker). (b)
    Probability of success versus $h_1$. Error bars are smaller than
    markers. }
\end{figure}

Figure~\ref{fig:dwave_p_vs_t_chi} shows a comparison of D-Wave data
and SVMC numerics with $\chi=0.0025$ (compare with
Fig.~\ref{fig:dwave_p_vs_t}). As explained in the text, we choose 128,000
sweeps for an algorithmic temperature of 15 mK. Crucially, the
temperature dependence is still the opposite for SVMC with $\chi$ than
for the D-Wave data, as expected. Figure~\ref{fig:p_vs_t_h1_0p44_all_chi}
shows the success probability as a function of temperature for D-Wave,
open system quantum numerics and the classical-path model (SVMC) for
$h_1 = 0.44$ in the Hamiltonian of Eq.~\eqref{eq:Hp_total}. We include
SVMC with $\chi=0.0025$. The probability of success is higher for SVMC
with than without $\chi$. The reason is that the problem modified with
$\chi$ is more ferromagnetic: it has an effectively lower ratio $h_1 /
J$. Nevertheless, the probability of success for SVMC with $\chi$ is
still lower than the probability of success for
D-Wave. Figure~\ref{fig:p_vs_h1_all_chi} shows the probability of success
versus $h_1 = [0.3, ..,0.48]$.

\twocolumngrid
\bibliography{experimental_tunneling}

\end{document}